\documentclass[twocolumn,showpacs,preprintnumbers,amsmath,amssymb,nofootinbib]{revtex4}

\usepackage{graphicx}

\def\met{\mbox{$E_T^{\rm{miss}}$}}

\newcommand{\SumPt}{\mbox{$\sum {p}_{T} $}}
\def\to{\rightarrow}

\def\et{\mbox{$E_T$}}
\newcommand{\nn}{\nonumber}
\newcommand{\be}{\begin{eqnarray}}
\newcommand{\ee}{\end{eqnarray}}

\newcommand{\gsim}{\mathrel{\raisebox{-.6ex}{$\stackrel{\textstyle>}{\sim}$}}}
\newcommand{\smur}{\tilde{\mu}_R}
\newcommand{\smul}{\tilde{\mu}_L}
\newcommand{\smurbar}{\bar{\tilde{\mu}}_R}
\newcommand{\smulbar}{\bar{\tilde{\mu}}_L}
\newcommand{\sq}{\tilde{q}}
\newcommand{\sqr}{\tilde{q}_R}
\newcommand{\sql}{\tilde{q}_L}
\newcommand{\sqbar}{\bar{\tilde{q}}}
\newcommand{\sqrbar}{\bar{\tilde{q}}_R}
\newcommand{\sqlbar}{\bar{\tilde{q}}_L}
\newcommand{\pslash}{\not{\hbox{\kern-2pt p}}}
\newcommand{\ppslash}{\not{\hbox{\kern-2pt p}}'}
\newcommand{\kslash}{\not{\hbox{\kern-2pt k}}}
\newcommand{\kpslash}{\not{\hbox{\kern-2pt k}}'}
\newcommand{\tdot}{\hspace{-2pt}\cdot\hspace{-2pt}}
\def\Dslash{\not{\hbox{\kern-4pt $D$}}}
\def\min#1{{\rm min} \kern -26pt \lower 15pt\hbox{$\scriptstyle #1$}\;}
\newcommand{\sig}{$\sigma\hspace{4pt}$}
\renewcommand{\toprule}{\hline}
\newcommand{\midrule}{\hline}
\newcommand{\bottomrule}{\hline}

\begin{document}

\preprint{FERMILAB-PUB-08-012-T}
\preprint{ANL-HEP-PR-08-30}

\title{Missing energy look-alikes with 100 pb$^{-1}$ at the LHC}

\author{Jay Hubisz}\email{hubisz@hep.anl.gov}
\affiliation{High Energy Physics Division,
Argonne National Laboratory,
Argonne, IL 60439, USA}

\author{Joseph Lykken}\email{lykken@fnal.gov}
\affiliation{Fermi National Accelerator Laboratory,
P.O. Box 500, Batavia, IL 60510, USA}

\author{Maurizio Pierini}\email{Maurizio.Pierini@cern.ch}
 
\author{Maria Spiropulu}\email{smaria@cern.ch}

\affiliation{Physics Department, CERN,
CH 1211 Geneva 23, Switzerland}

\date{\today}
\begin{abstract} 
A missing energy discovery is possible at the LHC with the first 100 pb$^{-1}$ 
of understood data. We present a realistic strategy to rapidly narrow the 
list of candidate theories at, or close to, the moment of discovery. The
strategy is based on robust ratios of inclusive counts of simple
physics objects. We study specific cases showing
discrimination of look-alike models in simulated data sets that
are at least 10 to 100 times smaller than used in previous studies. We discriminate
supersymmetry models from non-supersymmetric look-alikes with only 100 pb$^{-1}$
of simulated data, using combinations of observables that trace back to differences 
in spin.
\end{abstract}

\pacs{11.30.Pb,14.80.Ly,12.60.Jv} 

\maketitle

%

\section{Introduction}\label{sec:1}

\subsection{Twenty questions at the LHC}

Many well-motivated theoretical frameworks make dramatic
predictions for the experiments at the Large Hadron Collider. 
These frameworks are generally based upon assumptions about 
new symmetries, as is the case for supersymmetry (SUSY)
 \cite{Wess:1973kz,Ramond:1971gb} and  little Higgs 
(LH)\cite{ArkaniHamed:2001nc,ArkaniHamed:2002qx},  or upon 
assumptions about new degrees of freedom  such as extra
 large \cite{ArkaniHamed:1998rs} or warped \cite{Randall:1999ee} spatial
dimensions. Within each successful framework, one can construct a
large number of qualitatively different models consistent with all
current data. Collectively these models populate the ``theory space''
of possible physics beyond the Standard Model.  The BSM theory space
is many dimensional, and the number of distinct models within it is
formally infinite. 
Since the  data  will not provide a distinction between models that
differ by sufficiently tiny or experimentally irrelevant details,
infinity, in practice, becomes some large finite number $N$. The mapping of
these $N$ models into their experimental signatures at the LHC, though
still incomplete, has been explored in great detail.

As soon as discoveries are made at the LHC, physicists will face the
LHC Inverse Problem: given a finite set of measurements with finite
resolutions, how does one map back
 \cite{Binetruy:2003cy}-\cite{ArkaniHamed:2005px} 
to the underlying theory responsible for the new phenomena?  
So far, not enough progress has been made on this problem,
especially as it relates to the immediate follow-up of an
early LHC discovery.

When $N$ is large,
it is not a viable strategy to discriminate between $N$ alternative
explanations by performing $N$ tests. However, as the game ``twenty
questions'' illustrates, a well-designed series of simple tests can
identify the correct alternative in of order log($N$) steps,
proceeding along a decision tree such that, at each branching, of
order half of the remaining alternatives are eliminated.

Addressing the LHC Inverse Problem implies designing and implementing
this series of simple tests in the LHC experiments, so that with high
confidence a significant fraction of the remaining theory space is ruled out at
each step. The results of the first few tests will shape the
requirements for future tests, so the immediate need is to develop the
strategy for the early tests. In this paper we provide this strategy for
the case of an LHC discovery in the inclusive missing energy
signature.

\subsection{Missing energy at the LHC}
The existence of dark matter provides a powerful motivation to explore
missing energy signatures at the LHC, under the 
assumption that a significant fraction of dark matter may consist of
weakly interacting thermal relics. Missing energy at the LHC is
experimentally challenging. Most of the energy of 14 TeV $pp$
collisions is carried off by undetected remnants of the underlying
event, so missing energy searches actually look for missing transverse
energy (\met) of the partonic subprocess. \met~searches are plagued
by instrumental and spurious backgrounds, including cosmic rays,
scattering off beam halo and jet mismeasurement.  Standard Model
processes create an irreducible \met~background from processes such as
the $Z$ boson decay to neutrinos and $t\bar{t}$ production followed by
semileptonic decays of the top.  

In many theoretical frameworks with dark
matter candidates, there are heavy strongly interacting particles with
the same conserved charge or parity that makes the dark matter
particle stable. These colored particles will be pair-produced at the
LHC with cross sections roughly in the range 0.1 to 100 pb. Their
subsequent decays will produce Standard Model particles along with a
pair of undetected dark matter particles. Thus the generic
experimental signature is both simple and inclusive: 
large \met~accompanied by multiple energetic jets.  
A detailed strategy for early
discovery with the inclusive \met~signature was presented in the CMS
Physics Technical Design Report
\cite{Ball:2007zza}-\cite{Spiropulu:2008ug} and studied with full
simulation of the CMS detector. After a series of cleanup and analysis
cuts on a simulated \met~trigger sample targeting the reduction of
the instrumental and physics backgrounds, the signal efficiency
remained as high as 25\%. These results indicate that, for signal
cross sections as low as a few pb, an \met~discovery could be made
with the first 100 pb$^{-1}$ of understood LHC data\footnote{The first
100 pb$^{-1}$ of understood LHC data will not be the first 100 pb$^{-1}$
of data written to tape. The 10 TeV
data collected in the early running will be used for calibrations and
understanding of benchmark Standard Model processes.}.
In our study we
assume as the starting point that a greater than $5\sigma$ excess of events will
be seen in a 14 TeV LHC data sample of 100 pb$^{-1}$ with an inclusive missing
energy analysis.  For invariability and comparability we effectively
adopt the full analysis path and requirements used in
\cite{Ball:2007zza}.

\subsection{Look-alikes at the moment of discovery}
At the moment of discovery a large number of theory models will be
immediately ruled out because, within conservative errors, they give
the wrong excess.  However a large number of models will remain as
missing energy look-alikes, defined as models that predict the same
inclusive missing energy excess, within some tolerance, in the same
analysis in the same detector, for a given integrated luminosity. The
immediate challenge is then to begin discriminating the look-alikes.

The look-alike problem was studied in \cite{ArkaniHamed:2005px} as it
might apply to a later mature phase of the LHC experiments.  Even
restricted to the slice of theory space populated by a partial
scan of the MSSM, ignoring SM backgrounds and systematic errors, 
and applying an uncorrelated $\chi^2$-like statistical analysis to 
1808 correlated observables, this study found that a large number 
of look-alikes remained unresolved in a simulation equivalent to 10 fb$^{-1}$.
A more recent analysis \cite{Berger:2007ut} attempts to resolve these
look-alikes in a simulation of a future linear collider.

At the moment of an early discovery the look-alike problem will be
qualitatively different. The data samples will be much smaller, with a
limited palette of robust reconstructed physics objects.  For example,
$\tau$ or $b$ tagging in multijet final states will be in development 
during the 100 pb$^{-1}$ era. In many small data samples 
peaks and edges in invariant mass distributions
may not be visible, and most observables related to detailed features
of the events will be rate limited. The observables that \textit{are}
available to discriminate the look-alikes in the very early running
will be strongly correlated by physics and systematics making it 
imprudent to combine them in a multivariate analysis.

\subsection{Is it SUSY?}
By focusing on the discrimination of look-alikes, we are pursuing a
strategy of simple binary choices: is Model A a significantly better
explanation of the discovery data set than Model B? Each answer carries
with it a few bits of important fundamental information about the
new physics process responsible for missing energy. Obviously we will
need to make many distinct look-alike comparisons before we can hope
to build up a clear picture from these individual bits.

Consider how this strategy might play out for answering the basic
question ``is it SUSY?'' It may not be possible to
answer this question conclusively during the 100 pb$^{-1}$ era.  
Our strategy will consist of asking a series of more modest 
questions, some of them of the
form: ``does SUSY Model A give a significantly better explanation of
the discovery data set than non-SUSY Model B?''  None of these
individual bits of information by itself is equivalent to answering
``is it SUSY?''  However we demonstrate that we can build up a
picture from the data that connects back to features of the underlying theory.

Furthermore, we demonstrate a concrete method to obtain indirect
information about the spin of the new particles.  We establish how to
discriminate between a non-SUSY model and its SUSY look-alikes.
Even though we cannot measure the spins of the exotic particles 
directly, spin has significant effects on production cross sections, 
kinematic distributions and signal efficiencies. 
We are thus able to discriminate SUSY from non-SUSY using
combinations of observables that trace back to differences in spin.
Our study shows that in favorable cases this
can be accomplished with data sets as small as 100 pb$^{-1}$.

\subsection{Outline}
In section \ref{sec:2} we review in detail the missing energy discovery path, 
including the experimental issues and systematics that limit our ability 
to fully reconstruct events from the discovery data set. 
We explain how the missing energy signals are simulated,
and the uncertainties associated with these simulations.  
In section \ref{sec:3} we discuss the problem of populating
the parts of the theory space relevant to a particular missing energy
discovery. In section
\ref{sec:4} we introduce two groups of look-alike models
relative to two different missing energy signals. For models
differing only by spins, we discuss how cross
sections, kinematic distributions  and efficiencies can be used to distinguish them, 
drawing from formulae developed in the Appendix.
In section
\ref{sec:5} we define all of the robust observables that we use to
discriminate among the look-alikes, and in section \ref{sec:6}
we describe the look-alike analysis itself and how we compute the
significance of the discriminations. Sections \ref{sec:7} and 
\ref{sec:8} give a summary of our results, with details relegated 
to further Appendices. Finally section
\ref{sec:9} describes the steps we are following to improve this analysis for
use with real data.

\section{Discovery analysis for missing energy}\label{sec:2}
Missing energy hadron collider data has been used previously for successful
measurements of Standard Model processes with energetic neutrinos;
these include the\ $Z^{0}$ boson invisible decay rates,
the top quark cross section, searches for 
the Higgs boson~\cite{Acosta:2005ga} and a precise extraction of the $W$ mass 
from the reconstruction of the $W$ transverse mass \cite{Abe:1995np}.
Pioneering searches for new phenomena in missing energy data sets at the
Tevatron~\cite{Spiropulu:2000nb}-\cite{Affolder:2000ef}  led to the 
development and understanding of the basic techniques that will be used 
in missing energy searches at the LHC.

In an ideal detector, with hermetic 4$\pi$ solid angle coverage and
excellent calorimeter resolution, the measurement of missing energy
is the measurement of the neutrino energy and the energy of any other
neutral weakly interacting particles. In a real detector it is also 
a measurement of the energy that escapes detection due to
uninstrumented regions and other detector effects such as imperfect 
calorimeter response. 
Muons are sources of missing energy since a muon
typically deposits only of order
a few GeV of its total energy in the calorimeters\footnote{The energy
loss of muons is mostly due to ionization up to muon
energies of 100 GeV. Above 100 GeV bremsstrahlung and nuclear losses 
can cause a single ``catastrophic'' energy loss comparable to the
total muon energy.}.
QCD jets produce real \met~from semileptonic decays
of heavy flavor, and fake \met~from detector-induced mismeasurements.
Thus the \met~ distribution of a pure QCD multijet sample
has a long tail related to non-Gaussian tails in the detector response.
This gives rise to an important background to missing energy searches that is
difficult to estimate prior to data. At the Tevatron it has been 
shown that this background can be brought under control by exploiting 
the fact that the fake \met~from jet mismeasurements is highly correlated
with the azimuthal directions of the leading jets~\cite{Spiropulu:2000nb}.

There are other important sources of fake \met~at hadron colliders,
including beam halo induced \met , cosmic ray muons,
noise in the data acquisition system, and misreconstruction of
the primary vertex. Eliminating these sources requires unbiased
filters based on clean definitions of event quality.

To design a missing energy analysis, we need to have some idea
of the source of the \met~in the signal. The possibilities
include:
\begin{itemize}
\item The \met~is entirely from neutrinos. This could 
arise from the direct decay of new heavy particles to neutrinos,
or decays of new heavy particles to top, $W$'s, $Z$'s or $\tau$'s. 
One appropriate discovery strategy for this case is to 
look for anomalies in the energetic tails of data sets 
with reconstructed top, $W$'s or $Z$'s.
\item The \met~originates from a single weakly interacting exotic particle
in the final state. An example of this possibility is graviton
production in models with large extra dimensions~\cite{Hewett:2002hv}.
If strong production occurs, the signal will consist predominately
of monojets and large \met . Successful analyses for this case were
carried out at the Tevatron~\cite{Acosta:2003tz,Abazov:2003gp}.
Other signals that fit this case arise from unparticle 
models~\cite{Cheung:2007zza}
and from models with $s$-channel resonances that have invisible decays.
\item The \met~originates from many weakly interacting exotic particles.
This can be the case in hidden valley models~\cite{Strassler:2006im},
where the weakly interacting exotics are light pions of the hidden sector.
This case is experimentally challenging.
\item The \met~originates from two weakly interacting exotic particles
in the final state. This is the case for supersymmetry models
with conserved $R$ parity, where the weakly interacting particles
are neutralino LSPs. It also applies for more generic models
with WIMP dark matter candidates. 
\end{itemize}

We focus on a discovery analysis developed for the last case.
Thus we are interested in signal events with two heavy WIMPs in
the final state. For early discovery at the LHC, the signal events
should have strong production cross sections; we will assume that
each WIMP arises from the decay of a strongly interacting heavy
parent particle. The most generic signature is therefore large
\met~in association with at least two high $E_T$ jets. There will be
additional jets if the WIMP is not produced in a 2-body decay of
the parent particle. Furthermore, there is a
significant probability of an extra jet from QCD radiation, due to
the large phase space. Thus it is only slightly less generic
to design an inclusive analysis for large \met~in association
with three or more energetic jets. We will refer to this
as \textit{the inclusive missing energy signature}\footnote{
The requirement of a third energetic jet greatly reduces the
size and complexity of the Standard Model backgrounds. Thus while mature
LHC analyses will explore the fully inclusive \met~signature,
we assume here that an early discovery will be based on a
multijet + \met~ data sample.}. 

In the basic $2\to 2$ hard scattering, the heavy parent particles
of the signal will be produced back-to-back in the partonic
subprocess center-of-mass frame; typically they will have $p_T$ roughly
comparable to their mass $m_{\rm p}$. The WIMPs of mass
$m_{\rm dm}$ resulting from the parent particle decays will
fail to deposit energy $\ge m_{\rm dm}$ in the calorimeters;
if the WIMPS have fairly large $p_T$, a
significant fraction of this energy contributes to the \met .
Thus either large $m_{\rm p}$ or large $m_{\rm dm}$ leads to
large \met .
Note the azimuthal directions of the WIMPS are anticorrelated,
a feature inherited from their parents, so the magnitude
of the total \met~tends to be less than the magnitude of the
largest single contribution.

At the LHC, the most important Standard Model sources of large real 
\met~will be $t\bar{t}$, single top, $W$ and $Z$ plus jets associated
production, dibosons and heavy flavor decays. Most of these processes produce
a hard lepton in association with the \met~from an energetic
neutrino. The exception is $Z\to \nu\bar{\nu}$. Even with a perfect
detector, $Z\to \nu\bar{\nu}$ plus jets is an irreducible physics 
background.

\subsection{Analysis path}
In the real data this search will be performed starting from a primary
data set that includes  requirements of missing energy, jets and general 
calorimetric activity at the trigger path; the trigger efficiency should 
be measured in other data samples.

For the offline analysis, we will adopt the inclusive missing
energy benchmark analysis studied with the full detector simulation
for the CMS Physics Technical Design Report~\cite{Ball:2007zza,Yetkin:2007zz}.

The first phase is a preselection based on the event quality.
The purpose of this primary cleanup is to discard events with 
fake \met~from sources such as beam halo, data acquisition 
noise and cosmic ray muons.
To eliminate these types of backgrounds the benchmark
analysis uses jet variables, averages them over the event to define
corresponding event variables, and uses these to discriminate real
\met + multijet events from spurious backgrounds.
The event electromagnetic fraction (EEMF) is defined to be the
\et~weighted jet electromagnetic fraction.
We define an event charged fraction
(ECHF) as the event average of the jet charged fraction
(defined as the ratio of the \SumPt~ of the tracks
associated with a jet over the total calorimetric jet \et).
The preselection also has a quality requirement for the reconstructed
primary vertex. 

Events that are accepted by the preselection requirements proceed through
the analysis path if they have missing transverse energy
\met$\ge 200$ GeV and at least three jets with \et$\ge 30$ GeV
within pseudorapidity $|\eta|<3$. These requirements directly define the 
missing energy signal signature. 
In addition the leading jet is required to be
within the central tracker fiducial volume {\it i.e.} $|\eta|<1.7$.
Everywhere in this paper ``jets'' means uncorrected (raw) jets 
with \et$>30$ GeV and $|\eta |<3$ as measured
in the calorimeters; the jet reconstruction is with a simple
iterative cone algorithm with a 0.5 cone size in the $\eta - \phi$ space.
The missing energy is uncorrected for the presence of muons in the event.

The rest of the analysis path is designed based on elimination of the major
backgrounds. The QCD background from mismeasured jets is reduced
by rejecting events where the \met~is too closely correlated
with the azimuthal directions of the jets.
To reduce the large background from
$W(\rightarrow \ell\nu)+$jets, $Z(\rightarrow \ell\ell)+$jets and
$t\bar{t}$ production an {\it indirect lepton veto} (ILV)
scheme is designed that uses the tracker and the calorimeters.  The ILV
retains a large signal efficiency while achieving a factor of two
rejection of the remaining $W$ and $t\bar{t}$ backgrounds.
The veto is indirect because we don't identify leptons--
instead events are rejected if the electromagnetic
fraction of one of the two leading jets is too large, or if the
highest $p_T$ track of the event is isolated. The signals we are interested in
are characterized by highly energetic jets while leptons
in the signal originate from cascade decays of the parents or semileptonic
$B$ decays in the jets; thus even when a signal
event has leptons it is relatively unlikely to be rejected by the ILV. 
For the models in our study, approximately 85\% of all signal events and
70\% of signal events with muons or taus pass the ILV cut.

The final selections require that the leading jet has
$E_T>180$ GeV, and that the second jet has $E_T>110$ GeV.
We also require $H_{\mathrm{T}} > 500$ GeV, where
\be
H_{\mathrm{T}}=\sum_{i=2}^{4} E_{T}^{i} +
  E_{\mathrm{T}}^{\mathrm{miss}} \; , 
\ee
where the \et~ is summed over the second, third and
fourth (if present) leading jets.
These cuts select for highly energetic events, greatly
favoring events with new heavy particles over the Standard
Model backgrounds.

Table~\ref{tab:EvSel} summarizes the benchmark analysis path.
The table lists the cumulative efficiencies after each selection
for a benchmark signal model and the Standard Model backgrounds.
The signal model is the CMS supersymmetry benchmark model LM1,
which has a gluino with mass 611 GeV and squarks with masses
around 560 GeV. The last line of the table shows the expected
number of events that survive the selection in a data set
corresponding to 1000 pb$^{-1}$ of integrated luminosity.
For the QCD background and the single top background, which
are not shown in the table, the estimated number of remaining
events is 107 and 3, respectively. Thus the total estimated
Standard Model background after all selections is 245 events
per 1000 pb$^{-1}$.
 
\begin{table*}[htp]
\begin{center}
\caption{Cumulative selection efficiency after each requirement in the
  \met + multijets analysis path for a low mass SUSY signal and the
  major Standard Model backgrounds (EWK refers to
  $W/Z$,$WW/ZZ/ZW$), see~\cite{Ball:2007zza,Yetkin:2007zz}).}{\label{tab:EvSel}}
\begin{tabular}{lrrrr}
\toprule
Cut/Sample                  & Signal  & $t\bar{t}$   & $Z(\to\nu\bar{\nu})+$ jets  & EWK + jets \\ \midrule
All (\%)                    & 100     & 100          & 100                         & 100        \\ \midrule \midrule
Trigger                     & 92      & 40           & 99                          & 57         \\ \midrule 
\met$>200$ GeV              & 54      & 0.57         & 54                          & 0.9        \\ \midrule 
PV                          & 53.8    & 0.56         & 53                          & 0.9        \\ \midrule 
$N_{j}\geq$3                & 39      & 0.36         & 4                           & 0.1        \\ \midrule 
$|\eta_{d}^{j1}|\geq 1.7$   & 34      & 0.30         & 3                           & 0.07       \\ \midrule \midrule
EEMF $\geq$ 0.175           & 34      & 0.30         & 3                           & 0.07       \\ \midrule 
ECHF $\geq$ 0.1             & 33.5    & 0.29         & 3                           & 0.06       \\ \midrule \midrule 
QCD angular                 & 26      & 0.17         & 2.5                         & 0.04       \\ \midrule \midrule 
$Iso^{lead\;trk}=0$         & 23      & 0.09         & 2.3                         & 0.02       \\ \midrule 
$EMF(j1),$                  &         &              &                             &            \\
$EMF(j2) \geq 0.9$          & 22      & 0.086        & 2.2                         & 0.02       \\ \midrule \midrule 
$E_{T,1} >$ 180 GeV,        &         &              &                             &            \\
$E_{T,2} >$ 110 GeV         & 14      & 0.015        & 0.5                         & 0.003      \\ \midrule
$H_{T}> 500$ GeV            & 13      & 0.01         & 0.4                         & 0.002      \\ \midrule \midrule
\multicolumn{5}{c}{events remaining per 1000 pb$^{-1}$}                                         \\ \midrule \midrule
                            & 6319    & 54           & 48                          & 33         \\
\bottomrule
\end{tabular}
\end{center}
\end{table*}

\subsection{Triggers and ``boxes''}\label{sec:2.2}

\begin{figure}
\includegraphics[width=0.45\textwidth,height=0.45\textwidth,angle=0]{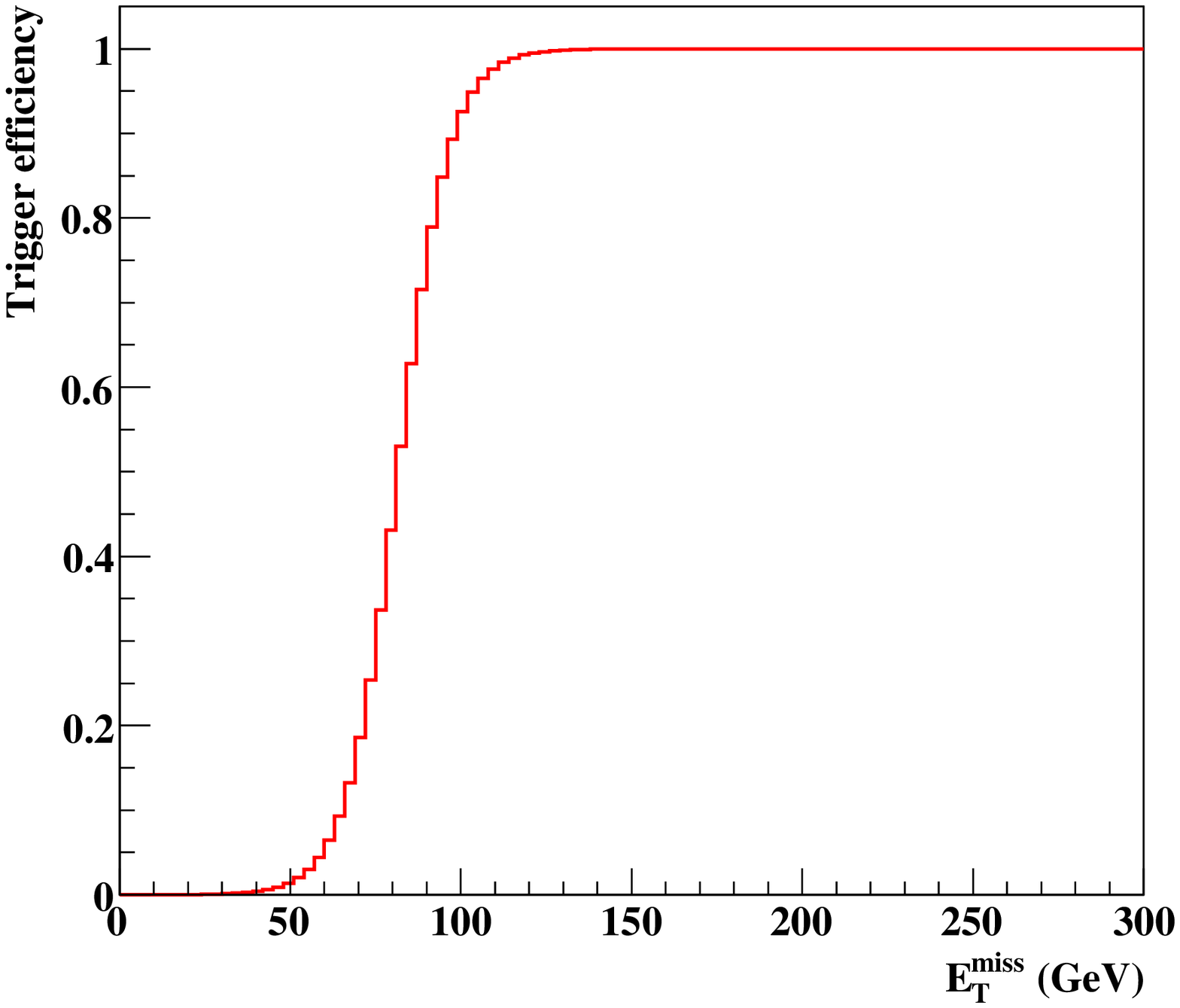}
\caption {\label{fig:met} The \met~trigger efficiency.}
\end{figure}

\begin{figure}
\includegraphics[width=0.45\textwidth,height=0.45\textwidth,angle=0]{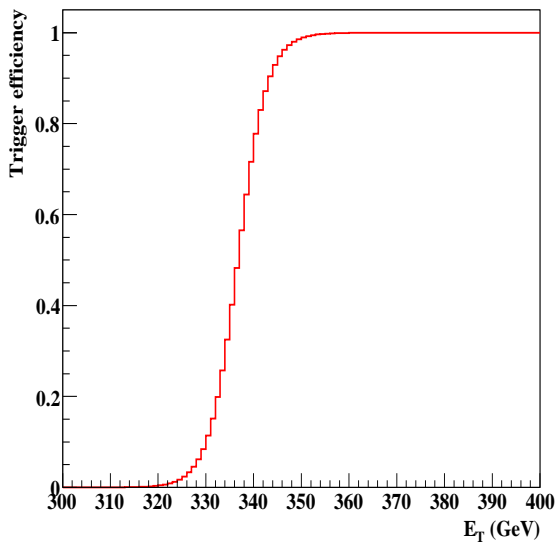}
\caption
{\label{fig:dijet} The DiJet trigger efficiency.}
\end{figure}

\begin{figure}
\includegraphics[width=0.45\textwidth,height=0.45\textwidth,angle=0]{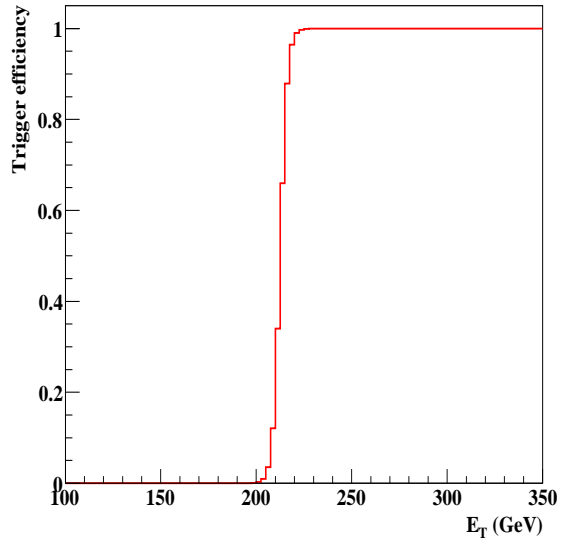}
\caption
{\label{fig:trijet} The TriJet trigger efficiency.}
\end{figure}

\begin{figure}
\includegraphics[width=0.45\textwidth,height=0.45\textwidth,angle=0]{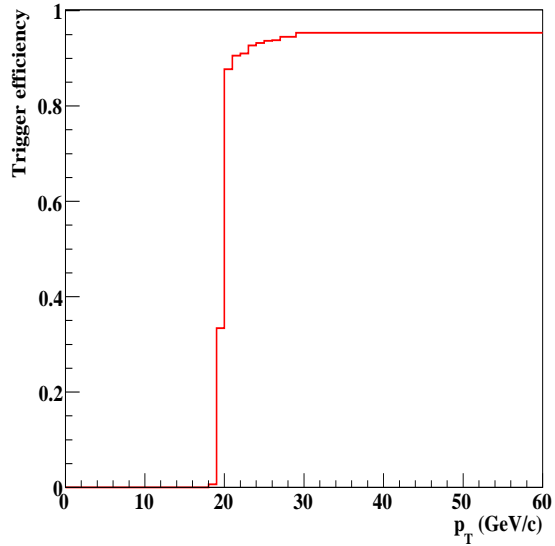}
\caption
{\label{fig:muon} The Muon20 trigger efficiency.}
\end{figure}

Having established a benchmark analysis path, we also need to define
benchmark data samples. With the real LHC data these will correspond
to data streams and data paths from various triggers. For the inclusive missing
energy signature relevant triggers are the \met~and 
jet triggers. A single lepton trigger is also of interest, since many
models produce energetic leptons in association with large \met .
For our study we have chosen simple but 
reasonable \cite{Sphicas:2002gg,Dasu:2000ge}
parametrizations of the trigger efficiencies defining our four benchmark
 triggers \footnote{These are made-up triggers for the purposes of our study. 
The guidance on our parametrizations is from the published trigger and physics 
reports of the CMS experiment. We expect that the trigger tables of 
the LHC experiments will include corresponding trigger paths, richer and 
better in terms of the physics capture.}
:
\begin{itemize}
\item The MET trigger is a pure inclusive \met~trigger. It is
50\% efficient for \met$>80$ GeV, as seen in Figure \ref{fig:met}.
\item The DiJet trigger requires two very high $E_T$ jets.
It is 50\% efficient for uncorrected jet \et$>340$ GeV,
as seen in Figure \ref{fig:dijet}.
\item The TriJet trigger requires three high $E_T$ jets.
It is 50\% efficient for uncorrected jet \et$>210$ GeV,
as seen in Figure \ref{fig:trijet}.
\item The Muon20 trigger requires an energetic
muon that is not necessarily isolated. The trigger
is 88\% efficient for muons with $p_T = 20$ GeV/$c$,
asymptoting to 95\% as seen in Figure \ref{fig:muon}.
\end{itemize}

After applying the selection requirements, these four triggers define
four potential discovery data sets.  In our simulation the DiJet, 
TriJet, and Muon20 data sets, after the inclusive missing energy analysis 
path is applied, are all subsets of the MET sample, apart from one or 
two events per 1000 pb$^{-1}$ \footnote{A perfectly designed 
trigger table will give rise to overlaps among datasets from different 
trigger paths due to both physics and slow/non-sharp trigger efficiency turn-ons (resolution).}. 
Thus the MET is the largest, 
most inclusive sample. We perform one complete analysis based on the MET
trigger. The other three triggers are then treated
as defining three more \textit{boxes}, i.e.
experimentally well-defined subsets of the MET discovery data set.
The simplest physics observables are the counts of events
in each box.

\subsection{Backgrounds and systematics}
In the CMS study the total number of Standard Model background events
remaining after all selections is 245 per 1000 pb$^{-1}$  
for an \met~trigger sample. The error on this estimate
is dominated by {\it i)} the uncertainty in how well the
detector simulation software simulates the response of the
actual CMS detector, and {\it ii)} the uncertainty on how well the
Standard Model event generators emulate QCD, top production,
and $W/Z$ plus jets production. Detailed studies of the real LHC
data will be required in order to produce reliable estimates
of these uncertainties. 

Prior to data we assign conservative error bars on these background
projections. We have checked that 100 pb$^{-1}$ of data in the MET trigger sample
is sufficient for a 5$\sigma$ discovery for the eight models in our
study, even if we triple
the backgrounds quoted above and include a 15\% overall systematic 
error. The look-alike analysis will be degraded, however,
in the event that the Standard Model backgrounds turn out
to be much larger than current estimates.

Prior to data, it is also difficult to make a reliable
estimate of the main systematic uncertainties that will
affect the inclusive missing energy analysis. Systematic
uncertainties will decrease over time, as the
detectors are better understood, calibration studies are
performed, and Standard Model physics is analyzed with
the LHC data. For our study we have assumed that, at the
moment of discovery, the dominant systematic errors in
the full discovery data set will 
come from three sources:
\begin{itemize}
\item Luminosity uncertainty: it affects the counting
of events. This systematic uncertainty is process independent.
\item Detector simulation uncertainty: it mainly affects
calorimetry-related variables in our study, in particular
jet counting and the missing energy. This systematic is
 partially process dependent.
\item QCD uncertainty: it includes the uncertainties from
the parton distribution functions, higher order
matrix elements, and large logarithms. This uncertainty
affects event counting, jet counting and the shapes
of kinematic distributions. It is partially process
dependent.
\end{itemize}

Note that, since we use uncorrected jets, we do not
have a systematic from the jet energy scale. This is
traded for a portion of the detector simulation
uncertainty, i.e. how well we can map
signal events into uncorrected jets as would be measured in
the real detector.

\subsection{Simulation of the signals}
A realistic study of look-alikes requires full detector simulation.
For the initial phase of this work a generator level analysis is
attractive, being computationally less intensive and providing a 
clear link between observables and the underlying 
theory models{\footnote{The full {\tt GEANT4}-based simulation is too slow 
to adequately sample the entire theory space. 
Having completed the first exploratory 
phase of this work, we are repeating the analysis
to validate these results with the full
experimental simulation.}}.

In a generator level analysis, jets are reconstructed by applying
a standard algorithm to particles rather than to calorimeter
towers. This obviously does not capture the effects of a
realistic calorimeter response, calorimeter segmentation, and
energy losses due to material in the tracker as well as magnetic field effects. 

A compromise between the full simulation and a generator level
analysis is a parameterized detector simulation.
For the LHC the publicly available software packages 
include {\tt AcerDET}~\cite{RichterWas:2002ch},
and {\tt PGS}~\cite{pgs}. In such a simulation,
electrons, muons and photons can be reconstructed using
parameterized efficiencies and resolutions based on 
abstract but educated rules-of-thumb
for modern multipurpose detectors.
Jets are reconstructed in a virtual calorimeter, from particle
energies deposited in cells that roughly mimic the segmentation
of a real calorimeter. Calorimeter response is approximated
by performing a Gaussian smearing on these energy deposits.
The \met~is reconstructed from the smeared energies in
these virtual towers.

We performed a preliminary study by comparing {\tt PGS} results to
the full simulation results reported for the SUSY
benchmark model LM1 \cite{Ball:2007zza,Yetkin:2007zz}.
We found that {\tt PGS} jets are not a good approximation of
uncorrected jets in the full simulation, even for the
most basic properties such as the \et~spectrum. Varying
the parameters and adding simple improvements, such as
taking into account the 4 Tesla field in the barrel, did
not change this conclusion. {\tt PGS} jets have a behavior,
not surprisingly, that is intermediate between generator
level jets and uncorrected full simulation jets.

We developed a modified simulation called
{\tt PGSCMS} with the geometry and approximate magnetic
field of the CMS detector. 
The {\tt PGS}
Gaussian smearing and uninstrumented effects in the calorimeters
are turned off. Electrons, muons and photons are extracted 
at generator level, and {\tt PGS} tau reconstruction is not used.
Track information is extracted as in the standard {\tt PGS}.
The calorimeter output improves on a generator level
analysis in that we include approximations to the
effects of segmentation and the 4 Tesla field, as well as 
an $\eta$ correction derived from the $z$ value of the 
primary vertex. We parameterized the detector response
in a limited set of look-up tables as a function of the
generator-level quantities.

At the analysis level we apply parameterized 
corrections and reconstruction efficiencies inspired by the
published CMS detector performance~\cite{Bayatian:2006zz}.
For the jets, we apply an \et~and $\eta$ dependent rescaling of
their \et, tuned to reproduce the full simulation
LM1 results in \cite{Ball:2007zza,Yetkin:2007zz}.
This rescaling makes the jets softer ({\it i.e.} takes into account the
detector reconstruction): a 50 GeV
generator level jet becomes an approximately 30 GeV raw
jet in our analysis. 

The \met~reconstructed from {\tt PGSCMS} is essentially identical,
modulo small calorimeter segmentation effects, to a
a generator level analysis, i.e. our \met~is 
virtually indistinguishable from the Monte Carlo truth
\met~obtained from minus the vector sum of the \et~of neutrinos, 
muons and the other weakly interacting particles (such as the LSP). 
We did not attempt to rescale the \met ; this
is a complicated task since \met~is a vector and 
in general energy losses, calorimeter response and 
mismeasurements tend to decrease
the real large \met~tails while increasing the
\met~tails in the distribution of non-real \met~events. 
Instead of attempting to rescale the \met~event by event, we raised the
\met~cut in our benchmark analysis to 220 GeV{\footnote{In a realistic
full simulation study with the
first jet data in hand, our \met~analysis will avoid such compromises.}}. 

Because of the limitations of our fast simulation,
we also simplified parts of the benchmark analysis.
The first phase primary cleanup is dropped since it is related
to supression of spurious processes that we do not simulate,
and it is nearly 100\% efficient for the signal.
We also drop the jet electromagnetic fraction cuts of
the ILV, because they are nearly 100\% efficient for the signal.

The resulting performance of our parameterized fast simulation
for the SUSY benchmark model LM1 is shown in 
Table~\ref{tab:LM1}. The agreement with the full simulation
study is very good. The largest single cut discrepancy is
2\%; this occurs for the QCD angular cuts, reflecting the
expected fact that our fast simulation does not accurately
reproduce jet mismeasurement effects.
Since the final efficiencies agree to within 7\%,
it is plausible that look-alikes defined in our
fast simulation study will remain look-alikes in our
upcoming full simulation study.

It is important to note that this fast simulation
\textit{does not} reproduce the Standard Model background
efficiencies shown in Table~\ref{tab:EvSel}.
In fact the discrepancies in the total efficiencies can
approach an order of magnitude. This is to be expected. 
We are cutting very hard on the Standard Model events, 
thus the events that pass
are very atypical. This is in contrast to the signal
events, where the fraction that pass are still fairly
generic, and their \et~and \met~spectra near the cuts
are less steeply falling than those of the background.
Since SM backgrounds cannot be estimated from a {\tt PGS}
level analysis, we take our backgrounds from the state-of-the-art
analysis in \cite{Ball:2007zza}; this approach only works
because we have also matched the analysis path used in \cite{Ball:2007zza}.

The full software chains we use in our study are summarized in
Table~\ref{tab:software}. All of the simulated data sets
include an average of 5 pileup events added to each signal
event, corresponding to low luminosity LHC running 
($\sim 10^{33}$ cm$^{-2}$s$^{-1}$).


\begin{table*}[htp]
\begin{center}
\caption{Comparison of cut-by-cut selection efficiencies
for our \met~analysis applied to the SUSY benchmark model LM1.
``Full'' refers to the full simulation study \cite{Ball:2007zza,Yetkin:2007zz};
``Fast'' is what we obtain from our parameterized fast simulation.}{\label{tab:LM1}}
\begin{tabular}{lrr}
\toprule
Cut/Software                & Full       & Fast\\ \midrule
Trigger and                 &            &     \\ 
\met$>200$ GeV              & 53.9\%     & 54.5\%\\ \midrule 
$N_{j}\geq$3                & 72.1\%     & 71.6\%\\ \midrule 
$|\eta_{d}^{j1}|\geq 1.7$   & 88.1\%     & 90.0\%\\ \midrule 
QCD angular                 & 75.6\%     & 77.6\%\\ \midrule  
$Iso^{lead\;trk}=0$         & 85.3\%     & 85.5\%\\ \midrule 
$E_{T,1} >$ 180 GeV,        &            &     \\
$E_{T,2} >$ 110 GeV         & 63.0\%     & 63.0\%\\ \midrule
$H_{T}> 500$ GeV            & 92.8\%     & 93.9\%\\\midrule\midrule
Total efficiency            & 12.9\%     & 13.8\%\\
\bottomrule
\end{tabular}
\end{center}
\end{table*}



\begin{table*}[htp]
\begin{center}
\caption{Summary of software chains used in this study.
The little Higgs spectrum is based on \cite{Hubisz:2005tx}.
{\tt PGSCMS} is a variation of {\tt PGS v4} \cite{pgs}.}{\label{tab:software}}
\begin{tabular}{lll}
\toprule
Software/Models             & Group 1 models                                & Group 2 models                    \\ \midrule
Spectrum generator          & {\tt Isajet v7.69} \cite{isajet} or           & private little Higgs              \\ 
                            & or {\tt SUSY-HIT v1.1} \cite{Djouadi:2006bz}   & or {\tt SuSpect v2.34} \cite{Djouadi:2002ze}\\ \midrule
Matrix element calculator   & {\tt Pythia v6.4} \cite{Sjostrand:2006za}     & {\tt MadGraph v4} \cite{madgraph} \\ \midrule
Event generator             & {\tt Pythia v6.4}                             & {\tt MadEvent v4} \cite{madevent}  \\
                            &                                               & with {\tt BRIDGE} \cite{bridge}   \\ \midrule
Showering and hadronization & {\tt Pythia v6.4}                             & {\tt Pythia v6.4}                 \\ \midrule
Detector simulation         & {\tt PGSCMS v1.2.5}                           & {\tt PGSCMS v1.2.5}               \\
                            & plus parameterized                             & plus parameterized                 \\
                            & corrections                                   & corrections                       \\ 
\bottomrule
\end{tabular}
\end{center}
\end{table*}



\section{Populating the theory space}
\label{sec:3}

In Section \ref{sec:2} we gave a partial classification of BSM models
according to how many new weakly interacting particles
appear in a typical final state. 
Our benchmark \met~analysis is optimized for the case of two
heavy weakly interacting particles per event, as applies to
SUSY models with conserved $R$ parity, little Higgs models with
conserved $T$ parity and Universal Extra Dimensions models with
conserved $KK$ parity.
This study is a first attempt at constructing groups of
look-alike models drawn from this rather large fraction of 
the BSM theory space, and developing strategies to discriminate
them shortly after an initial discovery.

One caveat is that models from other corners of the theory space
may also be look-alikes of the ones considered here. For example,
models with strong production of heavy particles that 
decay to boosted top quarks can produce higher 
$E_T$ jets and larger \met~from neutrinos than does Standard 
Model top production. Such look-alike possibilities also
require study, but they are not a major worry since
our results show that we have some ability to discriminate
heavy WIMPS from neutrinos even in small data sets. 

\subsection{SUSY}
In a large class of supersymmetry models with conserved $R$ parity,
not necessarily restricted to the MSSM,
the LSP is either the lightest neutralino or a right-handed sneutrino\footnote{Recent
analyses \cite{Lee:2007mt,Arina:2007tm} have argued for the 
phenomenological viability of sneutrino dark matter.}. 

In addition,
if the NLSP is a neutralino or sneutrino and the LSP is a gravitino,
the \met~signature is the same. Models based on gravity-mediated,
gauge-mediated or anomaly-mediated SUSY breaking all provide many
candidate models.

Because this relevant portion of SUSY theory space is already so
vast, there is a temptation to reduce the scope of the LHC Inverse
Problem by making explicit or implicit theoretical assumptions.
To take an extreme, one could approach an early LHC discovery
in the \met~channel having already made the assumptions that
{\it (i)} the signal is SUSY, {\it (ii)} it has a minimal Higgs
sector (MSSM), {\it (iii)} it has gravity-mediated SUSY breaking
(SUGRA), {\it (iv)} the breaking is minimal (mSUGRA) and {\it (v)} 100\%
of dark matter is thermal relic LSPs with an abundance given
by extrapolating standard cosmology back to the decoupling epoch.
We don't want to make any such assumptions; rather
we  want to \textit{test} theoretical hypotheses
in the LHC discovery data set combined with other measurements. 

For SUSY we have the benefit of more than one spectrum calculator
that can handle general models, more than one matrix element calculator
and event generation scheme, and a standardized interface via the
SUSY Les Houches Accord (SLHA) \cite{SLHA}. There are still a few bugs in this
grand edifice, but the existing functionality combined with the ability to
perform multiple cross-checks puts us within sight of where we
need to be when the data arrives.

\subsection{Little Higgs}

Little Higgs models are a promising alternative to weak 
scale supersymmetry \cite{ArkaniHamed:2002qy}-\cite{Carena:2006jx}. 
In little Higgs models, 
the Higgs is an approximate Goldstone boson, with global 
symmetries protecting its mass (which originates from a quantum level breaking of these symmetries) 
from large radiative 
corrections. Many of these 
LH models require an approximate $T$ parity discrete symmetry to reconcile LH 
with electroweak precision data.  
This symmetry is similar to $R$ parity in SUSY models.   
The new LH particles that would be produced at the LHC would be odd under this symmetry, 
enforcing the stability of the lightest particle that is odd under $T$ parity.  
This new particle is weakly interacting and would manifest itself as missing energy 
at the LHC\footnote{This symmetry may be inexact, or violated 
by anomalies~\cite{Hill:2007zv}. Such possibilities are 
model dependent \cite{Csaki:2008se,Krohn:2008ye}.}.  

Just as in SUSY, new colored particles are the
dominant production modes. These particles subsequently generate high 
multiplicity final states through decay chains that end with the 
lightest $T$ odd particle.  In LH models, the strongly coupled particles 
are $T$ odd quarks (TOQ's), analogous  to the squarks of SUSY.  The weakly 
coupled analogues of the gauginos are $T$ odd spin one vector bosons (TOV's).  
In the models considered to date, there is no analog of the gluino: this is an 
important consideration in constructing supersymmetric look-alikes of LH models.

In this study, we work with a minimal 
implementation of a little Higgs model with $T$ parity that is known 
as the littlest Higgs model with $T$ parity.  This model is 
based on a $SU(5)/SO(5)$ pattern of global symmetry breaking.  Each 
SM particle except the gluon has an associated LH partner 
odd under $T$ parity.
There is also an extra pair of top partners, one $T$ odd and the other $T$ even,
as well as singlets.
The lightest $T$ odd particle in this model we label $A_H$. It is a 
heavy gauge boson that is an admixture of a heavy copy of the hypercharge 
gauge boson and a heavy $W^3$ boson.

For event generation, we use a private implementation of the littlest
Higgs model within {\tt MadGraph}. There is a need to generalize this to a wider
class of models. 

\subsection{Universal extra dimensions}
Universal extra dimensions models are based on orbifolds of one or two
TeV$^{-1}$ size extra spatial dimensions~\cite{Appelquist:2000nn}-\cite{Dobrescu:2007ec}.
The five-dimensional version of UED is the simplest. 
At the first level of Kaluza-Klein ($KK$) excitations, each Standard Model
boson has an associated partner particle, and each Standard Model
fermion has two associated partner particles (i.e. a vector-like
pair). These $KK$ partners are odd under a $KK$ parity, the remnant
of the broken translational invariance along the fifth dimension.
This parity is assumed
to be an exact symmetry. After taking into account mass splittings
due to Standard Model radiative corrections, one 
finds that the lightest $KK$ odd
partner is naturally the weakly interacting partner of the
hypercharge gauge boson. A wide variety of spectra for the
$KK$ odd partners can be obtained by introducing additional
interactions that are localized
at the orbifold fixed points; these choices distinguish generic
UED from the original minimal model of \cite{Appelquist:2000nn}.
These models resemble SUSY.

A public event generation code based on a modification of {\tt Pythia}
is available for generic 5-dimensional UED models~\cite{Allanach:2006fy}.
There is a need to generalize this to a wider
class of models, e.g. 6-dimensional UED.
In our study we have not used any UED examples, but we will
include them in the future.


\begin{figure*}[thb]
\includegraphics[width=0.85\textwidth]{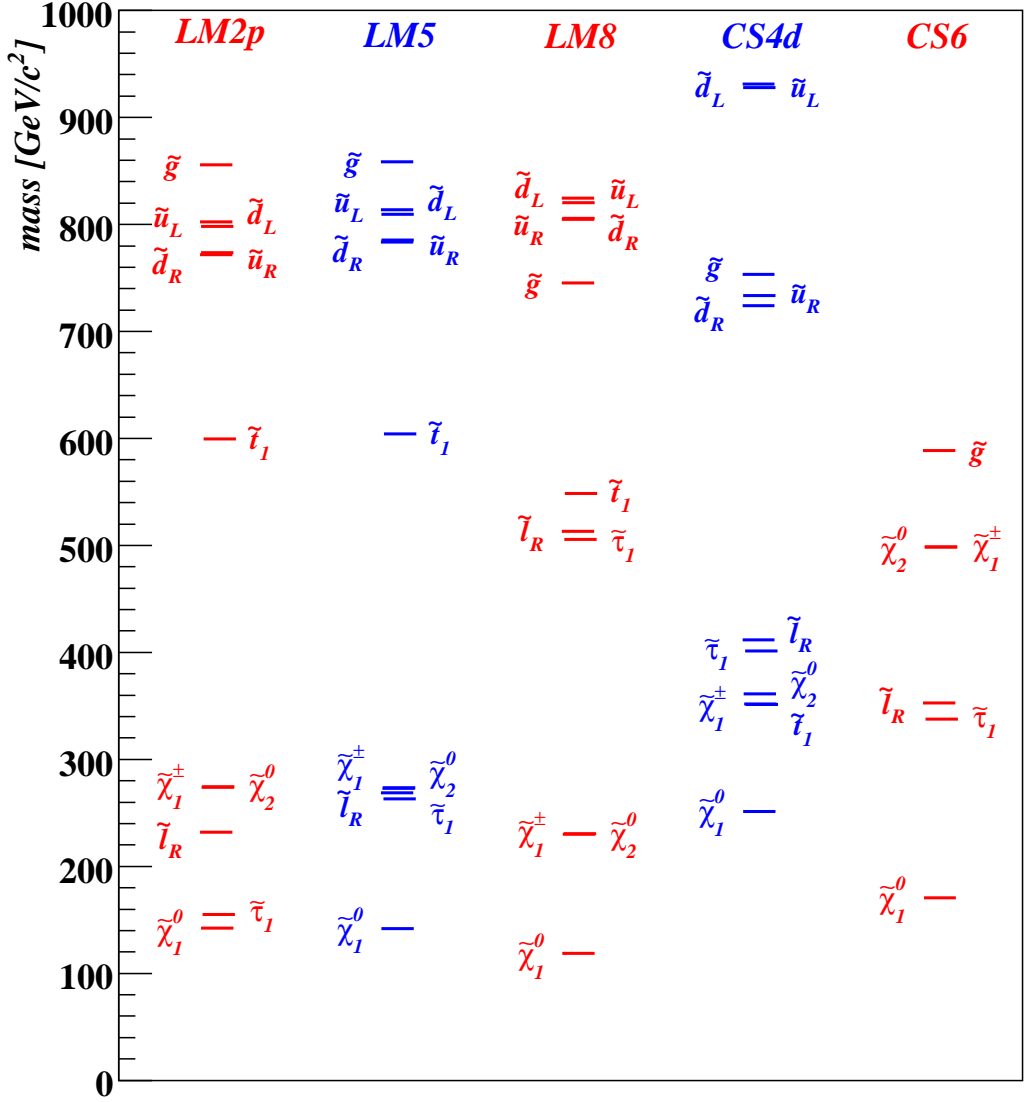}
\caption {\label{fig:group1spectra} The mass spectra of the MSSM
  models LM2p, LM5, LM8, CS4d and CS6.  Only the most relevant
  particles are shown: the lighter gauginos $\tilde{\chi}_1^0$,
  $\tilde{\chi}_2^0$ and $\tilde{\chi}_1^{\pm}$, the lightest stau
  $\tilde{\tau}_1$, the right-smuon and selectron denoted collectively
  as $\tilde{\ell}_R$, the lightest stop $\tilde{t}_1$, the gluino,
  and the left/right up and down squarks $\tilde{u}_L$, $\tilde{u}_R$,
  $\tilde{d}_L$ and $\tilde{d}_R$. The very heavy $\simeq 2$ TeV
  squarks of model CS6 lie outside the displayed range.  }
\end{figure*}


\section{Description of the models}\label{sec:4}
\subsection{Group 1}
The five look-alike models of Group 1 are all MSSM models.
Two of them (LM5 and LM8) are CMS SUSY benchmark models,
while another (LM2p) is a slight variation of a CMS benchmark.
It is a sobering coincidence that these are look-alikes of
the \met~analysis, since the benchmarks were developed by CMS to 
cover different experimental signatures, not produce look-alikes.
To round out Group 1 we found two other MSSM look-alikes whose
spectra and decay chains are as different from each other and
from the three CMS benchmarks as we could make them.

The models are consistent with all current experimental constraints,
but do not all give the ``correct'' relic density of
dark matter. Any comparison of relic densities to the so-called
WMAP constraints assumes at least three facts not yet in 
evidence: {\it (i)} that
dark matter is a thermal relic, {\it (ii)} that there is only one significant
species of dark matter and {\it (iii)} that cosmological evolution was 
entirely radiation-dominated from the time of dark matter decoupling
until the time of Big Bang nucleosynthesis. 
A missing energy discovery
at the LHC will help us test whether these assumptions have any validity.
For example, model LM8 produces a relic density an order of magnitude
larger than the WMAP upper bound; thus discriminating LM8 as a more
likely explanation of an early missing energy discovery would call
into question \cite{Gelmini:2006pq,Barenboim:2006jh}
assumptions (i) and (iii), or could be a hint
that the lightest neutralino is not absolutely stable.

LM2p, LM5
and LM8 are minimal supergravity models
\cite{Chamseddine:1982jx}-\cite{Hall:1983iz}.  They are specified by
the usual high scale mSUGRA input parameters as shown in Table
\ref{table:msugratable}; because the resulting superpartner spectra
depend strongly on RGE running from the high scale, a complete
specification of the models also requires fixing the top quark mass
and the particular spectrum generator program used. We have used
$m_{\rm top}$ = 175 GeV and the {\tt ISAJET v7.69} generator
\cite{isajet}, in order to maintain compatibility with
the CMS Physics TDR \cite{Ball:2007zza}. Models LM5 and LM8 are then
identical to the mSUGRA benchmark models of the CMS Physics TDR,
while LM2p is almost identical to benchmark model LM2; LM2p has a
slightly larger value of $m_{1/2}$ (360 versus 350 GeV) than LM2,
which makes it \textit{more} of a look-alike of the other Group 1 models.


\begin{table}[thb]
\begin{center}
\caption{\label{table:msugratable} Input parameters for the mSUGRA
  models LM2p, LM5 and LM8.  The notation comforms to
  \cite{isajet}.  The mass parameters and trilinear $A_0$
  parameter have units of GeV.}
\begin{tabular}{rrrr}
\toprule
&\multicolumn{1}{c}{LM2p}&\multicolumn{1}{c}{LM5}
&\multicolumn{1}{c}{LM8}
\\ \midrule
$m_0$        & 185 & 230 &  500
\\ \midrule
$m_{1/2}$    & 360 & 360 &  300
\\ \midrule
$A_0$        &   0 &   0 & -300
\\ \midrule
tan$\,\beta$ &  35 &  10 &   10
\\ \midrule
sign($\mu$)  &   + &   + &    +
\\ \bottomrule
\end{tabular}
\end{center}
\end{table}


\begin{table}[thb]
\begin{center}
\caption{\label{table:csinputs} Input parameters for the MSSM models
  CS4d and CS6.  The notation conforms to
  \cite{Djouadi:2006bz,Djouadi:2002ze}.  The mass parameters and
  trilinear $A$ parameters have units of GeV.}
\begin{tabular}{rrr}
\toprule
&CS4d & CS6
\\ \midrule
$M_1$                                     &  620    &  400
\\ \midrule
$M_2$                                     &  930    &  600
\\ \midrule
$M_3$                                     &  310    &  200
\\ \midrule
$A_{\tau},A_t,A_b,A_e,A_u,A_d$            & -400    & -300
\\ \midrule
$M_{Q_L},M_{t_R},M_{b_R}$                 &  340    & 2000
\\ \midrule
$M_{q_u},M_{u_R},M_{d_R}$                 &  340    & 2000
\\ \midrule
$M_{\tau_L},M_{\tau_R},M_{e_L},M_{e_R}$   &  340    &  340
\\ \midrule
$M_{h_u}^2,M_{h_d}^2$                     & 115600  & 115600
\\ \midrule
tan$\,\beta$                              &   10    &   10
\\ \midrule
sign($\mu$)                               &    +    &    + 
\\ \bottomrule
\end{tabular}
\end{center}
\end{table}


\begin{table*}[thb]
\begin{center}
\caption{\label{table:myproductiontable} Summary of LHC superpartner
  production for the Group 1 MSSM models LM2p, LM5, LM8, CS4d and CS6. The
  relative percentages are shown for each model, both before and after
  the event selection.  The squark--squark percentages shown are
  excluding the contributions from pair production of the lightest
  stops, which are shown separately. Note that squark--chargino
  includes the production of either chargino, and squark-neutralino
  includes all of the four neutralinos. The category ``other''
  includes weak production as well as the associated
  production of gluinos with charginos or neutralinos.
  The total NLO cross sections are from {\tt Prospino2}~\cite{prospino2}.}
\begin{tabular}{lrrrrrrrrrr}
\toprule
&\multicolumn{2}{c}{LM2p}&\multicolumn{2}{c}{LM5}
&\multicolumn{2}{c}{LM8}&\multicolumn{2}{c}{CS4d}
&\multicolumn{2}{c}{CS6}
\\ \midrule
NLO cross section (pb) & \multicolumn{2}{c}{8.6} & \multicolumn{2}{c}{8.1} 
& \multicolumn{2}{c}{12.7} & \multicolumn{2}{c}{14.5} & \multicolumn{2}{c}{12.6}
\\\midrule
&\multicolumn{1}{l}{before}&\multicolumn{1}{l}{after}
&\multicolumn{1}{l}{before}&\multicolumn{1}{l}{after}
&\multicolumn{1}{l}{before}&\multicolumn{1}{l}{after}
&\multicolumn{1}{l}{before}&\multicolumn{1}{l}{after}
&\multicolumn{1}{l}{before}&\multicolumn{1}{l}{after}
\\ [-3pt]
&\multicolumn{1}{l}{cuts}&\multicolumn{1}{l}{cuts}
&\multicolumn{1}{l}{cuts}&\multicolumn{1}{l}{cuts}
&\multicolumn{1}{l}{cuts}&\multicolumn{1}{l}{cuts}
&\multicolumn{1}{l}{cuts}&\multicolumn{1}{l}{cuts}
&\multicolumn{1}{l}{cuts}&\multicolumn{1}{l}{cuts}
\\ \midrule
squark--squark & 33\% & 36\% & 32\% & 38\% & 22\% & 33\% & 19\% & 34\% & 0.1\% & 0.1\%
\\ \midrule
squark--gluino & 45\% & 55\% & 46\% & 52\% & 48\% & 54\% & 41\% & 55\% & 3.7\% & 7.4\%
\\ \midrule
gluino--gluino & 7.2\% & 6.4\% & 7.4\% & 6.4\% & 14\% & 8.3\% & 11\% & 8\% & 95\% & 92\% 
\\ \midrule
stop--stop & 2.1\% & 1.1\% & 2.1\% & 0.9\% & 2.6\% & 1.5\% & 26\% & 1.4\% & - & - 
\\ \midrule
squark--chargino & 2.1\% & 0.5\% & 2.1\% & 0.7\% & 1.4\% & 0.7\% & 0.2\% & 0.2\% & - & -
\\ \midrule
squark--neutralino & 1.7\% & 0.4\% & 1.8\% & 0.4\% & 1.2\% & 0.6\% & 0.6\% & 0.2\% & - & -
\\ \midrule
other & 9.5\% & 0.7\% & 9.3\% & 0.8\% & 11\% & 0.8\% & 1.9\% & 0.3\% & 1.1\% & 0.1\%
\\ \bottomrule
\end{tabular}
\end{center}
\end{table*}


The Group 1 models CS4d and CS6 are not minimal supergravity; they are
more general high scale MSSM models based on the compressed
supersymmetry idea of Martin \cite{Martin:2007gf,Baer:2007uz}.  The
high scale input parameters are shown in Table \ref{table:csinputs}.
We have used $m_{\rm top}$ = 175 GeV and the spectrum generator
combination {\tt SuSpect v2.34} with {\tt SUSY-HIT v1.1}
\cite{Djouadi:2002ze,Djouadi:2006bz}.  Model CS4d is in fact part of
the compressed SUSY model line defined in \cite{Martin:2007gf}. Model
CS6 is a modification of compressed SUSY where all of the squarks have
been made very heavy, $\gsim 2$ TeV.

The superpartner mass spectra of the Group 1 models are displayed in
Figure \ref{fig:group1spectra}. One notes immediately that all of the
mSUGRA models are more similar to each other than they are to either
of the more general MSSM models CS4d and CS6; this shows the
limitations of the usual SUSY analyses that do not go beyond
mSUGRA. As their name implies, the compressed SUSY models CS4d and CS6
have a compressed gaugino spectrum relative to mSUGRA; this produces
either a light gluino (as in CS6) or a heavy LSP (as in CS4d).

The relative frequency of various LHC superpartner production
processes is summarized in Table \ref{table:myproductiontable}, for
the Group 1 models both before and after our event selection.
The production fractions are much more similar after the
event selection than before it; this is expected because the selection
shapes the kinematics of the surviving sample.
Gluino pair production dominates for
model CS6, while squark-gluino and squark-squark production dominate
for the other four models. Pair production of the lightest stop is
important for model CS4d before the selection cuts, 
but after the event selection very few of these events remain.

\begin{table}[thb]
\begin{center}
\caption{\label{brsummary} Summary of most relevant superpartner
  decays for the MSSM models LM2p, LM5, LM8, CS4d and CS6.}
\begin{tabular}{rrrrrr}
\toprule
&\multicolumn{1}{c}{LM2p}&\multicolumn{1}{c}{LM5}
&\multicolumn{1}{c}{LM8}&\multicolumn{1}{c}{CS4d}
&\multicolumn{1}{c}{CS6}
\\ \midrule
$\tilde{g} \to \tilde{q}q$              & 45\% & 45\% &  - &  - &  -
\\
$\to \tilde{b}_1b$                      & 25\% & 20\% & 14\% &  2\% &  -
\\
$\to \tilde{t}_1t$                      & 16\% & 23\% & 81\% & 94\% &  -
\\
$\to q\bar{q}\tilde{\chi}_1^0$          &  - &  - &  5\% &  - & 75\%
\\ \midrule
$\tilde{u}_L \to d\tilde{\chi}_1^{\pm}$ & 64\% & 64\% & 55\% &  - &  -
\\
$\to u\tilde{\chi}_2^0$                 & 32\% & 32\% & 27\% &  - &  -
\\
$\to u\tilde{g}$                        &  - &  - &  - & 83\% & 85\%
\\ \midrule
$\tilde{u}_R \to u\tilde{\chi}_1^0$     & 99\% & 99\% & 62\% & 92\% &  -
\\
$\to u\tilde{g}$                        &  - &  - & 38\% &  - & 85\%
\\ \midrule
$\tilde{b}_1 \to t\tilde{\chi}_1^-$     & 42\% & 36\% & 35\% & 20\% &  9\%
\\
$\to b\tilde{\chi}_2^0$                 & 29\% & 23\% & 22\% & 14\% &  5\%
\\
$\to b\tilde{\chi}_1^0$                 &  7\% &  2\% &  1\% & 50\% &  -
\\
$\to b\tilde{g}$                        &  - &  - &  - &  - & 85\%
\\ \midrule
$\tilde{t}_1 \to b\tilde{\chi}_1^+$     & 45\% & 43\% & 42\% &  - &  -
\\
$\to t\tilde{\chi}_1^0$                 & 22\% & 25\% & 30\% &  - &  4\%
\\
$\to t\tilde{g}$                        &  - &  - &  - &  - & 96\%
\\
$\to bW^+\tilde{\chi}_1^0$              &  - &  - &  - & 100\% & -
\\ \midrule
$\tilde{\chi}_1^{\pm} \to W^{\pm}\tilde{\chi}_1^0$ 
                                        &  5\% & 97\% & 100\% & 100\% & 2\%
\\
$\to \tilde{\tau}_1^{\pm}\nu_{\tau}$
                                        & 95\% &  - &  - &  - & 77\%
\\ \midrule
$\tilde{\chi}_2^0 \to Z\tilde{\chi}_1^0$ 
                                        &  1\% & 11\% & 100\% & 100\% & -
\\
$\to h\tilde{\chi}_1^0$                 &  3\% & 85\% &   - &   - & 2\%
\\
$\to \tilde{\tau}_1\tau$                & 96\% &  3\% &   - &   - & 77\%
\\ \midrule
$\tilde{\tau}_1 \to \tau\tilde{\chi}_1^0$
                                        & 100\% & 100\% & 88\% & 98\% & 100\%
\\ \bottomrule
\end{tabular}
\end{center}
\end{table}


Table \ref{brsummary} shows the most relevant superpartner decay
branching fractions. For models LM2p and LM5, gluino decay is
predominantly to quark$+$squark; for LM8 and CS4d it is dominantly to top
and the lightest stop, and gluinos decay in CS6 mostly through
the three-body mode $qq\tilde{\chi}_1^0$.  For models LM2p, LM5 and LM8,
left-squarks cascade through quark$+$chargino or quark$+$second neutralino; 
right-squarks have a two-body decay to quark$+$LSP; right-squarks in
model LM8 also have a large branching to quark$+$gluino. In model CS4d
left-squarks decay almost entirely to quark$+$gluino, while
right-squarks decay almost entirely to quark$+$LSP; for CS6 all squarks
except the stop decay dominantly to quark$+$gluino.

In models LM2p, LM5 and LM8 the decays of the lightest stop split between
$b+$chargino and top$+$LSP; for CS4d $\tilde{t}_1$ decays 100\% via the
three-body mode $bW^+\tilde{\chi}_1^0$, while for CS6 almost all of
the decays are to top$+$gluino.

Chargino decay is dominated by decays to the lightest stau and a
neutrino for models LM2p and CS6, and by decays to $W+$LSP for models
LM5, LM8 and CS4d.  The second neutralino $\tilde{\chi}_2^0$ decays
almost entirely to $\tau+$stau for models LM2p and CS6, and goes 100\% to
$Z+$LSP in models LM8 and CS4d.  The LM5 model has the distinct feature
that 85\% of $\tilde{\chi}_2^0$ decays are to Higgs$+$LSP.

Table \ref{fssummary} shows the most significant inclusive final
states for the Group 1 models.  By final state we mean that all
unstable superpartners have decayed, while Standard Model particles
are left undecayed.  We use $q$ to denote any first or second
generation quark or antiquark, but list bottom and top quarks
separately.  The percentage frequency of each final state is with
respect to the events passing our selection. The final states are
inclusive, thus e.g. the events in the $qqq\;\tilde{\chi}_1^0\tilde{\chi}_1^0$
final state are a subset of those in the
$qq\;\tilde{\chi}_1^0\tilde{\chi}_1^0$ final state, and the
total percentages in each column exceed 100\% .
By the same token, most exclusive final states actually
have more partons than are listed for the corresponding inclusive entries
in Table \ref{fssummary}, so even at   
leading order parton level they produce more jets.

For models LM2p, LM5 and CS6, the dominant inclusive final
state is $qq\tilde{\chi}_1^0\tilde{\chi}_1^0 + X$, i.e.
multijets plus missing energy from the two LSPs. This is the
motivation behind the design of our analysis. For model CS4d, the most
likely production is squark-gluino followed by squark decay to
quark$+$LSP; the gluino then decays to top$+$stop, with the stop decaying
via the three-body mode $bW^+\tilde{\chi}_1^0$.  The most popular
exclusive final state is thus $btWq\tilde{\chi}_1^0\tilde{\chi}_1^0$.
Similarly, for LM8 the most popular exclusive final states
are $btWq\tilde{\chi}_1^0\tilde{\chi}_1^0$ and $ttq\tilde{\chi}_1^0\tilde{\chi}_1^0$,
from squark-gluino production followed by gluino decay to top$+$stop.

Final states with $W$'s are prevalent in models LM5, LM8 and CS4d.
The LM2p model stands out because of the high probability of taus in the
final state.  Model LM5 produces a significant number of light Higgses
from superpartner decays.  Model LM8 has a large fraction of events
with $Z$ bosons in the final state.  Model CS4d is enriched in final
states with multiple tops and $W$'s, of which one representative example is
shown: $bbttWW\tilde{\chi}_1^0\;\tilde{\chi}_1^0$.


\begin{table}[thb]
\begin{center}
\caption{\label{fssummary} Summary of significant inclusive partonic final
  states for the Group 1 MSSM models LM2p, LM5, LM8, CS4d and CS6.  By final
  state we mean that all unstable superpartners have decayed, while
  Standard Model particles are left undecayed.  Here $q$ denotes any
  first or second generation quark or antiquark, and more generally
  the notation does not distinguish particles from antiparticles.  The
  percentage frequency of each final state is with respect to the events
  passing our selection.The final states are
  inclusive, thus e.g. the events in the $qqq\;\tilde{\chi}_1^0\tilde{\chi}_1^0$
  final state are a subset of those in the
  $qq\;\tilde{\chi}_1^0\tilde{\chi}_1^0$ final state, and the
  total percentages in each column exceed 100\% .}
\begin{tabular}{rrrrrr}
\toprule
&\multicolumn{1}{c}{LM2p}&\multicolumn{1}{c}{LM5}
&\multicolumn{1}{c}{LM8}&\multicolumn{1}{c}{CS4d}
&\multicolumn{1}{c}{CS6}
\\ \midrule
$qq\;\tilde{\chi}_1^0\;\tilde{\chi}_1^0$                                & 57\% & 61\% &  34\% &  38\% &  98\%
\\ \midrule
$qqq\;\tilde{\chi}_1^0\;\tilde{\chi}_1^0$                               & 20\% & 19\% &   3\% &   4\% &  79\%
\\ \midrule
$qqqq\;\tilde{\chi}_1^0\;\tilde{\chi}_1^0$                              &  1\% &  1\% &   1\% &   1\% &  77\%
\\ \midrule
$\tau\;\nu_{\tau}\;q\;\tilde{\chi}_1^0\;\tilde{\chi}_1^0$               & 39\% &  1\% &   -   &   - &     1\%
\\ \midrule
$\tau\tau\;q\;\tilde{\chi}_1^0\;\tilde{\chi}_1^0$                       & 25\% &  1\% &   -   &   - &     1\%
\\ \midrule
$b\;q\;\tilde{\chi}_1^0\;\tilde{\chi}_1^0$                              & 30\% & 25\% &  33\% &  69\% &  19\%
\\ \midrule
$b\;t\;W\;q\;\tilde{\chi}_1^0\;\tilde{\chi}_1^0$                        & 10\% & 19\% &  31\% &  67\% &   -
\\ \midrule
$W\;q\;\tilde{\chi}_1^0\;\tilde{\chi}_1^0$                              & 25\% & 52\% &  56\% &  93\% &   -
\\ \midrule
$h\;q\;\tilde{\chi}_1^0\;\tilde{\chi}_1^0$                              &  3\% & 20\% &    -  &   -   &   -
\\ \midrule
$tt\;q\;\tilde{\chi}_1^0\;\tilde{\chi}_1^0$                             &  9\% &  4\% &  40\% &  11\% &   2\%
\\ \midrule
$Z\;q\;\tilde{\chi}_1^0\;\tilde{\chi}_1^0$                              & 10\% &  8\% &  35\% &  11\% &   -
\\ \midrule
$Z\;W\;q\;\tilde{\chi}_1^0\;\tilde{\chi}_1^0$                           &  2\% &  6\% &  23\% &   6\% &   -
\\ \midrule
$bb\;tt\;WW\;\tilde{\chi}_1^0\;\tilde{\chi}_1^0$                        &  -   &   -  &   2\% &  18\% &   -
\\ \bottomrule
\end{tabular}
\end{center}
\end{table}


\begin{table}[thb]
\begin{center}
\caption{\label{table:G1counts} Estimated number of events passing our
selection per 100 pb$^{-1}$ of integrated luminosity, for
the Group 1 models LM2p, LM5, LM8, CS4d and CS6. These estimates use
NLO cross sections and the {\tt CTEQ5L} pdfs.}
\begin{tabular}{rrrrr}
\toprule
\multicolumn{1}{c}{LM2p}&\multicolumn{1}{c}{LM5}&\multicolumn{1}{c}{LM8}
&\multicolumn{1}{c}{CS4d}&\multicolumn{1}{c}{CS6}
\\ \midrule
211&200&195&195&212
\\ \bottomrule
\end{tabular}
\end{center}
\end{table}


Summarizing this discussion, we list the most significant features of
each model in Group 1:

{\bf Model LM2p:} 800 GeV squarks are slightly lighter than the
  gluino, and there is a 155 GeV stau. Dominant production is
  squark-gluino and squark-squark.  Left-squarks decay about
  two-thirds of the time to quark$+$chargino, and one-third to
  quark$+$LSP; right-squarks decay to quark$+$LSP. Gluino decay is mostly
  to quark$+$squark. Charginos decay to the light stau plus a neutrino, while
  the second neutralino decays to $\tau+$stau. Two-thirds of the final
  states after event selection have at least one $\tau$.

{\bf Model LM5:} 800 GeV squarks are slightly lighter than the
  gluino.  Dominant production is squark-gluino and squark pairs.
  Left-squark decays about two-thirds to quark$+$chargino,
  and one-third to quark$+$LSP; right-squarks decay to quark$+$LSP. Gluino
  decay is mostly to quark$+$squark. Charginos decay to a $W$ and an LSP, while the second
  neutralino decays to a light Higgs and an LSP. After selection more than half
  of final states have a $W$ boson, and a fifth have a Higgs.

{\bf Model LM8:} The 745 GeV gluino is slightly lighter than all
  of the squarks except $\tilde{b}_1$ and $\tilde{t}_1$. Dominant
  production is squark-gluino and squark pairs.  Left-squarks decay
  about two-thirds to quark$+$chargino, and one-third to
  quark$+$LSP; right-squarks decay two-thirds to quark$+$LSP and one-third
  to quark$+$gluino. Gluino decay is dominantly to top and a stop; the 548 GeV stops
  decay mostly to $b+$chargino or top$+$LSP. Charginos decay to $W+$LSP,
  and the second neutralino decays to $Z+$LSP. After selection 40\% of
  final states have two tops, which may or may not have the same
  sign. More than half of the final states have a $W$, more than a third
  have a $Z$, and a quarter have both a $W$ and a $Z$.

{\bf Model CS4d:} The 753 GeV gluino is in between the
  right-squark and left-squark masses. The LSP is relatively heavy,
  251 GeV, and the ratio of the gluino to LSP mass is small compared
  to mSUGRA models.  Dominant production is squark-gluino and
  squark-squark. Left-squarks decay to quark$+$gluino, and
  right-squarks decay to quark$+$LSP. Gluinos decay to top and a stop; the 352
  GeV stops decay 100\% to $bW^+\tilde{\chi}_1^0$. Two-thirds of the
  final states contain $btWq\tilde{\chi}_1^0\tilde{\chi}_1^0$, and a
  significant fraction of these contain more $b$'s, $t$'s and $W$'s.

{\bf Model CS6:} The 589 GeV gluino is much lighter than the 2
  TeV squarks, and the ratio of the gluino to LSP mass is small
  compared to mSUGRA models. Production is 92\% gluino-gluino, and
  gluinos decay predominantly via the three-body mode
  $qq\tilde{\chi}_1^0$.  The final states consist almost entirely of
  three or four quarks plus two LSPs, with a proportionate amount of
  the final state quarks being $b$'s.


\begin{figure*}[thb]
\begin{center}
\includegraphics[width=0.85\textwidth]{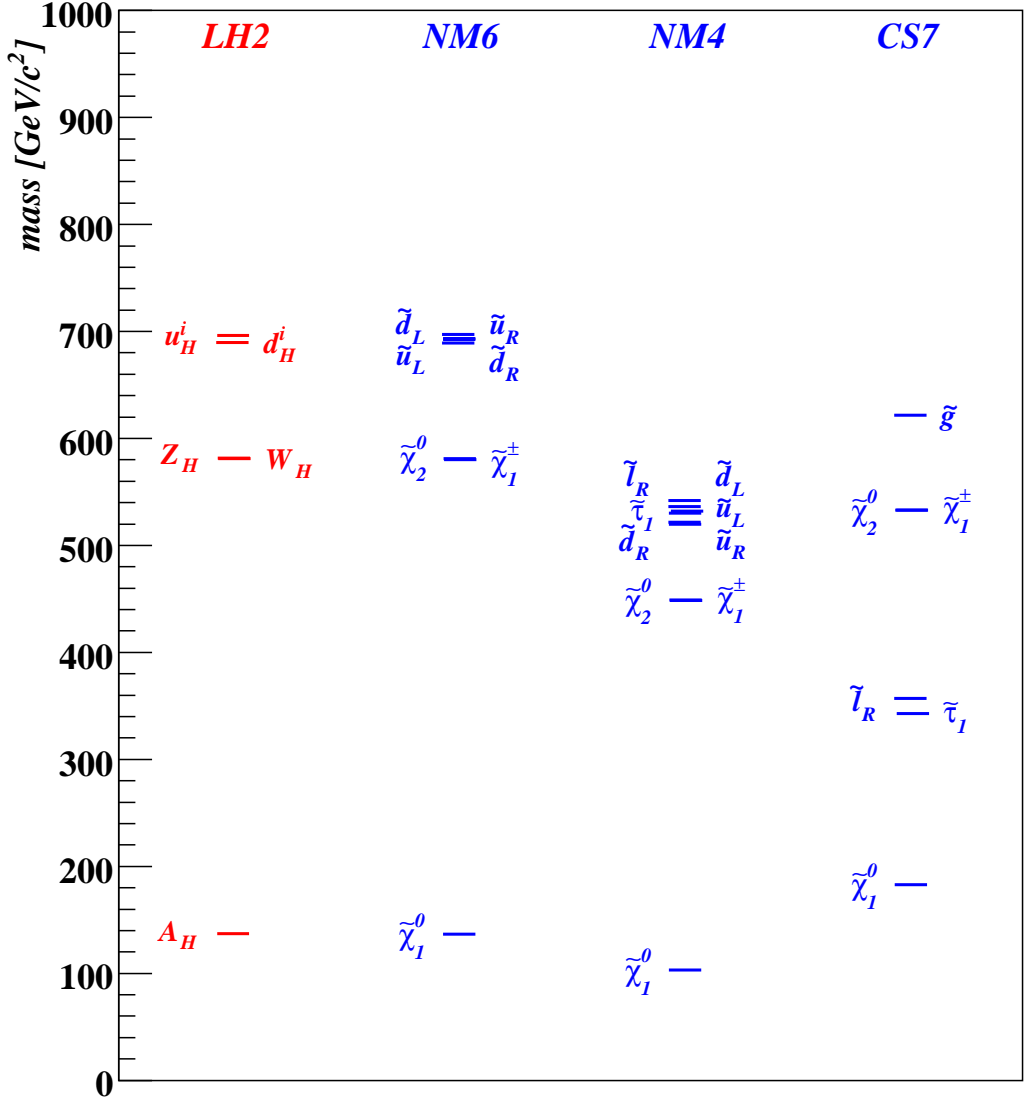}
\caption {\label{fig:group2spectra} The mass spectra of the
  models LH2, NM6, NM4 and CS7.  Only the most relevant
  partners are shown: the lighter gauginos $\tilde{\chi}_1^0$,
  $\tilde{\chi}_2^0$ and $\tilde{\chi}_1^{\pm}$, the lightest stau
  $\tilde{\tau}_1$, the right-smuon and selectron denoted collectively
  as $\tilde{\ell}_R$, the gluino,
  and the left/right up and down squarks $\tilde{u}_L$, $\tilde{u}_R$,
  $\tilde{d}_L$ and $\tilde{d}_R$. For the little Higgs model LH2,
  the relevant quark and vector partners are shown: the gauge boson partners
  $A_H$, $Z_H$, $W_H$, and the three generations of quark partners
  $u^i_H$, $d^i_H$, $i=1,2,3$.}
\end{center}
\end{figure*}


\begin{table}[htp]
\begin{center}
\caption{Parameter choices defining the little Higgs model LH2. 
We choose our conventions to agree with those found in~\cite{Hubisz:2005tx}:
$f$ is the symmetry-breaking scale, $\kappa_q^i$ is the $T$-odd quark
Yukawa coupling, $\kappa_l^i$ is the $T$-odd lepton Yukawa coupling
and $\sin \alpha$ is a mixing angle. CKM mixing has been suppressed for
our analysis.}{\label{tab:LHparams}.}
\begin{tabular}{lc}\toprule
$f$ & 700 GeV \\ \midrule
$\kappa_q^i$ & 0.55 \\ \midrule
$\kappa_l^i$ & 2.0 \\ \midrule
$\sin \alpha$ & 0.17 \\
\bottomrule
\end{tabular}
\end{center}
\end{table}


\begin{table}[thb]
\begin{center}
\caption{\label{table:G2inputs} Input parameters for the MSSM models
  NM6, NM4 and CS7.  The notation conforms to
  \cite{Djouadi:2006bz,Djouadi:2002ze}.  The mass parameters and
  trilinear $A$ parameters have units of GeV.}
\begin{tabular}{rrrr}
\toprule
& NM6 & NM4 & CS7
\\ \midrule
$M_1$                                     &   138    &  105    &  428 
\\ \midrule
$M_2$                                     &   735    &  466    &  642
\\ \midrule
$M_3$                                     &  2082    & 1600    &  214
\\ \midrule
$A_t$                                     &     0    &  -50    & -321
\\ \midrule
$A_b$                                     &  4000    &    0    & -321
\\ \midrule
$A_{\tau},A_t,A_b,A_e,A_u,A_d$            &     0    &    0    & -321
\\ \midrule
$M_{Q_L}$                                 &   755    &  590    & 2000
\\ \midrule
$M_{t_R}$                                 &   760    &  580    & 2000
\\ \midrule
$M_{q_u}$                                 &   770    &  590    & 2000
\\ \midrule
$M_{u_R},M_{b_R}$                         &   770    &  580    & 2000
\\ \midrule
$M_{d_R}$                                 &   765    &  580    & 2000
\\ \midrule
$M_{\tau_L},M_{\tau_R},M_{e_L},M_{e_R}$   &  2500    &  540    &  340
\\ \midrule
$M_{h_u}^2,M_{h_d}^2$                     & 115600  & 115600   & 115600
\\ \midrule
tan$\,\beta$                              &   10    &   10     & 10
\\ \midrule
sign($\mu$)                               &    +    &    +     & +
\\ \bottomrule
\end{tabular}
\end{center}
\end{table}


\subsection{Group 2}
Group 2 consists of three look-alike models: LH2, NM4 and CS7, and
a comparison model NM6.
LH2 is a littlest Higgs model with conserved $T$ parity.
The parameter choices defining this model are shown in Table~\ref{tab:LHparams}.
The mass spectrum of the lighter partners is shown in 
Figure~\ref{fig:group2spectra}; not shown are the heavier top 
partners $T_+$, $T_-$ with tuned masses 3083 and 3169 GeV
respectively, the charged lepton partners $\ell^1_H$, $\ell^2_H$, $\ell^3_H$ with mass 2522 GeV 
and the neutrino partners $\nu^1_H$, $\nu^2_H$, $\nu^3_H$ with mass 2546 GeV.
Model LH2 is consistent with all current experimental constraints \cite{Hubisz:2005tx,Asano:2006nr}.

NM6, NM4 and CS7 are all MSSM models. The high scale input parameters are listed
in Table~\ref{table:G2inputs}. 
We have used $m_{\rm top}$ = 175 GeV and the spectrum generator
{\tt SuSpect v2.34}.
The mass spectra are shown in Figure~\ref{fig:group2spectra};
not shown are the heavy gluinos of NM6 and NM4 with masses 2000 and 1536 GeV
respectively, and the $\gsim 2$ TeV squarks of model CS7.

The SUSY model NM6 was chosen to have a spectrum identical to that
of the little Higgs model LH2, apart from the heavy gluino that has no
counterpart in LH2. Thus to a good approximation these two models
differ only by the spins of the partners. While LH2 and NM6 are in this
sense twins, they are \textit{not} look-alikes of our benchmark inclusive
missing energy analysis. Models NM4 and CS7, by contrast, \textit{are} SUSY
look-alikes of the little Higgs model LH2. The superpartner spectrum of
NM4 is roughly similar to the partner spectrum of LH2, but the superpartners
are lighter. The spectrum of CS7 has no similarity to that of LH2.

The relative frequency of various LHC little Higgs partner production
processes are shown in
Table~\ref{tab:LHprod}, for the LH2 model both before and after our 
event selection.
For LH2 the predominant process is
$gg$ or $q\bar{q}$ partons initiating QCD production of a heavy partner 
quark-antiquark pair; this process is completely equivalent to $t\bar{t}$
production at the LHC. The most striking feature of 
Table~\ref{tab:LHprod} is that nearly
half of the total production involves weak interactions. For example the
second largest production mechanism, 14\% of the total, has two valence 
quarks in the initial state producing a pair of first 
generation heavy partner quarks;
at tree-level this is from $s$-channel annihilation into a $W$ and $t$-channel
exchange of a $Z_H$ or $A_H$ partner.

The superpartner production at the LHC for the SUSY models NM6, 
NM4 and CS7 is summarized in Table~\ref{table:G2productiontable}. 
For NM6 and NM4 a major contribution
is from $gg$ or $q\bar{q}$ partons initiating QCD production of a 
squark-antisquark pair. Production of a first generation squark
 pair from two initial state valence
quarks is also important; in contrast to the LH2 non-SUSY analog this is a QCD
process with $t$-channel exchange of the heavy gluino. For model CS7, 
which has a light gluino and very heavy squarks, 96\% of the production 
is gluino pairs.

The primary decay modes for the lighter LH2 partners are 
shown in Table~\ref{tab:LHbranch},
while those for the SUSY models are summarized in Table~\ref{table:G2brsummary}.
Tables~\ref{tab:LHfinalstates} and \ref{table:G2fssummary} display 
the most significant inclusive partonic final states for the Group 2 models.

For LH2, a large fraction of heavy partner quarks have a direct 2-body
decay to a quark and an $A_H$ WIMP. 
The other heavy partner quark decay mode is a two stage
cascade decay via the $W_H$ and $Z_H$ partner
bosons. Since the $W_H$ decays 100\% to $WA_H$ while the $Z_H$ decays 100\%
to $hA_H$, a large fraction of events have a $W$ or a Higgs in the final state.

Analogous statements apply to the SUSY models NM6 and NM4. We see
that 100\% of right-squarks and a significant fraction of left-squarks
undergo a direct 2-body decay to quark$+$LSP. 
The rest have mostly a two stage
cascade via the lightest chargino $\tilde{\chi}_1^{\pm}$ or the second
neutralino $\tilde{\chi}_2^0$. Since $\tilde{\chi}_1^{\pm}$ decays 100\%
to $W+$LSP, while the $\tilde{\chi}_2^0$ decays dominantly to a Higgs$+$LSP,
a significant fraction of events have a $W$ or a Higgs in the final state.

For the remaining SUSY model CS7, gluino pair production is followed by
3-body decays of each gluino to a quark-antiquark pair + LSP. As can be seen in
Table~\ref{table:G2fssummary}, this leads to high jet multiplicity
but nothing else of note besides a proportionate number of $b\bar{b}$
and $t\bar{t}$ pairs.


\begin{table}[ht]
\begin{center}
\caption{Production channels for little Higgs partners in the LH2 model,
both before and after the event selection. Here $Q$ stands for any of
the quark partners $u^i_H$, $d^i_H$, $i=1,2,3$.
The total LO cross section as reported by {\tt MadEvent} is 6.5 pb.}
\label{tab:LHprod}
\begin{tabular}{lcc}\toprule
& before cuts & after cuts \\ \midrule
$Q_i\bar{Q}_i$                                  & 55\%    & 64\%  \\ \midrule
$u^i_Hd^i_H$, $u^i_Hu^i_H$, $d^i_Hd^i_H$        & 14\%    & 16\%  \\ \midrule
$d^i_HW^+_H$, $u^i_HW^-_H$                      & 12\%    &  7\%  \\ \midrule
$u^i_HZ_H$, $u^i_HA_H$, $d^i_HZ_H$, $d^i_HA_H$  &  9\%    &  5\%  \\ \midrule
$Q_i\bar{Q}_j$, $i\ne j$                        &  3\%    &  3\%  \\ \midrule 
other                                           &  7\%    &  5\%  \\
\bottomrule
\end{tabular}
\end{center}
\end{table}


\begin{table*}[hbt]
\begin{center}
\caption{\label{table:G2productiontable} Summary of LHC superpartner
  production for the Group 2 MSSM models NM6, NM4 and CS7. The
  relative percentages are shown for each model, both before and after
  the event selection. Here $\tilde{q}_i$ denotes any of the 
  three generations of left and right squarks.
  Note that squark--chargino
  includes the production of either chargino, and squark-neutralino
  includes all of the four neutralinos. 
  The LO total cross sections are as reported by {\tt MadEvent}.}
\begin{tabular}{lrrrrrr}
\toprule
&\multicolumn{2}{c}{NM6}&\multicolumn{2}{c}{NM4}
&\multicolumn{2}{c}{CS7}
\\ \midrule
LO cross section (pb) & \multicolumn{2}{c}{2.3} & \multicolumn{2}{c}{10.3} 
& \multicolumn{2}{c}{5.0}
\\\midrule
&\multicolumn{1}{l}{before}&\multicolumn{1}{l}{after}
&\multicolumn{1}{l}{before}&\multicolumn{1}{l}{after}
&\multicolumn{1}{l}{before}&\multicolumn{1}{l}{after}
\\ [-3pt]
&\multicolumn{1}{l}{cuts}&\multicolumn{1}{l}{cuts}
&\multicolumn{1}{l}{cuts}&\multicolumn{1}{l}{cuts}
&\multicolumn{1}{l}{cuts}&\multicolumn{1}{l}{cuts}
\\ \midrule
$\tilde{q}_i\bar{\tilde{q}}_i$           & 31\%   & 29\%   & 34\%  & 26\%   & -    & -
\\ \midrule
$\tilde{u}\tilde{d}$, $\tilde{u}\tilde{u}$, $\tilde{d}\tilde{d}$ 
                                         & 32\%   & 28\%   & 29    & 23\%   & -    & -
\\ \midrule
squark--gluino                           & 3\%    & 10\%   & 5\%   & 23\%   & 4\%  & 8\%
\\ \midrule
gluino--gluino                           & -      & -      & -     & -      & 96\% & 91\%
\\ \midrule
squark--chargino                         & 2\%    & 2\%    & 3\%   & 1\%    & -    & -
\\ \midrule
squark--neutralino                       & 4\%    & 1\%    & 4\%   & -      & -    & -
\\ \midrule
$\tilde{q}_i\bar{\tilde{q}}_j$, $i\ne j$ & 15\%   & 17\%   & 17\%  & 14\%   & -    & -
\\ \midrule
other                                    & 13\%   & 13\%   & 8\%   & 13\%   & -    & -
\\ \bottomrule
\end{tabular}
\end{center}
\end{table*}



\begin{table}[htp]
\begin{center}
\caption{Decay modes for the lighter little Higgs partners of model LH2.}
\label{tab:LHbranch}
\begin{tabular}{rlc}\toprule
$d^i_H$, $i=1,2$                   & $\rightarrow u^i W_H$             &  52\%    \\ \midrule
\                                  & $\rightarrow d^i Z_H$             &  26\%    \\ \midrule
\                                  & $\rightarrow d^i A_H$             &  22\%    \\ \midrule
$d^3_H$                            & $\rightarrow b Z_H$               &  54\%    \\ \midrule
\                                  & $\rightarrow b A_H$               &  46\%    \\ \midrule
$u^i_H$, $i=1,2$                   & $\rightarrow d^i W_H$             &  31\%    \\ \midrule
\                                  & $\rightarrow u^i Z_H$             &  15\%    \\ \midrule
\                                  & $\rightarrow u^i A_H$             &  54\%    \\ \midrule
$u^3_H$                            & $\rightarrow b W_H$               &  41\%    \\ \midrule
\                                  & $\rightarrow t A_H$               &  59\%    \\ \midrule
$W_H$                              & $\rightarrow W A_H$               & 100\%    \\ \midrule
$Z_H$                              & $\rightarrow h A_H$               & 100\%    \\
\bottomrule 
\end{tabular}  
\end{center}
\end{table}


\begin{table}[thb]
\begin{center}
\caption{\label{table:G2brsummary} Summary of most relevant superpartner
  decays for the MSSM models NM6, NM4 and CS7.}
\begin{tabular}{rrrr}
\toprule
&\multicolumn{1}{c}{NM6}&\multicolumn{1}{c}{NM4}
&\multicolumn{1}{c}{CS7}
\\ \midrule
$\tilde{g} \to \tilde{q}q$              & 66\% & 67\% &  - 
\\
$\to \tilde{b}b$                        & 17\% & 17\% &  - 
\\
$\to \tilde{t}_1t$                      & 17\% & 16\% &  -
\\
$\to q\bar{q}\tilde{\chi}_1^0$          &  -   &  -   & 99\% 
\\ \midrule
$\tilde{u}_L \to d\tilde{\chi}_1^{\pm}$ & 39\% & 59\% & 12\%
\\
$\to u\tilde{\chi}_1^0$                 & 44\% & 12\% &  - 
\\ \midrule
$\tilde{u}_R \to u\tilde{\chi}_1^0$     & 100\%& 100\%&  4\%
\\ \midrule
$\tilde{b}_1 \to b\tilde{\chi}_2^0$     & 24\% & 14\% &  6\% 
\\
$\to b\tilde{\chi}_1^0$                 & 70\% & 86\% &  - 
\\ \midrule
$\tilde{t}_1 \to b\tilde{\chi}_1^+$     & 40\% & 27\% &  - 
\\
$\to t\tilde{\chi}_1^0$                 & 60\% & 73\% &  5\% 
\\ \midrule
$\tilde{\chi}_1^{\pm} \to W^{\pm}\tilde{\chi}_1^0$ 
                                        & 100\%& 100\%&  1\%
\\
$\to \tilde{\tau}\nu_{\tau}$
                                        &  -   &  -   & 35\% 
\\ 
$\to \tilde{\nu}\ell$
                                        &  -   &  -   & 28\% 
\\ 
$\to \tilde{\nu}\tau$
                                        &  -   &  -   & 17\% 
\\ \midrule
$\tilde{\chi}_2^0 \to Z\tilde{\chi}_1^0$ 
                                        & 22\% & 19\% &  -
\\
$\to h\tilde{\chi}_1^0$                 & 78\% & 81\% &  - 
\\
$\to \tilde{\tau}\tau$                  &  -   &  -   & 39\% 
\\ 
$\to \tilde{\nu}\nu$                    &  -   &  -   & 45\% 
\\ 
$\to \tilde{\ell}\ell$                  &  -   &  -   & 16\% 
\\ \bottomrule
\end{tabular}
\end{center}
\end{table}


\begin{table}[htp]
\begin{center}
\caption{Significant inclusive partonic final states for the little Higgs model LH2.
The
percentage frequency of each final state is with respect to the events
passing our selection.}
\label{tab:LHfinalstates}
\begin{tabular}{lr}\toprule
$qq\;A_HA_H$             & 64\% \\ \midrule
$W\;qq\;A_HA_H$          & 39\% \\ \midrule
$h\;qq\;A_HA_H$          & 22\% \\ \midrule
$bb\;A_HA_H$             & 14\% \\ \midrule
$WW\;qq\;A_HA_H$         & 8\% \\ \midrule
$hh\;bb\;A_HA_H$         & 4\% \\ \midrule
$hh\;qq\;A_HA_H$         & 3\% \\ \midrule
$tt\;A_HA_H$             & 3\% \\
\bottomrule
\end{tabular}
\end{center}
\end{table}


\begin{table}[thb]
\begin{center}
\caption{\label{table:G2fssummary} Significant inclusive partonic final
  states for the Group 2 MSSM models NM6, NM4 and CS7. The
  percentage frequency of each final state is with respect to the events
  passing our selection.}
\begin{tabular}{rrrr}
\toprule
&\multicolumn{1}{c}{NM6}&\multicolumn{1}{c}{NM4}
&\multicolumn{1}{c}{CS7}
\\ \midrule
$qq\;\tilde{\chi}_1^0\;\tilde{\chi}_1^0$                                & 84\% & 83\% & 100\%   
\\ \midrule
$qqq\;\tilde{\chi}_1^0\;\tilde{\chi}_1^0$                               &  8\% & 16\% & 100\%   
\\ \midrule
$qqqq\;\tilde{\chi}_1^0\;\tilde{\chi}_1^0$                              &  -   &  -   &  95\%   
\\ \midrule
$bb\;q\;\tilde{\chi}_1^0\;\tilde{\chi}_1^0$                             &  2\% &  5\% &  11\%   
\\ \midrule
$W\;qq\;\tilde{\chi}_1^0\;\tilde{\chi}_1^0$                             & 26\% & 35\% &   -   
\\ \midrule
$h\;q\;\tilde{\chi}_1^0\;\tilde{\chi}_1^0$                              & 14\% & 19\% &   -    
\\ \midrule
$tt\;q\;\tilde{\chi}_1^0\;\tilde{\chi}_1^0$                             &  1\% &  1\% &  11\%    
\\ \midrule
$Z\;q\;\tilde{\chi}_1^0\;\tilde{\chi}_1^0$                              &  4\% &  5\% &   -   
\\ \midrule
$WW\;q\;\tilde{\chi}_1^0\;\tilde{\chi}_1^0$                             &  4\% &  9\% &   -   
\\ \bottomrule
\end{tabular}
\end{center}
\end{table}


\begin{table}[thb]
\begin{center}
\caption{\label{table:G2counts} Estimated number of events passing our
selection per 100 pb$^{-1}$ of integrated luminosity, for
the Group 2 models LH2, NM6, NM4 and CS7. These estimates use
leading order cross sections and the {\tt CTEQ5L} pdfs.}
\begin{tabular}{rrrr}
\toprule
\multicolumn{1}{c}{LH2}&\multicolumn{1}{c}{NM6}&\multicolumn{1}{c}{NM4}
&\multicolumn{1}{c}{CS7}
\\ \midrule
94&43&97&91
\\ \bottomrule
\end{tabular}
\end{center}
\end{table}


\subsection{Comparison of models differing only by spin}
We have already noted that SUSY model NM6 has a superpartner spectrum
almost identical to the heavy partner spectrum of the non-SUSY little
Higgs model LH2. The only relevant difference, other than the spins of
the partners, is that model NM6 has a very 
heavy 2 TeV gluino that has no analog in LH2.
Despite being very heavy, the gluino does make a significant contribution
to squark-squark production via $t$-channel exchange.

This pair of models provides the opportunity for a 
comparison of realistic models that within a good approximation differ
only by the spins of the partner particles.
If it turned out that these two models were look-alikes in our
benchmark inclusive missing energy analysis, then discriminating them
would be physically equivalent to determining the spins of at least
some of the heavy partners.

It is an ancient observation (see e.g. \cite{Farrar:1980uy})
that models differing only by the
spins of the new heavy exotics have significant differences in
total cross section. The most familiar example is the comparison
of pair production of heavy leptons near threshold with pair
production of spinless sleptons. For mass $m$ and total
energy $\sqrt{s}$, the lepton cross section is proportional
to $\beta$,
\be
\label{eq:firstbetadef}
\textstyle{
\beta \equiv \sqrt{1-\frac{4
m^2}{s}}
} \; ,
\ee 
while the slepton cross section is proportional to $\beta^3$.
Thus slepton production is suppressed near threshold
compared to production of heavy leptons with the same mass.
A somewhat less familiar fact is that the sleptons never
catch up: even if we introduce both left and right sleptons, to
match the degrees of freedom of a Dirac lepton, the total
cross section for left+right slepton pairs is one-half that of 
Dirac lepton pairs
in the high energy limit $\beta\to 1$.

For the hadroproduction relevant to our models LH2 and NM6,
the discussion is more complicated: the most relevant details 
and references are presented in our Appendix.
From Tables~\ref{tab:LHprod} and \ref{table:G2productiontable},
we see a large difference in the leading order total LHC cross sections
for these models: 6.5 pb for LH2 versus only 2.3 pb for NM6.
Thus the non-SUSY twin has almost a factor of three cross section
enhancement, in spite of the fact that the SUSY model benefits from
some extra production mediated by gluinos.

The possibility of distinguishing SUSY from non-SUSY twins at the
LHC using total cross section was first suggested by
Datta, Kane and Toharia~\cite{Datta:2005vx}, and studied in
more detail in \cite{Kane:2008kw}.
To implement this idea,
we must also compare the relative efficiencies of the SUSY and
non-SUSY twins in a real analysis, since what is measured in
an experiment is not total cross section but rather
cross section times efficiency.

An important observation is that the $p_T$ distributions, in addition
to the total cross sections, have large differences due solely to
differences in spin. As an example, consider the LHC production
of a pair of 500 GeV heavy quarks, versus the production of a
pair of 500 GeV squarks. We can compare the $p_T$ distributions
by computing
\be\label{eq:dlogsig}
\frac{d{\rm log}\,\sigma}{dp_T} = \frac{1}{\sigma}\frac{d\sigma}{dp_T}
\,
\ee
where we factor out the difference in the total cross sections.
Using the analytic formulae reviewed in the Appendix, we
have computed (\ref{eq:dlogsig}) for two relevant partonic
subprocesses. The first is gluon-gluon initiated production,
for which the fully differential cross sections are given in
(\ref{eq:ggQdiffcross}) and (\ref{eq:ggsqdiffcross}); at
leading order this arises from an $s$-channel annihilation diagram,
a gluon seagull for the squark case,
and $t$ and $u$ channel exchanges of either
the spin 1/2 heavy quark or the spin 0 squark. 
The second example
is quark-antiquark initiated production, in the simplest case
where the quark flavor does not match the quark/squark partner flavor;
at leading order there is only one diagram: $s$-channel annihilation.
The fully differential cross sections are given in
(\ref{eq:Qdiffcross}) and (\ref{eq:sqdiffcross}).

\begin{figure}[thb]
\includegraphics[width=0.46\textwidth]{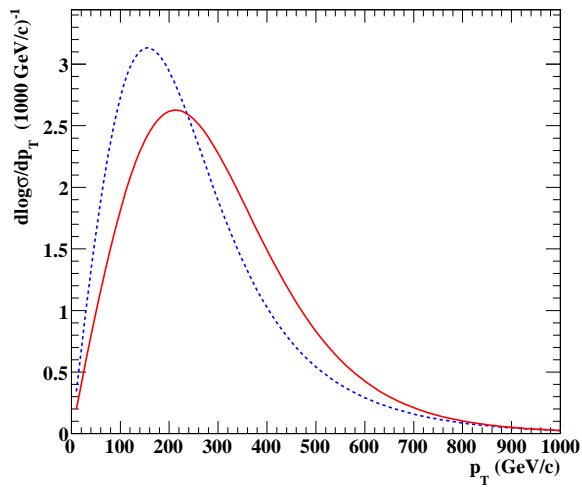}
\caption {\label{fig:ggptdist} Comparison of the normalized $p_T$ distributions
for leading order $gg$ initiated production of a pair of 500 GeV particles.
The solid (red) line corresponds to quark-antiquark pair; the dot-dashed (blue)
line to a squark-antisquark pair. The distributions have been integrated
over the parton fluxes using the {\tt CTEQ5L} pdfs.}
\end{figure}

\begin{figure}[thb]
\includegraphics[width=0.46\textwidth]{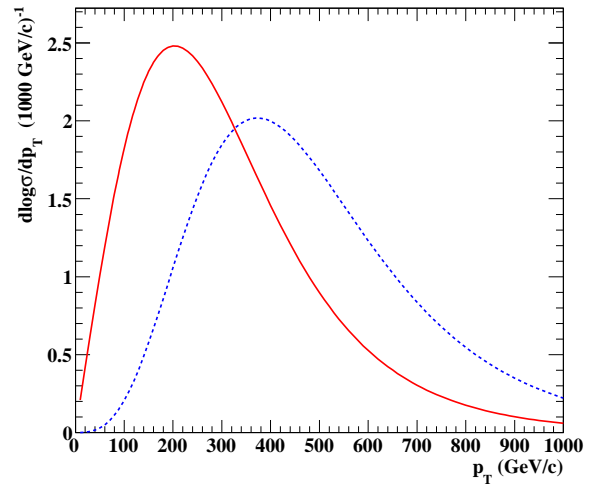}
\caption {\label{fig:qqbarptdist} Comparison of the normalized $p_T$ distributions
for leading order $q\bar{q}$ initiated production of a pair of 500 GeV particles,
for the case that the initial parton flavor does not match the final parton flavor.
The solid (red) line corresponds to quark-antiquark pair; the dot-dashed (blue)
line to a squark-antisquark pair.
The distributions have been integrated
over the parton fluxes using the {\tt CTEQ5L} pdfs.}
\end{figure}

For this simple example, we have integrated the fully differential
cross sections over the parton fluxes, using the {\tt CTEQ5L} parton
distribution functions. The resulting normalized $p_T$
distributions are shown in Figures \ref{fig:ggptdist} and \ref{fig:qqbarptdist}.
For the $gg$ initiated production the SUSY case has a significantly
softer $p_T$ distribution, while for the $q\bar{q}$ initiated production 
the SUSY case has a significantly harder $p_T$ distribution.

\begin{figure}[thb]
\includegraphics[width=0.46\textwidth]{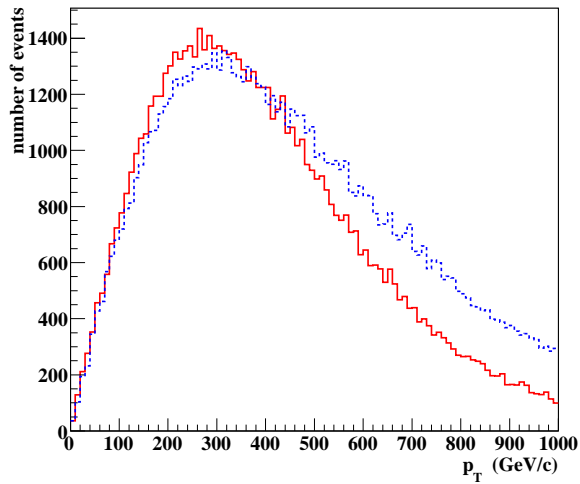}
\caption {\label{fig:LH2NM6ptdist} Comparison of the $p_T$ distributions
for heavy quarks from the little Higgs model LH2 and squarks from
the ``twin'' SUSY model NM6.
The solid (red) line corresponds to heavy quark partners from model LH2; the dashed (blue)
line to squarks from model NM6.
For both models 100,000 events were generated using {\tt MadGraph} and the {\tt CTEQ5L} pdfs;
no selection was applied.}
\end{figure}

We see similar differences in the complete models LH2 and NM6.
Figure \ref{fig:LH2NM6ptdist} shows a comparison of the $p_T$
distributions for heavy quark partner production from LH2
and squark production for NM6. All of the leading order partonic
subprocesses are combined in the plot, and no event selection
has been performed. The $p_T$ distribution for the
SUSY model is harder than for the non-SUSY model.
Part of this net effect is due to intrinsic spin differences e.g.
as depicted in Figure \ref{fig:qqbarptdist}, and part is due to
SUSY diagrams with virtual gluino exchange.
One would expect SUSY events to have a higher efficiency to
pass our missing energy selection than non-SUSY events.
Indeed this is the case: 19\% of NM6 events overall pass the
selection, whereas only 14\% of LH2 events do.
The higher efficiency of SUSY NM6 events in passing the selection
somewhat compensates for the smaller total cross section
compared to the non-SUSY LH2. 

The event counts can be obtained by
multiplying each total cross section times the total efficiency
times the integrated luminosity. For a 100 pb$^{-1}$ sample,
the total signal count is 94 events for model LH2 and 43 events
for model NM6. The net result is that although LH2 and NM6
are twins in the sense of their spectra, they are \textit{not}
missing energy look-alikes in our benchmark analysis.
Thus the good news is that, for models that differ only
(or almost only) by spin, the event count in the discovery data
sample is already good enough to discriminate them.
This is one of the important conclusions of our study.

However this is also something of an academic exercise, since
in the real experiment we will need to discriminate a
large \textit{class} of SUSY models from a large \textit{class}
of non-SUSY models. In this comparison a SUSY model can
be a look-alike of a non-SUSY model even though the spectra
of partner particles don't match. This is what happens with
SUSY models NM4 and CS7, which are both look-alikes
of LH2. Model NM4 looks particularly challenging, since its
superpartner spectrum is basically just a lighter version
of NM6. Compared to NM6, the total cross section of NM4
is more than 4 times larger (10.3 pb) while the efficiency to pass our
missing energy selection is only half as good (9\%).
This gives a total count of 97 events for 100 pb$^{-1}$,
making NM4 a look-alike of LH2.


\section{Observables}\label{sec:5}
Having in mind an early discovery at the LHC, e.g. in the first $100$
pb$^{-1}$ of understood data, we have made conservative assumptions
about the physics objects that will be sufficiently well-understood for
use in our look-alike analysis of a missing energy discovery.

We assume that we can reconstruct and count high $E_T$ jets and hard tracks, 
as is required for our benchmark missing energy selection.
We do not assume that validated jet corrections for multijet
topologies will be available. 
We assume it will be possible to
use the uncorrected (raw) \met (without subtracting 
the momentum of muons or correcting for other calorimetric effects).

We assume the ability to reconstruct and count
high $p_T$ muons; a study
of $Z\to \mu^+\mu^-$ events is a necessary precursor to understanding
the Standard Model \met~backgrounds. It will also be possible
to count high $E_T$ electrons, however we are not yet including
electrons in our study because of the high ``fake'' rate
expected at start-up.  Multiflavor multilepton signatures are 
of great importance as model discriminators, though challenging
with small data sets; this is worthy of
a separate dedicated study~\cite{Moortgat-Pape,Baer:2008kc}.

In our study instead of applying sophisticated $b$ and $\tau$ tagging
algorithms we
isolate enriched samples of $b$ quarks and
hadronic $\tau$'s, by defining simple variables similar to the typical
components of the complete tagging algorithms: leptons in jets, track counting,
and impact parameter of charged tracks, to mention a few. 
This ``poor man's'' tagging is not sufficient to
obtain pure samples of $b$'s or $\tau$'s, but allows the discrimination of  
look-alike models with large differences in the $b$ or $\tau$ multiplicity.

\subsection{Inclusive counts and ratios}
We build our look-alike analysis strategy using simple ingredients.
We start with the four trigger boxes defined in section \ref{sec:2.2}:
MET, DiJet, TriJet and Muon20. For the simulated data samples
corresponding to each box, we compute the following inclusive counts
of jets and muons:

\begin{itemize}
\item N, the number of events in a given box after our benchmark selection.
\item N(nj), the number of events with at least n jets (n=3,4,5).
Note that N(3j) = N because of our selection.
\item N(m$\mu$-nj), the number of events with at least n jets
  and m muons (n=3,4 and m=1,2).
\item N(ss$\mu$), the number of events with at least two same-sign
  muons.
\item N(os$\mu$), the number of events with at least two opposite-sign
  muons.
\end{itemize}
In these counts a muon implies a reconstructed muon with $p_T > 20$ GeV/$c$
and $|\eta |<2.4$ with no isolation requirement.

From these inclusive counts we can define various interesting ratios.
All of the ratios are of correlated observables. Examples of ratios 
are:
\begin{itemize}
\item r(nj)(3j)$\equiv$N(nj)/N(3j), with n=4,5, a measure of jet multiplicity.
\item r(2$\mu$-nj)(1$\mu$-nj)$\equiv$N(2$\mu$-nj)/N(1$\mu$-nj), with n=3,4, a measure of muon multiplicity.
\end{itemize}

In appropriately chosen ratios of inclusive counts, important
systematic effects cancel partially or completely. For example,
if the detector simulation does not precisely reproduce the
\et~spectrum of signal jets as seen in the real detector, this
introduces a systematic error in the N(nj) counts, since
jets are only counted if they have \et$> 30$ GeV. However
we expect partial cancellation of this in the
ratio of inclusive counts r(4j)(3j). Another large
systematic in the jet counts, the pdf uncertainty, also
cancels partially in the ratios. The luminosity uncertainty
cancels completely in the jet ratios.
As we discuss below,
ratios of correlated observables are also
less sensitive to statistical fluctuations.

In order to enhance the robustness and realism of our study,
we have cast \textit{all} of our physical observables into the form
of inclusive counts and ratios thereof, and our look-alike analysis
only uses the ratios.
This has the added advantage of allowing us to compare different
discriminating variables on a more even footing. 
In the next five subsections we  explain how this
casting into counts and ratios is done.

\subsection{Kinematic observables}
As noted in Section \ref{sec:2}, the
distribution of the missing transverse energy in the signal events
is related to $m_{\rm dm}$, the mass of the WIMP, as well as
to $m_{\rm p}$, the mass of the parent particles produced in the
original $2\to 2$ partonic subprocess.
In the benchmark \met~selection, we used the
kinematic variable $H_T$, as well as the $E_T$ of the
leading and second leading jets. The distributions of these
kinematic variables are also related to the underlying mass
spectrum of the heavy partners. 

We have employed two other kinematic variables in our study.
The first is $M$, the total invariant mass of all of the
reconstructed jets and muons in the event. The second is
$M_{\rm eff}$, the scalar sum of the \met~with the \et~of all the 
jets in the event.

\begin{figure}[thb]
\includegraphics[width=0.46\textwidth]{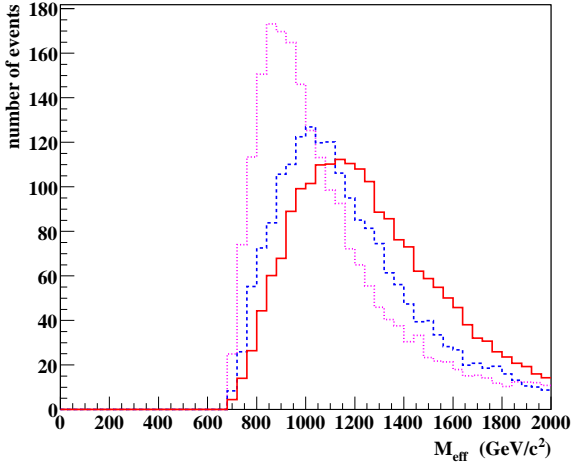}
\caption {\label{fig:Meffdist} Comparison of the $M_{\rm eff}$ distributions
for Group 1 MSSM models LM2p (solid red line), CS4d (dashed blue line) and CS6
(dotted magenta line).
For each model 100,000 events were generated then rescaled to 1000 pb$^{-1}$.}
\end{figure}

\begin{figure}[thb]
\includegraphics[width=0.46\textwidth]{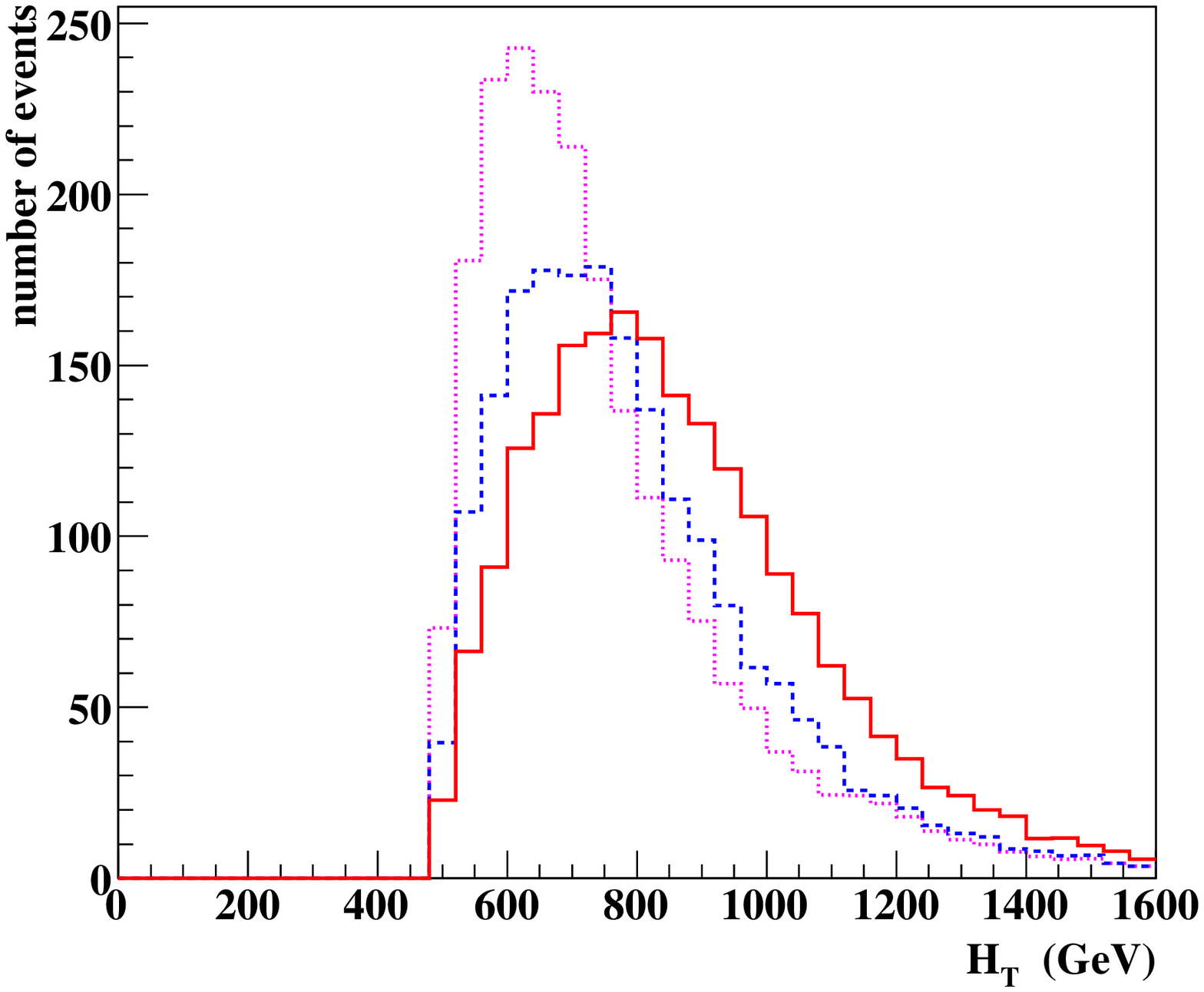}
\caption {\label{fig:Htdist} Comparison of the $H_T$ distributions
for Group 1 MSSM models LM2p (solid red line), CS4d (dashed blue line) and CS6
(dotted magenta line).
For each model 100,000 events were generated then rescaled to 1000 pb$^{-1}$.}
\end{figure}

\begin{figure}[thb]
\includegraphics[width=0.46\textwidth]{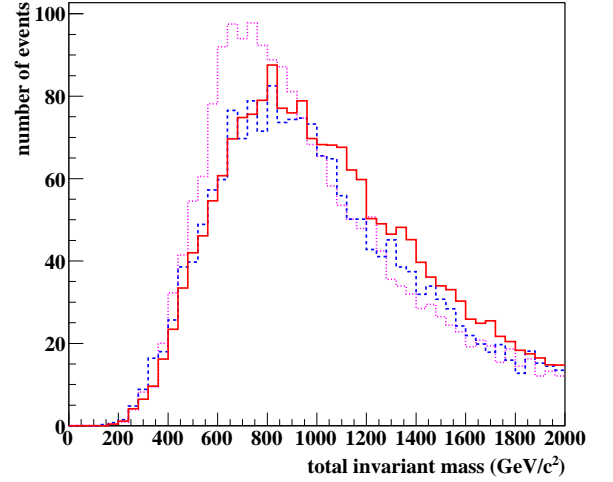}
\caption {\label{fig:Mdist} Comparison of the distributions of the total invariant mass
of jets and muons per event
for Group 1 MSSM models LM2p (solid red line), CS4d (dashed blue line) and CS6
(dotted magenta line).
For each model 100,000 events were generated then rescaled to 1000 pb$^{-1}$.}
\end{figure}

\begin{figure}[thb]
\includegraphics[width=0.46\textwidth]{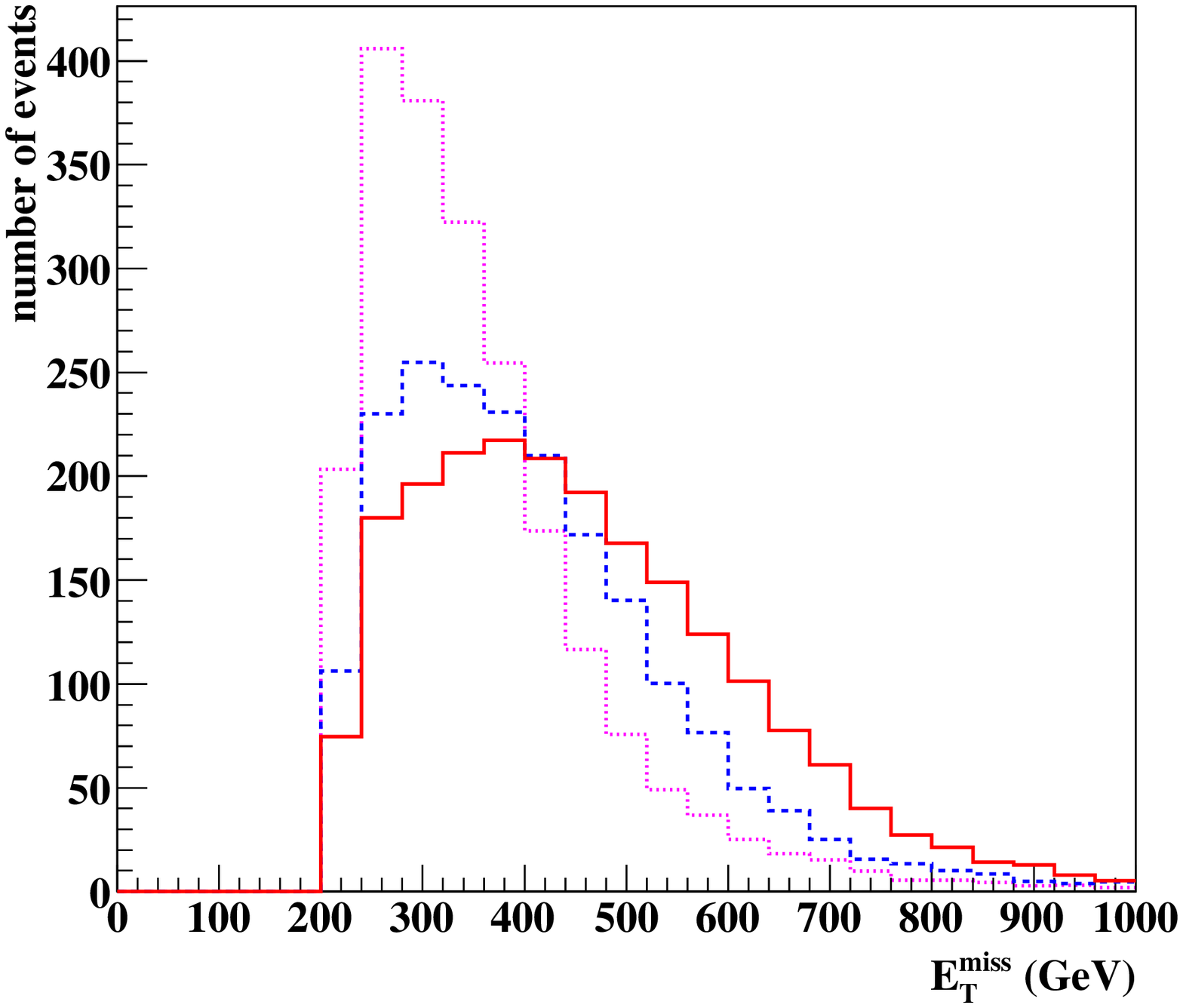}
\caption {\label{fig:Metdist} Comparison of the \met~distributions
for Group 1 MSSM models LM2p (solid red line), CS4d (dashed blue line) and CS6
(dotted magenta line).
For each model 100,000 events were generated then rescaled to 1000 pb$^{-1}$.}
\end{figure}

Figures \ref{fig:Meffdist}-\ref{fig:Metdist} show
a comparison of the $M_{\rm eff}$, $H_T$, $M$ and \met~distributions,
after selection,
for models LM2p, CS4d and CS6.
These models, though look-alikes of our \met~analysis,
have a large spread in their superpartner spectra, as is
evident from Figure \ref{fig:group1spectra}.
For sufficiently large data samples, this leads to
kinematic differences that are apparent, as seen in
Figures \ref{fig:Meffdist}-\ref{fig:Metdist}.

All of the distributions exhibit broad peaks and long
tails. 
In principle one could use the shapes of these distributions as a
discrimination handle. This would require a deep understanding of
the detector or a very conservative systematic error related to the
knowledge of the shapes.
The location of the peaks is correlated with the
mass $m_p$ of the parent particles in the events, but
there is no practical mapping from one to the other.
By the same token, this implies that these kinematic
distributions are highly correlated, and
it is not at all clear how to combine the information
from these plots.


Our approach to kinematic observables in small data sets (low luminosity) is
to define inclusive counts based on large bins. 
The dependence
on the details of the detector simulation is strongly reduced by limiting
the number of bins and using a bin width much larger than the expected
detector resolution. 

For $M_{\rm eff}$ we define two bins and one new inclusive count for
the kinematic distributions in each box:
\begin{itemize}
\item N(Meff1400) the number of events after selection with 
$M_{\rm eff} > 1400$ GeV/$c^2$.
\end{itemize}
For $H_T$ we also define two bins and one new inclusive count:
\begin{itemize}
\item N(HT900) the number of events after selection with $H_T > 900$ GeV/$c^2$.
\end{itemize}
Recall that the \met~selection already required $H_T > 500$ GeV.
For the invariant mass $M$ we
define three bins and two new inclusive counts:
\begin{itemize}
\item N(M1400) the number of events after selection with $M > 1400$ GeV/$c^2$;
\item N(M1800) the number of events after selection with $M > 1800$ GeV/$c^2$;
\end{itemize}
For \met~we define four bins and three new inclusive counts:
\begin{itemize}
\item  N(MET320), the number of events after selection having \met$> 320$ GeV.
\item  N(MET420), the number of events after selection having \met$> 420$ GeV.
\item  N(MET520), the number of events after selection having \met$> 520$ GeV.
\end{itemize}
Note that the \met~selection already required \met$>220$ GeV.

\subsection{Kinematic peaks and edges}
With large signal samples, kinematic edges involving leptons will be
a powerful tool for model discrimination and to eventually extract
the mass spectrum of the heavy partners. With small samples,
in the range of 100 pb$^{-1}$ to 1000 pb$^{-1}$ considered in our study,
this will only be true in favorable cases. In fact for the 8 models
studied here, we find no discrimination at all based on kinematic
edges with leptons. This is due mostly to the 
small number of high $p_T$ muons in our signal 
samples \footnote{Low $p_T$ lepton and dilepton 
trigger paths as well as cross-triggers combining 
leptons with jets and leptons with missing energy  
requirements are needed for Standard Model 
background calibration and understanding; these will
be important for signal appearance and edge/threshold 
studies even at start-up. The LHC experiments are preparing 
rich trigger tables along these lines \cite{HLT-LHCC}.}, 
as well as a lack of especially favorable decay chains.

\begin{figure}[thb]
\includegraphics[width=0.46\textwidth]{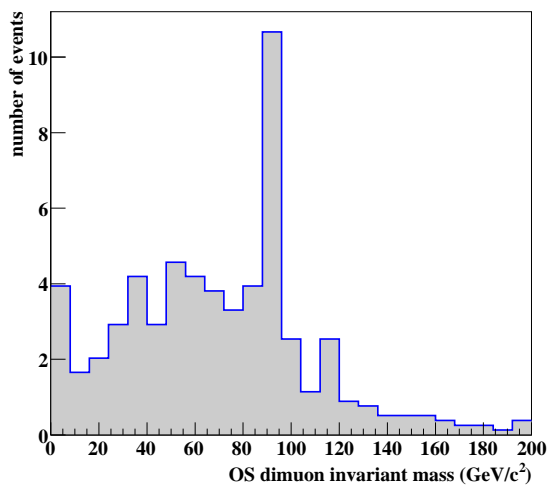}
\caption {\label{fig:dimuon} The invariant mass distribution
for opposite sign dimuon pairs passing the missing energy selection,
from MSSM model LM8.
The plot is from 100,000 generated events rescaled to 1000 pb$^{-1}$.
}
\end{figure}

Although we do not observe any dimuon edges, we do see
a dimuon peak for model LM8. This is shown in Figure (\ref{fig:dimuon}),
for signal events after our missing energy selection rescaled to 1000 pb$^{-1}$
of integrated luminosity. A $Z$ peak is clearly visible, arising from
squark decays to quark  $+\tilde{\chi}_2^0$, with the $\tilde{\chi}_2^0$
decaying 100\% to $Z+$ LSP.


\subsection{Event shapes, hemispheres and cones}
Event shapes are a way to extract information about the underlying
partonic subprocess and the resulting decay chains. This information
is not uncorrelated with kinematics, but it does have the potential
to provide qualitatively new characteristics and properties
of the event topology.
For our signal the partonic subprocess consists of the production of two heavy
partner particles, so each event has a natural separation into two
halves that we will call ``hemispheres'', consisting of all of the
decay products of each partner. Associated to each hemisphere of the event
should be one WIMP plus some number of reconstructed jets and muons.
A perfect hemisphere separation is thus an associative map of each 
reconstructed
object into one of the two constituent decay chains.

We can define two hemisphere axes in $\eta$-$\phi$ as the directions
of the original parent particles; these axes are not back-to-back
because of the longitudinal (and transverse) boosts from the subprocess
center-of-mass frame back to the lab frame. Even if we knew event by 
event how to boost to the center-of-mass frame, the hemisphere
separation would still not be purely geometrical, since in some events
a decay product of one parent will end up in the \textit{geometrical}
hemisphere of the other parent.

Having  defined a perfect hemisphere separation, we need a
practical algorithm to define reconstructed hemispheres.
We will follow a study of 6 hemisphere algorithms presented in
the CMS Physics TDR \cite{Ball:2007zza}. These algorithms
are geometrical and kinematic, based on the large though
imperfect correlation between the $\eta$-$\phi$ vector of the 
initial parent and the reconstructed objects from this parent's
decay chain. The algorithms consist of two steps: a seeding
method that estimates the two hemisphere axes, and an association
method that uses these two axes to associate each reconstructed
object to one hemisphere. 

For the Group 1 model LM5, the CMS study found that 
jets were correctly associated to parent squarks 87\% of the
time, and to parent gluinos 70\% of the time. The differences
in the performance of the 6 algorithms were small; the best-performing
hemisphere algorithm combines the following methods:
\begin{itemize}
\item Seeding method: The two hemisphere axes are chosen as
the $\eta$-$\phi$ directions of the pair of reconstructed objects
with the largest invariant mass. The hardest of these objects defines
the \textit{leading} hemisphere.
\item Association method: A reconstructed object is assigned
to the hemisphere that minimizes the Lund distance \cite{Ball:2007zza}.
\end{itemize}

\begin{figure}[thb]
\includegraphics[width=0.46\textwidth]{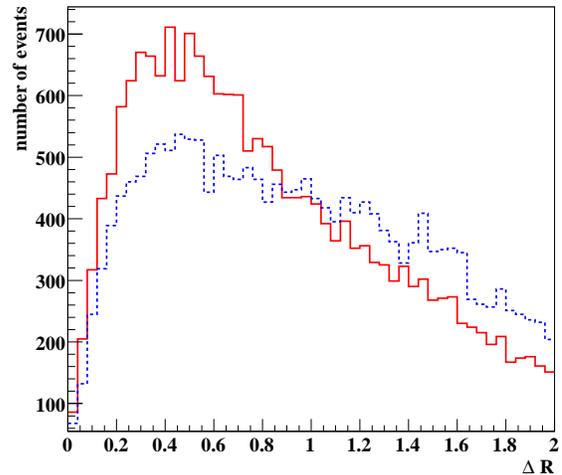}
\caption {\label{fig:HemideltaR} The distribution of the $\Delta R$
separation between the $\eta$-$\phi$ direction of the parent superpartner
and the reconstructed hemisphere axis. This is from 24,667 events
of model LM5 passing our selection. The solid red line
is for the leading hemisphere, while the dashed blue line is for the
second hemisphere.
 }
\end{figure}

\begin{figure}[thb]
\includegraphics[width=0.46\textwidth]{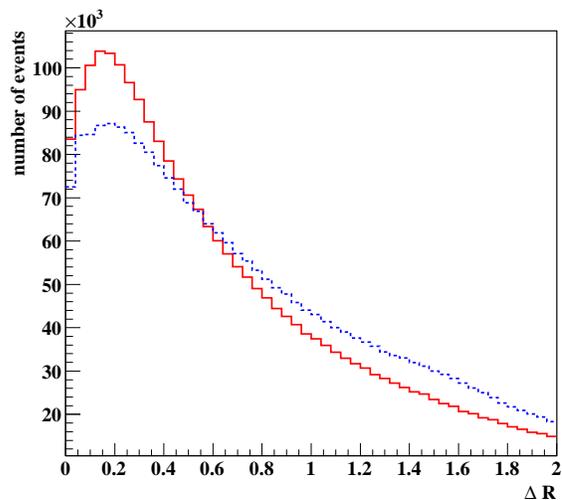}
\caption {\label{fig:HemideltaRtop} The distribution of the $\Delta R$
separation between the $\eta$-$\phi$ direction of the parent top quark 
and the reconstructed hemisphere axis. This is from 3,000,000 {\tt Pythia}
$t\bar{t}$ events with no selection. The solid red line
is for the leading hemisphere, while the dashed blue line is for the
second hemisphere.
 }
\end{figure}

Figure \ref{fig:HemideltaR} shows how well the seeding method
produces hemisphere axes that match the actual axes defined by the
two parent particles. The separation is shown in units of $\Delta R$,
defined as
\be
\Delta R \equiv \sqrt{(\Delta\eta )^2 + (\Delta\phi )^2}
\; .
\ee
The agreement is not overwhelmingly good, and is substantially
worse than that obtained for $t\bar{t}$ events, as shown in
Figure \ref{fig:HemideltaRtop}. We have checked that all 6 
hemisphere algorithms
produce very similar results.
Since the agreement is better for the leading hemisphere,
all of our single hemisphere derived observables are based just
on the leading hemisphere.

We define three inclusive counts based on comparing the two
reconstructed hemispheres:
\begin{itemize}
\item N(Hem$j$) the number of events for which the reconstructed object
multiplicity (jets + muons)
in the two hemispheres differs by at least j, j=1,2,3.
\end{itemize}

Once the two hemispheres are identified, we can break the degeneracy
among the models by looking at the topology in the events. For a given
mass of the parent particles, the events will look more jet-like 
rather than isotropic
if the decays are two-body rather than multibody cascades.
In the
case of jet-like events, the projection of the observed track
trajectories on the transverse plane will cluster along the
transverse boost of the particles generating the cascade. 

In order to quantify this behaviour, we start from the central axes of
each hemisphere and draw slices in the transverse plane with increasing 
opening angles in $\phi$
symmetric around the hemisphere axis. We refer
to the slices as {\it cones}, reminiscent of the cones used in CLEO analyses \cite{cleo}
to discriminate between jet-like QCD background and isotropic decays
of $B$ meson pairs. 

We build five cones of opening angle $2\alpha$ ($\alpha$ $=$
$30^\circ$, $45^\circ$, $60^\circ$, $75^\circ$ and $90^\circ$) in each hemisphere.
In terms of these cones we define variables:
\begin{itemize}
\item N(nt-c$\alpha$), the number of events having at least n
  tracks (n=10,20,30,40) in the leading hemisphere cone of opening
  angle $2\alpha$.
\item N(ntdiff-c$\alpha$), the number of events having a
  difference of at least n tracks (n=10,20,30,40)
  between the cones of
  opening angle $2\alpha$ in each hemisphere.
\end{itemize}
Tracks are counted if they have $p_T > 1$ GeV/$c$ and $|\eta| < 2.4$.
Since the cone of opening angle $2\alpha$ includes the one of
opening angle $2\beta$ for $\alpha > \beta$, these variables have an
inclusive nature. 


\subsection{The stransverse mass $m_{T2}$} 

A potentially powerful observable 
for model discrimination and mass extraction is
the stransverse mass variable $m_{T2}$~\cite{Lester:1999tx}-\cite{Barr:2003rg}.
Let us briefly review how this is supposed to work for our missing energy
signal.
Ignoring events with neutrinos, our signal events have two heavy parent particles
of mass $m_{\rm p}$, each of which contributes to the final state a WIMP 
of mass $m_{\rm dm}$ plus some number of visible
particles. Supposing also that we have a perfect hemisphere separation,
we can reconstruct each set of visible particles into a 4-vector $p^X_{\mu}$.
If we also knew the mass and the $p_T$ of each WIMP, we could
reconstruct a transverse mass for each hemisphere from the formula
\be
m_T^2 = m_X^2 + m_{\rm dm}^2 + 2(E^X_T\,E^{\rm dm}_T - \mathbf{p}^X_T\cdot
\mathbf{p}^{\rm dm}_T) \; .
\ee
This transverse mass is always less than or equal to the mass $m_{\rm p}$
of the parent particle. Thus the \textit{largest} of the two transverse masses
per event is also a lower bound on $m_{\rm p}$.

Of course we do not know the $p_T$ of each WIMP, only the combined \met .
Let $p_T^{(1)}$ and $p_T^{(2)}$ denote a possible decomposition of
the total $p_T^{\rm miss}$ into two azimuthal vectors, one for each WIMP.
Note that this decomposition ignores initial state radiation, the underlying
event and detector effects.
Then we can define the stransverse mass of an event as
\be
&&
m_{T2}^2(m_{\rm dm}) \equiv \\
&&
\min{p_T^{(1)}+p_T^{(2)}=p_T^{\rm miss}}
\hspace{-30pt}
\left[ {\rm max}\left[ 
m_{T2}^2(m_{\rm dm};p_T^{(1)}),m_{T2}^2(m_{\rm dm};p_T^{(2)})
\right] \right]
.
\nn
\ee
Since (with the caveats above) one of these partitions
is in fact the correct one, this
quantity is also a lower bound on the parent mass $m_{\rm p}$.

For a large enough data sample, with the caveats above and ignoring
finite decay widths,
the upper endpoint of the stransverse
mass distribution saturates at the parent mass $m_{\rm p}$, provided
we somehow manage to input the correct value of the
WIMP mass $m_{\rm dm}$.
In the approximation that the invariant mass $m_X$ of the visible
decay products is small, the lower endpoint of the stransverse
mass distribution is at $m_{\rm dm}$.

These impressive results seem to require that we know \textit{a priori}
the correct input value for $m_{\rm dm}$. However it has 
been shown~\cite{Cho:2007qv}-\cite{Barr:2007hy} that
in principle there is a kink in the plot of the 
upper endpoint value of $m_{T2}$ as a function of the assumed $m_{\rm dm}$,
precisely when the input value of $m_{\rm dm}$ equals its true value.
Thus it may be possible to extract both $m_{\rm p}$ and $m_{\rm dm}$
simultaneously.

Summarizing the remaining caveats, the stransverse mass method requires
a large data sample, and is polluted by effects from incorrect
hemisphere separation, ISR, the underlying event, finite decay widths and 
detector effects.
We can compare this to the precision extraction of the $W$ mass at
CDF~\cite{CDF:2007ypa,Aaltonen:2007ps},
using the transverse mass distribution of the charged lepton and neutrino
from the $W$ decay. Here the WIMP mass is essentially zero, the data samples are huge,
and hemisphere separation is not applicable since there is only one WIMP.
In the CDF analysis the ISR uncertainty was traded for an FSR uncertainty,
by interpreting the vector sum of all the calorimetric \et~not associated
with the charged lepton as coming from ISR plus the underlying event; this then pollutes
the reconstructed neutrino $p_T$ with final state radiation from the lepton.
The measured transverse mass distribution has a considerable tail extending
above the putative endpoint at $m_W$, but a precision mass (and width) extraction is
still possible by modeling the distribution.


\begin{figure}[thb]
\includegraphics[width=0.46\textwidth]{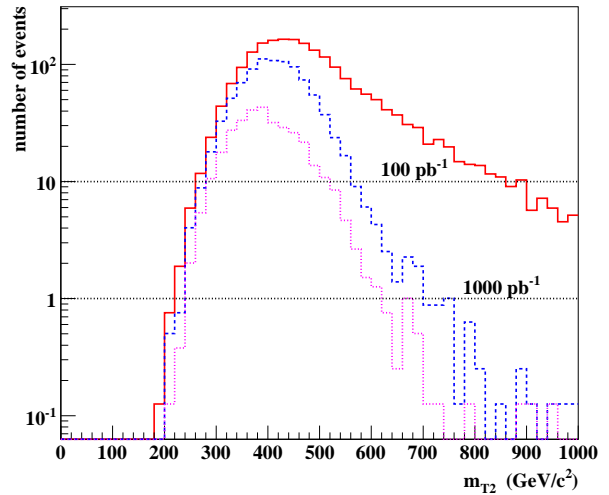}
\caption {\label{fig:Mt2cutcomp} Comparison of the stransverse mass $m_{T2}$ distributions
for model CS6, varying the cut on the maximum hemisphere invariant mass.
The solid red line shows the $m_{T2}$ distribution when no cut is applied.
The dashed blue line shows the $m_{T2}$ distribution rejecting events
when the total reconstructed invariant mass of either hemisphere exceeds 300 GeV.
The dotted magenta line shows the $m_{T2}$ distribution when this cut is lowered
to 200 GeV.
In each case 100,000 events were generated then rescaled to 1000 pb$^{-1}$.
The dotted horizontal lines are to guide the eye for where the distribution
cuts off for 100 pb$^{-1}$ and 1000 pb$^{-1}$.}
\end{figure}


\begin{figure}[thb]
\includegraphics[width=0.46\textwidth]{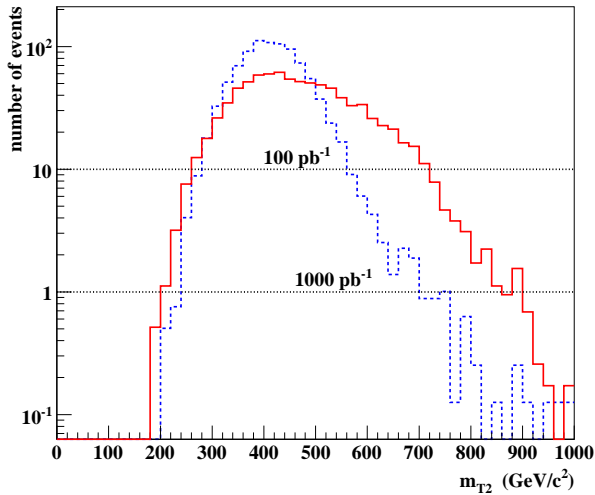}
\caption {\label{fig:Mt2dist} Comparison of the stransverse mass $m_{T2}$ distributions
for look-alike models LM2p (solid red line) and CS6 (dashed blue line).
For each model 100,000 events were generated then rescaled to 1000 pb$^{-1}$.
The dotted horizontal lines are to guide the eye for where the distribution
cuts off for 100 pb$^{-1}$ and 1000 pb$^{-1}$.}
\end{figure}


In \cite{Nojiri:2008hy} it was shown that an imperfect hemisphere
separation greatly degrades the $m_{T2}$ distribution for simulated 
LHC SUSY events. Two approaches to solve this problem have been suggested.
The first, used in \cite{Nojiri:2008hy}, is to reject events when the
total invariant mass of the reconstructed objects in either hemisphere
exceeds some value. This strategy is based on the fact that for a
correct hemisphere separation with $m_{T2}$ near the endpoint the hemisphere
invariant mass is small, while incorrect hemisphere assignments naturally
lead to large hemisphere invariant masses. The second approach,
used in \cite{Lester:2007fq}, is to replace $m_{T2}$ with a new variable
$m_{TGen}$. The new variable minimizes not only over 
all partitions of the $p_T^{\rm miss}$
into $p_T^{(1)}$ and $p_T^{(2)}$, but also over all possible hemisphere
assignments. Since one of the hemisphere separations is in fact the
correct one, $m_{TGen}$ has the same endpoint as $m_{T2}$ would have with
a perfect hemisphere separation.

Figure~\ref{fig:Mt2cutcomp} shows the degradation of the $m_{T2}$ distribution
for our model CS6. The $m_{T2}$ endpoint should be at 
589 GeV, the value of the gluino mass in CS6. However the solid red line
shows that a large fraction of events are above this endpoint, due to
the imperfect hemisphere separation. Applying the strategy of \cite{Nojiri:2008hy},
the dotted magenta line shows that we regain the correct endpoint by making
a hard cut of 200 GeV on the maximum reconstructed hemisphere invariant mass.
However with such a high requirement we take a big hit in statistics.
The dashed blue line shows that we still do pretty well with a 
300 GeV requirement, while gaining a lot in statistics. 
For this study we have used
the 300 GeV requirement in all of our $m_{T2}$ analysis. The value is 
unoptimized but also unbiased, since it was determined by 
asking for approximately
the most stringent cut that retains reasonable statistics for the
$m_{T2}$ distributions of the entire set of models considered.

Figure \ref{fig:Mt2dist} shows a comparison of the $m_{T2}$ distributions
for two of our Group 1 look-alike models, LM2p and CS6. For LM2p, the parents
are gluinos of mass 856 GeV and squarks of mass approximately 800 GeV;
for model CS6, by contrast, the parents are gluinos of mass 589 GeV.
In each case we have input the correct LSP mass, 142 GeV for LM2p and
171 GeV for CS6. 

Each plot is rescaled to
1000 pb$^{-1}$. With this many events notice that we are just starting to saturate
the appropriate endpoints at $m_{T2}=m_{\rm p}$, and notice the onset
of tails above the endpoints. The dotted lines in the figure guide the eye
to where the distributions cut off for data samples of 
100 pb$^{-1}$ and 1000 pb$^{-1}$.
Obviously for 100 pb$^{-1}$ we are not close to populating
the endpoints.

However even for 100 pb$^{-1}$ 
there are significant differences between the $m_{T2}$ distributions
of the two models. These differences only become larger if we use the
same input mass for the LSP.
Thus $m_{T2}$ is at least as interesting for
look-alike discrimination as the more traditional kinematic variables
discussed above. Furthermore, even if we are not close to populating
the endpoint, it might be possible to extract a direct estimate of
$m_{\rm p}$ by fitting or extrapolating the distributions.

For our study we 
define five bins and four new inclusive counts from $m_{T2}$:
\begin{itemize}
\item N(mT2-300) the number of events after selection with $m_{T2} > 300$  GeV/$c^2$,
\item N(mT2-400) the number of events after selection with $m_{T2} > 400$  GeV/$c^2$,
\item N(mT2-500) the number of events after selection with $m_{T2} > 500$  GeV/$c^2$.
\item N(mT2-600) the number of events after selection with $m_{T2} > 600$  GeV/$c^2$.
\end{itemize}

When
comparing a model M1, playing the role of the data, with a model M2,
playing the role of the model to test, we will use the mass of the
WIMP in model
M2 as the input mass in calculating $m_{T2}$ for both models.

\subsection{Flavor enrichment}

In order to have some model discrimination based on the $\tau$ or
$b$ content,  we need simple algorithms to create
subsamples enriched with $b$ quarks and $\tau$'s.
We  refer to these algorithms as ``tagging'',
despite the fact that the tagging efficiencies and the purity of
the subsamples are rather poor.  

Without attempting any detailed optimizations, we have designed
two very simple tagging algorithms. We expect these algorithms to be
robust, since they only require a knowledge of uncorrected high $E_T$
jets, high $p_T$ muons, and basic counting of high $p_T$ tracks inside 
jets.

{\bf $\mathbf{\tau}$ enrichment:} For each jet we define a 
0.375 cone centered around the jet axis. Inside this cone we 
count all reconstructed
charged tracks with $p_T > 2$ GeV/$c$. If only one such track is found,
and if this track has $p_T > 15$ GeV/$c$, we tag the jet as a $\tau$ jet.

The $\tau$ algorithm is based on single-prong hadronic $\tau$ decays,
which as their name implies produce a single charged track.
In addition, leptonic decays of a $\tau$ to an electron and two neutrinos
can be tagged, since some fraction of electrons reconstruct as jets.
Soft tracks with $p_T \le 2$ GeV/$c$ are not counted, a fact that makes
the algorithm much more robust. The $p_T > 15$ GeV/$c$ requirement
on the single track reduces the background from non-$\tau$ jets.
Increasing the cone size decreases the efficiency to tag genuine
$\tau$'s, because stray tracks are more likely to be inside the cone;
decreasing the cone size increases the fake rate.
A genuine optimization of this algorithm can only be done
with the real data.

Table \ref{tab:tautagging} shows the results of applying
our $\tau$ tagging algorithm to simulations of
the Group 1 models LM2p, LM5, LM8, CS4d and CS6.
The efficiency, defined as the number of $\tau$ tags
divided by the number of generator-level $\tau$'s that
end up reconstructed as jets, varies between 12\% and
21\%. The efficiency is lowest for models LM8 and CS4d,
models where $\tau$'s come entirely from $W$ and $Z$
decays. The efficiency is highest for model LM2p,
which has a large final state multiplicity of $\tau$'s
from decays of charginos, second neutralinos and staus.

The purity, defined as the fraction of $\tau$ tagged jets
that actually correspond to generator-level $\tau$'s,
is quite low for models LM8 and CS4d, and is only
8\% for model CS6, which contains very few $\tau$'s.
We obtain a reasonably high purity of 55\% for
LM2p, the model with by far the largest $\tau$ multiplicity.

We conclude that it is possible to obtain significantly
enriched samples of $\tau$'s from our simple algorithm, but
only for models that do have a high multiplicity of
energetic $\tau$'s to begin with. From the counts in
Table \ref{tab:tautagging}, it is clear that this tagging
method is not viable with 100 pb$^{-1}$ of integrated
luminosity.


\begin{table}[thb]
\begin{center}
\caption{\label{tab:tautagging} Results of our $\tau$ tagging
algorithm applied to
the Group 1 models LM2p, LM5, LM8, CS4d and CS6.
Counts are rescaled to 1000 pb$^{-1}$ from 100,000 events per model.
The listing for $\tau$ jets counts generator level $\tau$'s
that are reconstructed as jets in events that pass our selection.
}
\begin{tabular}{lrrrrr}
\toprule
&\multicolumn{1}{c}{LM2p}&\multicolumn{1}{c}{LM5}
&\multicolumn{1}{c}{LM8}&\multicolumn{1}{c}{CS4d}
&\multicolumn{1}{c}{CS6}
\\ \midrule
$\tau$ jets per fb$^{-1}$   & 409  & 144  & 171   &  112  &  34
\\ \midrule
tags per fb$^{-1}$          & 157  & 110  & 122   &  102  &  59 
\\ \midrule
correct                     &       &       &         &         &
\\
tags per fb$^{-1}$          &  86  & 25   &  21   &   14  &  5
\\ \midrule
efficiency                  & 21\% & 18\% &  12\% &  13\% &  16\%
\\ \midrule
purity                      & 55\% & 23\% &  17\% &  14\% &  8\%
\\ \bottomrule
\end{tabular}
\end{center}
\end{table}


{\bf $\mathbf{b}$ enrichment:} For each jet we search for a reconstructed
muon inside the jet (recall that our muons have
$p_T > 20$ GeV/$c$ and $|\eta |<2.4$). If a muon is found within $\Delta R < 0.2$
of the jet axis we tag it as a $b$ jet.

This $b$ algorithm is based on tagging muons from semileptonic $B$
decays inside the $b$ jet. This is inspired by the ``soft muon''
tagging that was used in the top quark discovery 
at the Tevatron~\cite{Abe:1995hr,Abachi:1995iq}.
In our case ``soft'' is a misnomer, since in fact we
only count reconstructed muons with $p_T > 20$ GeV/$c$.
This requirement makes the tagging algorithm more robust,
but reduces the efficiency.

Table \ref{tab:btagging} shows the results of applying
our $b$ tagging algorithm to simulations of
the Group 1 models LM2p, LM5, LM8, CS4d and CS6.
The tagging efficiency is defined as the number of $b$ tags
divided by the number of generator-level $b$'s that
are within $\Delta R < 0.3$ of the center of a reconstructed
jet.
Although all of these models have a high multiplicity
of generator-level $b$'s, the tagging efficiency is poor: only
about 5\% for all models. However the purity of the samples is
rather good: above 70\% for every model except CS6.

We conclude that it is possible to obtain significantly
enriched samples of $b$'s from our simple algorithm, but
with low efficiency.
From the counts in
Table \ref{tab:btagging}, it is clear that this tagging
method is not viable with 100 pb$^{-1}$ of integrated
luminosity, but should become useful as we approach
1000 pb$^{-1}$.

In our study, discrimination based on $\tau$'s and $b$'s is
obtained from ratios that involve the following two inclusive counts:

\begin{itemize}
\item N($\tau$-tag), the number of events after selection having at least one
$\tau$ tag.
\item N($b$-tag), the number of events after selection having at least one
soft muon $b$ tag.
\end{itemize}


\begin{table}[thb]
\begin{center}
\caption{\label{tab:btagging} Results of our $b$ tagging
algorithm applied to
the Group 1 models LM2p, LM5, LM8, CS4d and CS6.
Counts are rescaled to 1000 pb$^{-1}$ from 100,000 events per model.
The listing for $b$ jets counts
generator level $b$ quarks matched to reconstructed jets
that pass our selection.
}
\begin{tabular}{lrrrrr}
\toprule
&\multicolumn{1}{c}{LM2p}&\multicolumn{1}{c}{LM5}
&\multicolumn{1}{c}{LM8}&\multicolumn{1}{c}{CS4d}
&\multicolumn{1}{c}{CS6}
\\ \midrule
$b$ jets per fb$^{-1}$      & 1547  & 1693  &  2481   &   1596  &   748
\\ \midrule
tags per fb$^{-1}$          &  115  &  112  &   148   &    105  &   106 
\\ \midrule
correct                     &       &       &         &         &
\\
tags per fb$^{-1}$          &  82   &   81  &   112   &     75  &    41
\\ \midrule
efficiency                  &  5\%  &  5\%  &   5\%   &    5\%  &   5\%
\\ \midrule
purity                      & 72\%  & 72\%  &  75\%   &   71\%  &  39\%
\\ \bottomrule
\end{tabular}
\end{center}
\end{table}


\section{The look-alike analysis}\label{sec:6}
The look-alike analysis proceeds in four steps: 

1. We choose one of the models to play the part of the data.  
We run the inclusive \met +jets
analysis on the MET trigger and verify that the predicted yield
establishes an excess (at $ > 5 \sigma$) above the SM
background with $100$ pb$^{-1}$. We call the number of events selected
in this way the {\it observed yield} $N_{\rm data}$. In what follows, we
assume that a subtraction of the residual Standard Model background has already been
performed. We assume large signal over background ratios
for the models considered so that the statistical error on the background
has a small impact on the total error.

2. We identify a set of models giving a predicted yield $N$
compatible with $N_{\rm data}$. The compatibility is established if the
difference in the two counts is less than twice the total error,
{\it i.e} if the pull
\begin{equation}
\frac{|N_{\rm data}-N|}{\sigma(N)}
\label{eq:pull_def}
\end{equation}
is smaller that two. In the formula $\sigma(N)$ represents the error
associated to the expected number of events $N$. We calculate it as
the sum in quadrature of several contributions:
\begin{itemize}
\item A Poissonian error which takes into account the statistical
  fluctuations associated to the event production (statistical
  component of the experimental error).
\item An error associated to the detector effects (systematic
  component of the experimental error).
\item Theoretical error on the predicted number of events $N$
  (including a statistical and a systematic component).
\end{itemize}
We discuss the origin of each contribution below.

3. For each additional observable $N^i$ previously listed, we
consider the value on the data ($N^i_{\rm data}$) and the predicted
value $N^i_j$ for the model $j$. We calculate the pull as in
eqn.~\ref{eq:pull_def} and we identify the variable with the largest
pull as the best discriminating counting variable. We ignore all
the variables for which both the model and the data give a yield below a
fixed threshold $N_{\rm min}$. We use $N_{\rm min}=10$, i.e. we
require a minimum yield that is more than three times its Poisson
error $\sqrt{N^i}$; for the data this corresponds to excluding at 3$\sigma$ the
possibility that the observed yield is generated
by a fluctuation of the background.

4. We form ratios of some of the observables used above and we repeat
the procedure of step 3. Since part of the uncertainties cancel out
in the ratio, these variables allow a better discrimination than the
counting variables.  In addition, provided that the two
variables defining the ratio are above the threshold $N_{\rm min}$, the
ratios of two correlated variables (such as $N(4j)/N(3j)$) are
less sensitive to the statistical fluctuations. Details on the
calculation of the errors on the ratios are given below.

In each of the four trigger boxes we define the following ratios of correlated inclusive counts:
\begin{itemize}
\item r($n$j)(3j), with n=4,5
\item r(MET320)
\item r(MET420)
\item r(MET520)
\item r(HT900)
\item r(Meff1400)
\item r(M1400)
\item r(M1800)
\item r(Hem$j$) with $j$=1,2,3
\item r(2$\mu$-nj)(1$\mu$-nj) with n=3,4
\item r($\tau$-tag)
\item r($b$-tag)
\item r(mT2-300) with the theory LSP mass
\item r(mT2-400) with the theory LSP mass
\item r(mT2-500) with the theory LSP mass
\item r(mT2-600) with the theory LSP mass
\item r(mT2-400/300) with the theory LSP mass 
\item r(mT2-500/300) with the theory LSP mass
\item r(mT2-600/300) with the theory LSP mass
\item r(nt-c$\alpha$) for n=10,20,30,40 and
$\alpha=30^\circ$,$ 45^\circ$, $60^\circ$, $75^\circ$, $90^\circ$
\item r(ntdiff-c$\alpha$) for for n=10,20,30,40 and
  $\alpha=30^\circ$, $45^\circ$, $60^\circ$, $75^\circ$, $90^\circ$
\end{itemize}
For most of these ratios the numerator is the corresponding count
defined in Section \ref{sec:5} and the denominator is the total event count
in the trigger box. The exceptions are r($n$j)(3j)$=$N($n$j)/N(3j),
r(2$\mu$-nj)(1$\mu$-nj)$=$N(2$\mu$-nj)/N(1$\mu$-nj),
and r(mT2-$n$/300)$=$N(mT2-$n$)/N(mT2-300), $n = 400,500,600$.
We also use the ratios of the counts in the DiJet, TriJet and Muon20
boxes to the count in the MET box:
\begin{itemize}
\item r(DiJet)
\item r(TriJet)
\item r(Muon20)
\end{itemize}
As mentioned previously, it turns
out that the DiJet, TriJet and Muon20 boxes are subsamples of the
MET box to an excellent approximation, thus these ratios are also
ratios of inclusive counts.

Finally we iterate and perform the transpose comparisons (the model that was considered as data takes the role of the model).

\subsection{Theoretical uncertainty}

We take into account several sources of uncertainty. First of all,
there is an error associated to the knowledge of the parton
probability density functions (pdfs) that are used to generate the
event samples. In order to evaluate this error, we produce and analyze 
all samples with three different sets of pdfs: {\tt CTEQ5L}~\cite{cteq}, {\tt CTEQ6M}~\cite{cteq},
and {\tt MRST2004nlo}~\cite{MRST} or {\tt MRST2002nlo}~\cite{MRST} for Group 1 and Group 2
respectively. We quote as central value
the average of the three values; for the pdf uncertainty we crudely estimate it by
taking half the spread of the three values.  This uncertainty, 
as we will show, has important effects on the results.

An additional error is given by the relative QCD scale uncertainty
when we compare different look-alike models.
This is an overall systematic
on the relative cross sections that we take to be $5\%$. It is actually
larger than this in our study, at least for the Group 2 models
where we use LO cross sections, but we 
are assuming some improvement by the time of the real discovery.

There is an additional uncertainty for
each observable from the missing higher order matrix elements. It is
not included in the analysis shown here. It could be included
crudely by running {\tt Pythia} with different values of the ISR
scale controlled by MSTP(68), similar to how we evaluate the pdf
uncertainties. A better way is to include, for the signals, the higher order matrix
elements for the emission of extra hard jets. The ideal approach  would be
a full NLO generator for the signals.

The sum in quadrature of all these effects gives the systematic error
associated to the theoretical prediction. In the case of ratios, the
error on the cross section cancels out. In a similar way, the
correlated error on the pdfs cancels out by calculating the
ratios for the three sets of pdfs and then averaging them.

In the case of mSUGRA models, the result of the simulation also
depends on which RGE evolution code we use\footnote{For the
CMS benchmark models, we used {\tt
Isajet v7.69} but compared the spectra results with
{\tt SuSpect v2.34} + {\tt SUSY-HIT v.1.1} and {\tt SoftSusy v.2.0.14}~\cite{softsusy}.  
The differences in the computed spectra led to differences in our observed yield of 3 to 10\%.} 
to go from the parameters at the high scale to the
SUSY spectrum at the Terascale. Rather than including an error associated to
such differences we take one of the codes ({\tt Isajet v7.69} or {\tt SuSpect v2.34})
as part of the definition of the theory model we are considering.

The theory predictions are also affected by a statistical error,
related to the fact that the value of each observable is evaluated on
a sample of limited size. Generating the same sample with a
different Monte Carlo seed one obtains differences on the predicted
values of the observable. The differences, related to statistical
fluctuations, are smaller for larger generated data sets. Considering
that each number of events $N_i^j$ for observable $i$ and model $j$
can be written as $N_i^j = \epsilon_i^j \times \sigma^j$ and that the
error on $\sigma^j$ is already accounted for in the systematic
contribution to the theoretical error, the efficiency
$\epsilon_i^j$ has an associated binomial error:
\begin{equation}
\sigma(\epsilon_i) = \sqrt{\frac{\epsilon_i^j \times
    (1-\epsilon_i^j)}{N_{\rm GEN}}}
\label{eq:teostat}
\end{equation}
where $N_{\rm GEN}$ is the size of the generated sample before any
selection requirement. This error can be made negligible by generating
data sets with large values of $N_{\rm GEN}$. We include the contribution
of the statistical error summing it in quadrature to the systematic
error.

When the variables defining the ratio are uncorrelated, the error on
the ratio is obtained by propagating the errors on the numerator and
denominator, according to the relation
\begin{equation}
\sigma(r) = \sqrt{\left(\frac{\sigma(N_{\rm num})}{N_{\rm den}} \right)^2 +
\left(\frac{N_{\rm num}\sigma(N_{\rm den})}{N_{\rm den}^2} \right)^2}
\label{eq_uncorratioerror}
\end{equation}
where $r=N_{\rm num}/N_{\rm den}$.

This is not the correct formula in our case, since all of the
counts on our ratios are correlated.
For instance, $N(4j)$
and $N(3j)$ are correlated, since all the events with at least four jets have
also three jets. Only a fraction of the events defining $N(3j)$
will satisfy the requirement of an additional jet, {\it i.e.} applying the
requirement of an additional jet on the $\geq 3$ jets sample
corresponds to a binomial process, with the ratio
$r(4j)(3j)=N(4j)/N(3j)$ the associated efficiency. The error on $r$ is
then given by eqn.~\ref{eq:teostat}, replacing $\epsilon_i^j$ with the
$r$ and $N_{\rm GEN}$ with $N(4j)$. The same consideration applies to
all the ratios built from correlated variables. In order to
use eqn.~\ref{eq:teostat} for the error, we always define the ratios
such that they are in the range [0,1].

\subsection{Statistical uncertainty}

The production of events of a given kind in a detector is a
process ruled by Poisson statistics. The error on a counting
variable $N$ is given by $\sqrt{N}$. 

In analogy to the statistical error on the theoretical predictions,
the statistical error on a ratio of two correlated variables is
calculated according to eqn.~\ref{eq:teostat}, replacing $\epsilon_i^j$
with the $r$ and $N_{\rm GEN}$ with the value of the denominator variable
for the reference data luminosity.  Unlike the case of the theoretical
error, this error is associated to the statistics of the data set
and not to the size of the generated sample; this error is intrinsic
to the experimental scenario we are considering and cannot be
decreased by generating larger Monte Carlo samples.

\subsection{Systematic uncertainty}

We consider two main sources of systematic error, the knowledge
of the collected luminosity and the detector effects on the counting
variables. Estimating the two contributions to be of the order of
$10\%$, we assign a global systematic error of $15\%$ to the counting
variables. When calculating the ratios, we expect this error to be
strongly reduced. On the other hand the cancellation is not exact,
and it will be less effective at the start-up, because of potentially 
poorly-understood features related to the reconstruction. Hence we
associate a residual systematic error of $5\%$ to the ratios.



\begin{table*}[htp]
\begin{center}
\caption{\label{tab:g1resultssummary} Summary of the best discriminating ratios
for model comparisons in Group 1. The models listed in rows are
taken as simulated data, with either 100 or 1000 pb$^{-1}$ of
integrated luminosity assumed, and uncertainties as described in the
text. The models listed in columns are then compared pairwise with
the ``data''. In each case, the three best distinct discriminating ratios
are shown, with the estimated significance. By distinct we mean that we
only list the best ratio of each type; thus if r(5j)(4j) is listed,
then r(4j)(3j) is not, etc.
The asterix on the ratio
r($b$-tag) indicates that it is defined in the Muon20 box; all other
ratios are defined in the MET box, and r(Muon20) denotes the ratio of
the number of events in the Muon20 box to the number in the MET box.
The $m_{T2}$ ratios are computed
using the LSP mass of the relevant ``theory'' model, not the ``data'' model.}
\setlength{\tabcolsep}{0.5mm}
\begin{tabular}{llrlrlrlrlr}
\toprule
&\multicolumn{2}{c}{LM2p}&\multicolumn{2}{c}{LM5}
&\multicolumn{2}{c}{LM8}&\multicolumn{2}{c}{CS4d}
&\multicolumn{2}{c}{CS6}
\\ \midrule
LM2p &&&&&&
\\
100  &                   &         &r(5j)(3j)          &1.6\sig  &r(5j)(3j)          &4.4\sig &r(MET520)         &4.1\sig &r(mT2-600/300)    &11.4\sig
\\ 
     &                   &         &r(mT2-300)         &1.4\sig  &r(MET520)          &3.7\sig &r(HT900)          &3.6\sig &r(MET520)         &10.6\sig
\\  
     &                   &         &r($\tau$-tag)      &1.2\sig  &r(10t-c45)         &2.9\sig &r(Meff1400)       &3.0\sig &r(HT900)          &6.8\sig
\\ \cline{2-11}   
1000 &                   &         &r($\tau$-tag)      &3.1\sig  &r(MET520)          &8.2\sig &r(MET520)         &9.4\sig &r(mT2-600/300)    &33.0\sig
\\
     &                   &         &r(5j)(3j)          &2.8\sig  &r(mT2-500)         &6.7\sig &r(HT900)          &6.4\sig &r(MET520)         &26.6\sig
\\
     &                   &         &r(mT2-400)         &2.6\sig  &r(5j)(3j)          &6.5\sig &r(mT2-600)        &6.0\sig &r(HT900)          &14.6\sig 
\\\midrule
LM5  &&&&&&
\\ 
100  &r(5j(3j)           &1.8\sig  &                   &         &r(5j)(3j)          &2.9\sig &r(HT900)          &3.6\sig &r(mT2-600/300)    &11.6\sig
\\
     &r(mT2-300)         &1.5\sig  &                   &         &r(MET520)          &2.7\sig &r(Meff1400)       &3.2\sig &r(MET520)         &9.2\sig
\\
     &r(10t-c30)         &1.4\sig  &                   &         &r(Muon20)          &2.5\sig &r(MET520)         &3.1\sig &r(HT900)          &6.8\sig
\\ \cline{2-11}
1000 &r(5j)(4j)          &3.4\sig  &                   &         &r(MET520)          &6.0\sig &r(MET520)         &7.1\sig &r(mT2-600/300)    &33.7\sig
\\
     &r($\tau$-tag)      &2.7\sig  &                   &         &r(Muon20)          &4.9\sig &r(HT900)          &6.4\sig &r(MET520)         &22.9\sig
\\
     &r(mT2-400)         &2.6\sig  &                   &         &r(5j)(3j)          &4.3\sig &r(mT2-600/400)    &6.1\sig &r(HT900)          &14.6\sig
\\\midrule
LM8  &&&&&&
\\ 
100  &r(5j)(3j)          &5.5\sig  &r(5j)(3j)          &3.3\sig  &                   &        &r(5j)(3j)         &3.1\sig &r(Muon20)         &10.1\sig
\\
     &r(10t-c30)         &3.7\sig  &r(Muon20)          &3.1\sig  &                   &        &r(mT2-400)        &2.2\sig &r(mT2500/300)     &5.2\sig
\\ 
     &r(Muon20)          &3.6\sig  &r(MET520)          &2.4\sig  &                   &        &r(20t-c45)        &2.1\sig &r(Hem3)           &4.1\sig
\\ \cline{2-11}
1000 &r(5j)(3j)          &10.1\sig &r(Muon20)          &7.2\sig  &                   &        &r(5j)(3j)         &5.4\sig &r(Muon20)         &25.8\sig
\\
     &r(Muon20)          &8.0\sig  &r(Hem3)            &5.7\sig  &                   &        &r(Hem3)           &5.3\sig &r(mT2-600/300)    &20.1\sig
\\
     &r(Hem3)            &7.3\sig  &r(5j)(3j)          &5.6\sig  &                   &        &r(Muon20)         &4.1\sig &r(Hem3)           &14.2\sig
\\\midrule
CS4d &&&&&&
\\
100  &r(MET520)          &3.5\sig  &r(HT900)           &3.0\sig  &r(5j)(3j)          &2.8\sig &                  &        &r(Muon20)         &6.8\sig
\\
     &r(HT900)           &3.2\sig  &r(MET520)          &2.7\sig  &r(mT2-300)         &2.1\sig &                  &        &r(MET420)         &5.5\sig
\\
     &r(Meff1400)        &2.6\sig  &r(Meff1400)        &2.6\sig  &r(10t-c30)         &1.9\sig &                  &        &r(mT2-500/300)    &5.2\sig     
\\ \cline{2-11}
1000 &r(MET520)          &6.5\sig  &r(MET520)          &5.1\sig  &r(5j)(3j)          &4.2\sig &                  &        &r(Muon20)         &17.3\sig
\\
     &r(mT2-600)         &5.3\sig  &r(mT2-600/400)     &4.8\sig  &r(10tdiff-c30)     &3.6\sig &                  &        &r(mT2-500)        &12.8\sig
\\
     &r(HT900)           &5.2\sig  &r(HT900)           &4.5\sig  &r(Hem3)            &3.6\sig &                  &        &r(MET520)         &11.5\sig
\\\midrule
CS6  &&&&&&
\\  
100  &r(MET420)          &7.0\sig  &r(MET420)          &6.0\sig  &r($b$-tag)$^*$     &6.5\sig &r(MET420)         &4.3\sig &                  &  
\\
     &r(mT2-500/300)     &5.1\sig  &r(mT2-500/300)     &4.6\sig  &r(Muon20)          &5.2\sig &r(Muon20)         &4.0\sig &                  &  
\\
     &r(HT900)           &4.8\sig  &r(HT900)           &4.5\sig  &r(MET420)          &4.0\sig &r(mT2-500/300)    &2.9\sig &                  &  
\\ \cline{2-11}
1000 &r(MET520)          &11.5\sig &r($b$-tag)$^*$     &11.0\sig &r($b$-tag)$^*$     &15.6\sig &r($b$-tag)$^*$   &14.9\sig&                  &  
\\ 
     &r($b$-tag)$^*$     &11.2\sig &r(MET520)          &10.3\sig &r(Muon20)          &10.2\sig &r(Muon20)        &8.4\sig &                  &  
\\
     &r(mT2-500)         &10.2\sig &r(mT2-500)         &9.2\sig  &r(MET520)          &7.6\sig  &r(MET420)        &7.6\sig &                  &  
\\ \bottomrule
\end{tabular}
\end{center}
\end{table*}


\section{Summary of Group 1 results}\label{sec:7}

Table~\ref{tab:g1resultssummary} summarizes the results from the five MSSM models of Group 1.
There are 20 pairwise model comparisons. One model is taken as the simulated data, with
the observed yield scaled to either the 100 pb$^{-1}$ discovery data set or
a 1000 pb$^{-1}$ follow-up. The actual number of events generated in each case was 100,000,
thus the ``data'' has smaller
fluctuations than would be present in the real experiment; we are interested in 
identifying the best discriminators and their approximate significance, not simulating
the real experiment. Note, however, that in our analysis we include the correct statistical
uncertainty arising from the assumed 100 pb$^{-1}$ or 1000 pb$^{-1}$ of integrated
luminosity in the ``data'' sample, as described in Section \ref{sec:6}.B.
When the rescaled counts are below a minimum value, the corresponding ratio is not used
in our analysis, as described in step 3 of analysis procedure outlined in Section \ref{sec:6}.

Given a ``data'' model, we can compare it to four theory models. We want to understand
to what extent we can reject each theory model, based on discrepancies in our
discriminating ratios. 
With the real LHC data, the test needs to be performed as follows:
given Model A and Model B, we will ask if Model A is a better description
of the data than Model B; this is a properly posed hypothesis test.
Every time we reject a theory model as an explanation of the ``data'',
we learn something about the true properties of the model underlying the ``data''.
When several discriminating ratios have high significance, we may learn more than one
qualitative feature of the underlying model from a single pairwise comparison;
this is not always the case however, since many ratios probe very similar features
of the models and are thus highly correlated.

In general the discriminating power of the robust ratios is quite impressive.
For 8 of the pairwise comparisons at least one ratio discriminates at
better than 5$\sigma$ with only 100 pb$^{-1}$ of 
simulated data\footnote{As is common practice in high energy physics,
we take 5$\sigma$ and 3$\sigma$ as reference values for discovery and evidence,
repsectively.}. 
In only three cases (discussed
in more detail below) do we fail
to discriminate by at least 5$\sigma$ with 1000 pb$^{-1}$. In 14 out of 20
cases 1000 pb$^{-1}$ gives $> 5\sigma$ discrimination with 3 or more
different ratios, giving multiple clues about the underlying data model.


\begin{figure}[thb]
\includegraphics[width=0.46\textwidth]{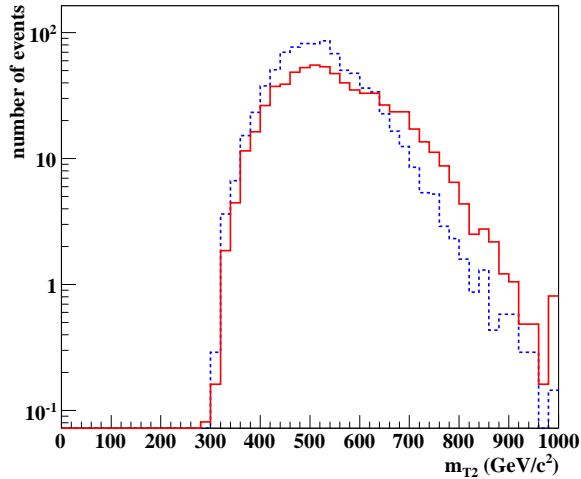}
\caption {\label{fig:Mt2LM5CS4d} Comparison of the stransverse mass $m_{T2}$ distributions
for look-alike models LM5 (solid red line) and CS4d (dashed blue line).
For each model 100,000 events were generated then rescaled to 1000 pb$^{-1}$.}
\end{figure}


\subsection{LM5 vs CS4d}\label{sec:7.1}

We begin with one of the simplest pairwise comparisons: LM5 is treated as the data and
compared to theory model CS4d. Averaging over the three pdfs used in our study, we
find that LM5 would produce 1951 signal events in the 100 pb$^{-1}$ discovery sample.
This is only 7\% more than the 1817 events predicted by theory model CS4d, so the
data model and theory model are indeed look-alikes. 
If we peek at the features of the two models we see that they have a number
of phenomenological similarities.
Both models have about the same
proportion of squark-gluino and squark-squark production. 
In CS4d gluinos decay to stop-top, followed by the three-body
stop decay $\tilde{t}_1\to bW^+\tilde{\chi}_1^0$; this resembles LM5
gluinos decaying to left-squark-quark, followed by left-squark
cascade to chargino-quark, with the chargino decaying to $W\tilde{\chi}_1^0$.
Both models have a large fraction of events with $W$'s.

At the moment of discovery CS4d is excluded as an explanation of the 
LM5 ``data'' by more than 3$\sigma$ in three kinematic ratios: 
r(HT900), r(Meff1400) and r(MET520). These
ratios discriminate based on the proportion of highly energetic events; their
values are about 50\%~larger for LM5 than for CS4d. This indicates that
the LM5 signal arises from production of heavier parent particles.
From the superpartner spectra in Figure~\ref{fig:group1spectra}
we see that indeed the gluino mass is about 100 GeV heavier in LM5 than
in CS4d, and the lightest squarks are also somewhat heavier.
Note that LM5 has a harder \met~distribution even though its LSP mass
is $\sim 100$ GeV lighter than that of CS4d.

Since the $m_{T2}$ endpoint is a direct measure of the (largest) parent
particle mass, we would expect the $m_{T2}$ ratios to be good
discriminators. However as can be seen in Figure~\ref{fig:Mt2LM5CS4d}
with 100 pb$^{-1}$
we are hampered by poor statistics near the endpoint. The best $m_{T2}$
ratio is r(mT2-600/300) in the MET box, computed with the LSP mass of CS4d;
it is defined as the number of events in the MET box with $m_{T2} > 600$ GeV 
divided by the number of events with $m_{T2} > 300$ GeV. This ratio 
has 2.8$\sigma$ significance with 100 pb$^{-1}$.

Making the same comparison at 1000 pb$^{-1}$, the significance of r(MET520), r(HT900)
and r(Meff1400)
as discriminators improves to 7.1, 6.4 and 5.9$\sigma$ respectively.
More importantly, two of the $m_{T2}$ ratios, r(mT2-600/300)
and r(mT2-600/400), now discriminate at better than 6$\sigma$.
We have not attempted to perform a direct extraction of the
endpoint, but it is clear that the $m_{T2}$ ratios can compete
with the kinematic ratios as discriminators while simultaneously providing more direct
information about the underlying $2\to 2$ parent subprocess.

Although we can reject CS4d conclusively, some qualitative differences
between LM5 and CS4d are not resolved.
Model CS4d
produces more tops, while LM5 produces a lot of Higgs; however
Table~\ref{tab:btagging} shows that the number of $b$ jets, $b$ tags and
tagged $b$ jets are all about the same. Lacking an explicit
reconstruction of tops or Higgs, we do not discriminate these models based
on these features.

\subsection{LM2p vs LM5}\label{sec:7.2}

This is the most difficult pair of look-alikes in our study. From
Figure~\ref{fig:group1spectra} we see that the superpartner spectra
are almost identical; the only significant difference is that LM2p
has a much lighter stau. As a result, LM2p events are much more
likely to contain $\tau$'s, while LM5 events are much more likely
to contain $W$'s (mostly from chargino decays).

As Table~\ref{tab:g1resultssummary} shows, at the moment of discovery
LM5 cannot be ruled out as the explanation of LM2p ``data''.
Without $\tau$ tagging, we would not have 3$\sigma$ discrimination
even with 1000 pb$^{-1}$; the best we could do is the 
jet multiplicity ratio r(5j)(3j) with 2.8$\sigma$: LM5 produces
more high multiplicity jet events after selection, because
we can get two jets from a $W$ decay in LM5 compared to only
one hadronic $\tau$ from a stau decay in LM2p.


\begin{figure}[htp]
\includegraphics[width=0.46\textwidth]{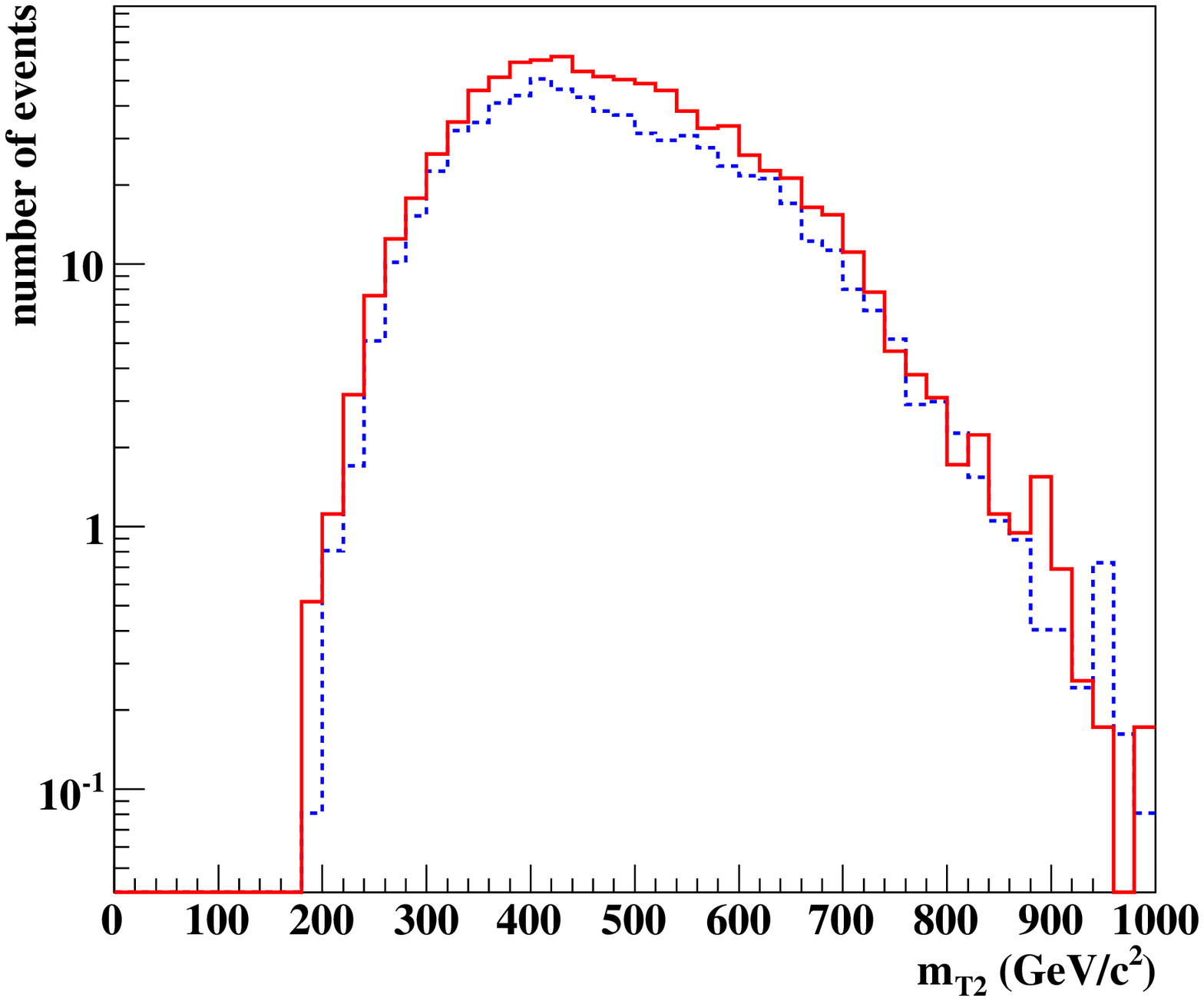}
\caption {\label{fig:mt2LM2pvLM5} Comparison of the stransverse mass $m_{T2}$ distributions
for look-alike models LM2p (solid red line) and LM5 (dashed blue line).
For each model 100,000 events were generated then rescaled to 1000 pb$^{-1}$.}
\end{figure}


\begin{figure}[htp]
\includegraphics[width=0.46\textwidth]{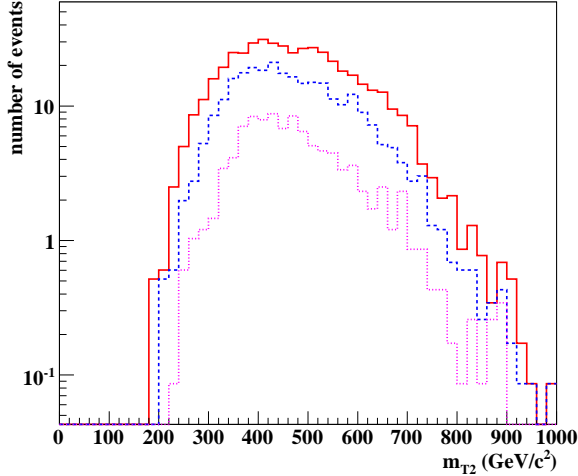}
\caption {\label{fig:mt2multiLM2p} Comparison of the stransverse mass $m_{T2}$ distributions
for model LM2p with a fixed number of reconstructed objects
(jets or muons): 3 objects (solid red line), 4 objects (dashed blue line)
and 5 objects (dotted magenta line).
For each case 100,000 events were generated then rescaled to 1000 pb$^{-1}$.}
\end{figure}


\begin{figure}[htp]
\includegraphics[width=0.46\textwidth]{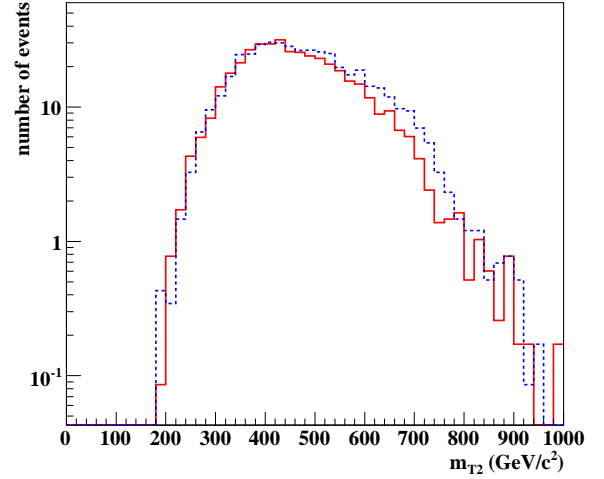}
\caption {\label{fig:mt2neutrinos} Comparison of the stransverse mass $m_{T2}$ distributions
for two subsamples of model LM2p: events with neutrinos (solid red line) and events
without neutrinos (dashed blue line).
A total of 100,000 events were generated, rescaled to 1000 pb$^{-1}$,
then sorted into the two subsamples.}
\end{figure}



\begin{table*}[htp]
\begin{center}
\caption{\label{tab:g2resultssummary} Summary of the best discriminating ratios
for model comparisons in Group 2. The models listed in rows are
taken as simulated data, with either 100 or 1000 pb$^{-1}$ of
integrated luminosity assumed, and uncertainties as described in the
text. The models listed in columns are then compared pairwise with
the ``data''. In each case, the three(five) best distinct discriminating ratios for 100(1000) pb$^{-1}$
are shown, with the estimated significance. 
By distinct we mean that we
only list the best ratio of each type; thus if r(5j)(4j) is listed,
then r(4j)(3j) is not, etc.
Square brackets denote ratios
defined in the DiJet, TriJet or Muon20 boxes; all other
ratios are defined in the MET box, and r(DiJet), r(TriJet) denotes the ratio of
the number of events in the DiJet/TriJet boxes to the number in the MET box.
The $m_{T2}$ ratios are computed
using the LSP mass of the relevant ``theory'' model, not the ``data'' model.}
\setlength{\tabcolsep}{1.6mm}
\begin{tabular}{llrlrlr}
\toprule
&\multicolumn{2}{c}{LH2}&\multicolumn{2}{c}{NM4}
&\multicolumn{2}{c}{CS7}
\\ \midrule
LH2 &&&&
\\
100  &                    &         &r(mT2-500)          &4.9\sig  &r(mT2-500)         &6.7\sig 
\\ 
     &                    &         &r(Meff1400)         &3.0\sig  &r(MET420)          &6.5\sig 
\\  
     &                    &         &r(M1400)            &2.7\sig  &r(4j)(3j)          &4.0\sig 
\\ \cline{2-7}    
1000 &                    &         &r(mT2-500)          &14.1\sig &r(mT2-500)         &18.9\sig 
\\
     &                    &         &r(mT2-300) [TriJet] &11.0\sig &r(MET420)          &16.7\sig 
\\
     &                    &         &r(mT2-400) [DiJjet] &7.9\sig  &r(mT2-500) [TriJet]&8.8\sig 
\\
     &                    &         &r(Meff1400)         &7.2\sig  &r(4j)(3j) [DiJet]  &7.3\sig
\\
     &                    &         &r(M1400)            &6.6\sig  &r(mT2-300) [DiJet] &6.7\sig
\\\midrule
NM4  &&&&
\\ 
100  &r(Meff1400)         &4.2\sig  &                    &         &r(Meff1400)        &4.3\sig 
\\
     &r(M1400)            &4.0\sig  &                    &         &r(DiJet)           &4.1\sig 
\\
     &r(mT2-400)          &3.8\sig  &                    &         &r(MET420)          &4.0\sig 
\\ \cline{2-7}
1000 &r(Meff1400)         &10.8\sig &                    &         &r(Meff1400)        &11.2\sig 
\\
     &r(TriJet)           &10.4\sig &                    &         &r(MET520)          &10.6\sig 
\\
     &r(M1400)            &9.8\sig  &                    &         &r(DiJet)           &10.6\sig 
\\
     &r(DiJet             &8.2\sig  &                    &         &r(HT900)           &9.0\sig
\\
     &r(HT900)            &8.0\sig  &                    &         &r(4j)(3j)          &6.1\sig
\\\midrule
CS7  &&&&
\\ 
100  &r(MET420)           &4.9\sig  &r(4j)(3j)           &4.4\sig  &                   &       
\\
     &r(4j)(3j)           &4.6\sig  &r(MET420)           &3.3\sig  &                   &       
\\ 
     &r(mT2-400)          &4.1\sig  &r(Hem1)             &3.2\sig  &                   &       
\\ \cline{2-7}
1000 &r(5j)(3j) [DiJet]   &16.8\sig &r(4j)(3j)           &9.4\sig  &                   &       
\\
     &r(TriJet)           &10.4\sig &r(5j)(3j) [DiJet]   &7.4\sig  &                   &      
\\
     &r(MET420)           &9.6\sig  &r(Meff1400)         &7.4\sig  &                   &   
\\
     &r(4j)(3j)           &9.5\sig  &r(DiJet)            &6.9\sig  &                   & 
\\
     &r(mT2-500)          &8.3\sig  &r(HT900)            &6.2\sig  &                   & 
\\ \bottomrule
\end{tabular}
\end{center}
\end{table*}


With our crude $\tau$ tagging algorithm we manage to discriminate
LM2p and LM5 at 3.1$\sigma$ with 1000 pb$^{-1}$. As seen in
Table~\ref{tab:tautagging}, the LM2p sample has almost three times
as many hadronic $\tau$'s that reconstruct as jets than does the
LM5 sample. This results in four times as many jets being
correctly tagged as hadronic $\tau$'s, and almost 50\% more
total $\tau$ tags for LM2p versus LM5.

The only other ratios that give even 2$\sigma$ discrimination
are based on the stransverse mass $m_{T2}$.
The ratio r(mT2-400) counts the number of events with
$m_{T2} > 400$ GeV divided by the total number of events
after selection; here $m_{T2}$ is computed using the LSP
mass of the theory model LM5. Figure~\ref{fig:mt2LM2pvLM5} compares the
$m_{T2}$ distributions of LM2p and LM5. The shapes of 
the distributions are very similar; the only obvious difference is
that there are fewer events in the LM5 bins.

It is not surprising that the shapes are similar, since
LM2p and LM5 are very similar models with nearly identical
gluino and squark masses. Thus ratios like r(mT2-600/400)
should not be good discriminators, and indeed they are not.
The ratio r(mT2-400), which is a good discriminator, is
obviously just picking up the fact that there are
fewer events in the LM5 bins than in the LM2p bins.

Figure~\ref{fig:mt2multiLM2p} compares the
$m_{T2}$ distributions of LM2p subsamples with fixed
multiplicity of jets$+$muons. We see that events with higher
multiplicity are significantly more likely to fail the
300 GeV hemisphere invariant mass upper bound that we
imposed to make up for the effect on $m_{T2}$ of imperfect
hemisphere separation. This makes sense since events with
higher multiplicity are more likely to have mistakes
in the hemisphere assignments.

Thus the discriminating power of r(mT2-400)
in this case is correlated with r(5j)(3j), not with
kinematic ratios like r(HT900) and r(Meff1400).

It is important to note that the $m_{T2}$ ratios
have some ability to discriminate based on neutrinos in the
final state: Figure~\ref{fig:mt2neutrinos} shows a comparison of the
$m_{T2}$ distributions for LM2p events containing
neutrinos versus those without neutrinos.
The events with neutrinos have a softer $m_{T2}$
distribution, i.e. the subsample with
neutrinos is less efficient at populating the $m_{T2}$ upper endpoint.
Models LM2p and LM5 differ greatly in the proportion of events
after selection that have neutrinos: about 50\% for LM2p
but only about 10\% for LM5. The neutrino content effect on the
$m_{T2}$ distributions actually reduces the discrimination
of LM2p versus LM5, because the neutrino effect works in the
opposite direction from the dominant effect of jet multiplicity.

This example shows that the interpretation of the $m_{T2}$ ratios
requires a comparison with other discriminators.
If the $m_{T2}$ ratios r(mT2-{\tt{xxx/yyy}})
have a high significance positively correlated
with e.g. r(HT900) and r(Meff1400), then the
$m_{T2}$ ratios are predominantly indicating kinematics.
If the $m_{T2}$ ratios r(mt2-{\tt{xxx}}) have a high significance 
but r(mT2-{\tt{xxx/yyy}}) do not (as occurred here), 
we expect they will be positively correlated with the jet ratios, 
indicating a difference in the multiplicity of reconstructed objects. 
If the $m_{T2}$ ratios r(mT2-{\tt{xxx/yyy}}) have a high significance 
uncorrelated or negatively
correlated with either kinematics or jet multiplicity, this could
signal the presence of three unseen particles
(e.g. two LSPs and a neutrino) in the final state of
a large fraction of events.

\subsection{CS4d vs LM8}\label{sec:7.3}

This is the second most difficult pair of look-alikes in our study. From
Figure~\ref{fig:group1spectra} we see that the gluino and squark superpartner 
spectra
are roughly similar. The gluino masses agree to within 10 GeV;
in LM8 the left and right squarks are nearly degenerate and
slightly heavier than the gluino, while in CS4d the right squarks are slightly
lighter than the gluino and the left-squarks are about 180 GeV heavier.

Both models have about the same fractions of squark pair, squark-gluino
and gluino pair production after selection. In both models the gluino
decays predominately to top-stop. Both models produce many tops,
$b$'s and $W$'s. LM8 also produces a lot of $Z$'s from $\tilde{\chi}_2^0$
decays.

The main difference between these models is that CS4d has a much lighter stop 
and a much heavier LSP.
With $m_{\tilde{t}_1}$$=$$352$ GeV and $m_{\tilde{\chi}^0_1}$$=$$251$ GeV
a two-body light stop decay cannot occur, and this
stop can just barely manage the three-body decay
$\tilde{t}_1\to b W^+ \tilde{\chi}_1^0$.

Consider the case that we perform our look-alike analysis 
taking CS4d as the ``data'' model
and LM8 as the theory model. At the moment of discovery, LM8 will
explain not only the overall size of the signal but also its kinematics:
with 100 pb$^{-1}$ we find no kinematic observable that discriminates
better than r(Meff1400) with 1.7$\sigma$, and even with 1000 pb$^{-1}$
no kinematic observables discriminate at even the 3$\sigma$ level.
Given that LM8 is an mSUGRA  benchmark used  at the LHC experiments
it may be tempting to falsely conclude that LM8 is the probable explanation 
of the discovery!  Since LM8 has 230 GeV charginos, even a preliminary 
result in this direction could be used, for example, as a justification 
to start  building a 500 GeV linear collider. Since the actual chargino
 mass of the underlying compressed SUSY model is 352 GeV, this would 
lead to embarrassment.

Fortunately our look-alike analysis gives additional discriminating
handles:
\begin{itemize}
\item The ratio r(Muon20)
has a 3.4$\sigma$ significance with 1000 pb$^{-1}$, reflecting a larger
fraction of reconstructed muons in LM8 events over CS4d. 
\item With 1000 pb$^{-1}$
there are enough dimuons to reconstruct the $Z$ peak as shown in
Figure~\ref{fig:dimuon} for LM8, while no peak appears for CS4d.
\item The ratio r(5j)(3j) in the MET box differs by 2.8$\sigma$ at 100 pb$^{-1}$,
increasing to 4.2$\sigma$ with 1000 pb$^{-1}$, reflecting a larger
jet multiplicity in LM8 events versus CS4d. 
\item With 1000 pb$^{-1}$
we also see discrepancies in the event shape variables. One of
these, r(Hem3), is the fraction of events where the object counts
in the two hemispheres differ by at least 3; the other, r(10tdiff-c300),
is the fraction of events where the track count in a 30$^{\circ}$ cone
around each hemisphere axis differs by at least 10. With 1000 pb$^{-1}$
both these ratios have a significance of 3.6$\sigma$, reflecting more
symmetrical object counts in CS4d events than for LM8. 
\end{itemize}
From Table~\ref{tab:btagging} we would have hoped for a significant
difference in r($b$-tag) between LM8 and CS4d. However even with
1000 pb$^{-1}$ we obtain only a 2.4$\sigma$ significance with this ratio;
this is due to the combination of low efficiency in the $b$ tagging and
bad luck in that this ratio happens to have a rather large uncertainty
from the pdf spread.

These handles exclude LM8 with reasonable confidence as the explanation
of the ``data''. They also give clues on how to modify LM8 (still
within the hypothesis of SUSY) to better fit the data. The $Z$ peak
in LM8 comes from left-squarks decaying to $\tilde{\chi}_2^0$; making
the left-squarks heavier will cause them to decay instead to quark-gluino.
To keep the parent kinematics and observed yield constant, this also
suggests lowering the right-squark masses. This has the added benefit
of favoring the 2-body squark decay $\tilde{q}_R\to q\tilde{\chi}^0_1$
over the decay $\tilde{q}_R\to q\tilde{g}$, which lowers the
jet multiplicity.
The large hemisphere asymmetries in LM8 are derived from gluino cascades
resulting in two top quarks, and thus up to six jets, in a single hemisphere.
The obvious way to reduce this without drastically changing the model is to
squeeze out the phase space for the 2-body stop decay $\tilde{t}_1\to t\tilde{\chi}^0_1$,
and further squeeze the 3-body stop decay. By thus reducing the amount of
visible energy reconstructed in the events, this
reduces the hemisphere asymmetries, the jet multiplicity, and the
muon counts.

Thus with 1000 pb$^{-1}$ we might not only exclude LM8 but also come close to 
guessing CS4d from this simple look-alike comparison.
Without the benefit of additional model comparisons this guess would be relying
on the strong assumptions
that the ``data'' was SUSY (an assumption that
we will relax in the Group 2 analysis) and that the data
was full of $b$'s and tops despite lacking explicit confirmation from
$b$ tagging or top reconstruction.


\begin{figure}[htp]
\includegraphics[width=0.48\textwidth]{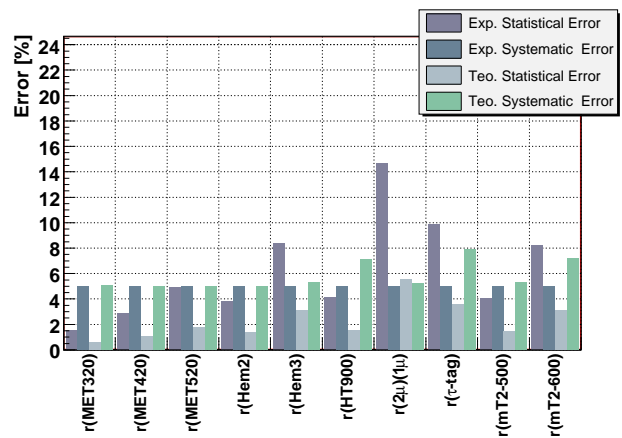}
\caption {\label{fig:CS61000pbCS4dErr_MET_1} Breakdown of estimated uncertainties for discriminating
ratios with 1000 pb$^{-1}$, in the comparison of look-alike models CS6 and CS4d, with
CS6 treated as the ``data''.}
\end{figure}



\begin{figure}[htp]
\includegraphics[width=0.48\textwidth]{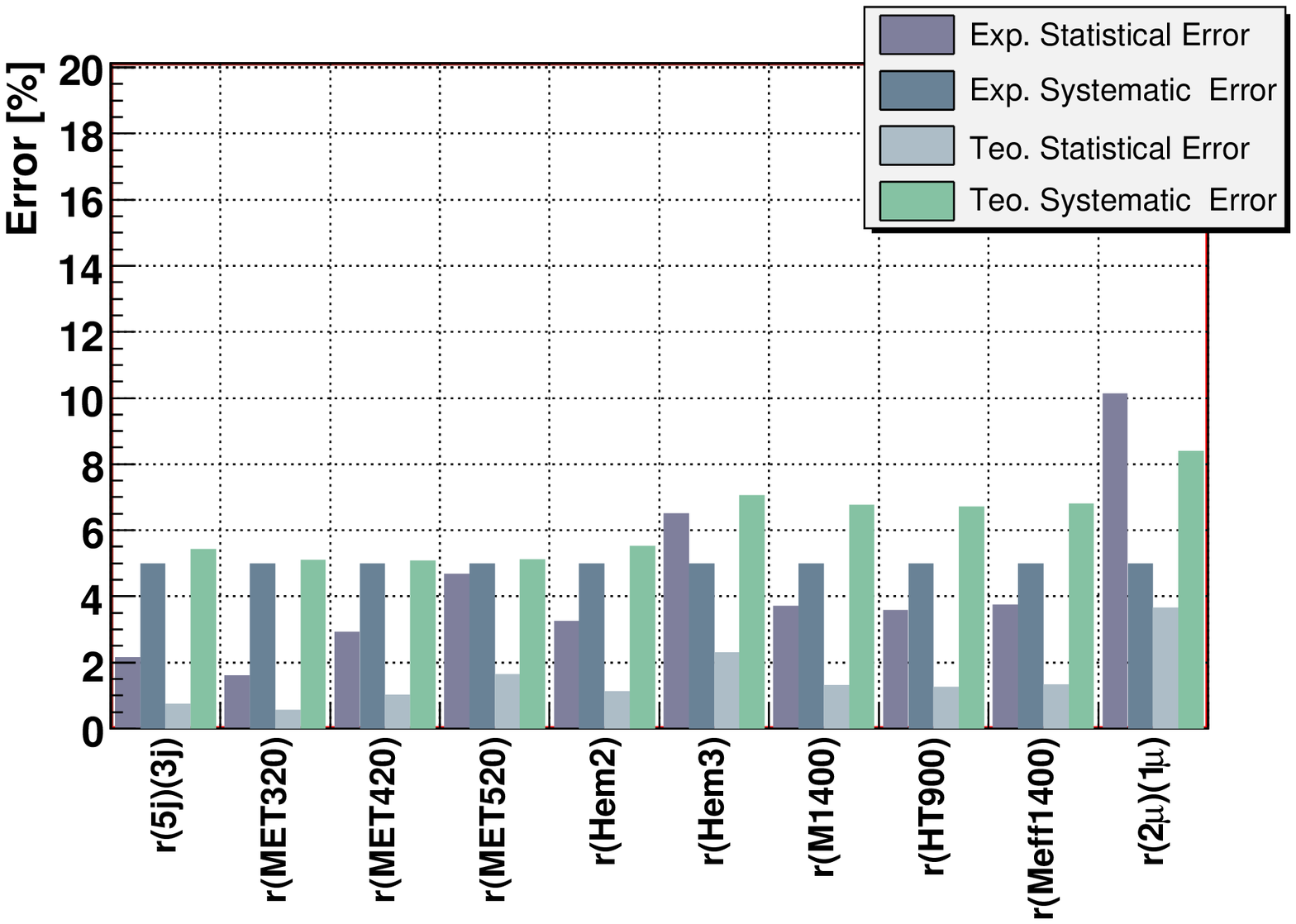}
\caption {\label{fig:CS61000pbLM8Err_MET_1} Breakdown of estimated uncertainties for discriminating
ratios with 1000 pb$^{-1}$, in the comparison of look-alike models CS6 and LM8, with
CS6 treated as the ``data''.}
\end{figure}



\begin{figure}[htp]
\includegraphics[width=0.48\textwidth]{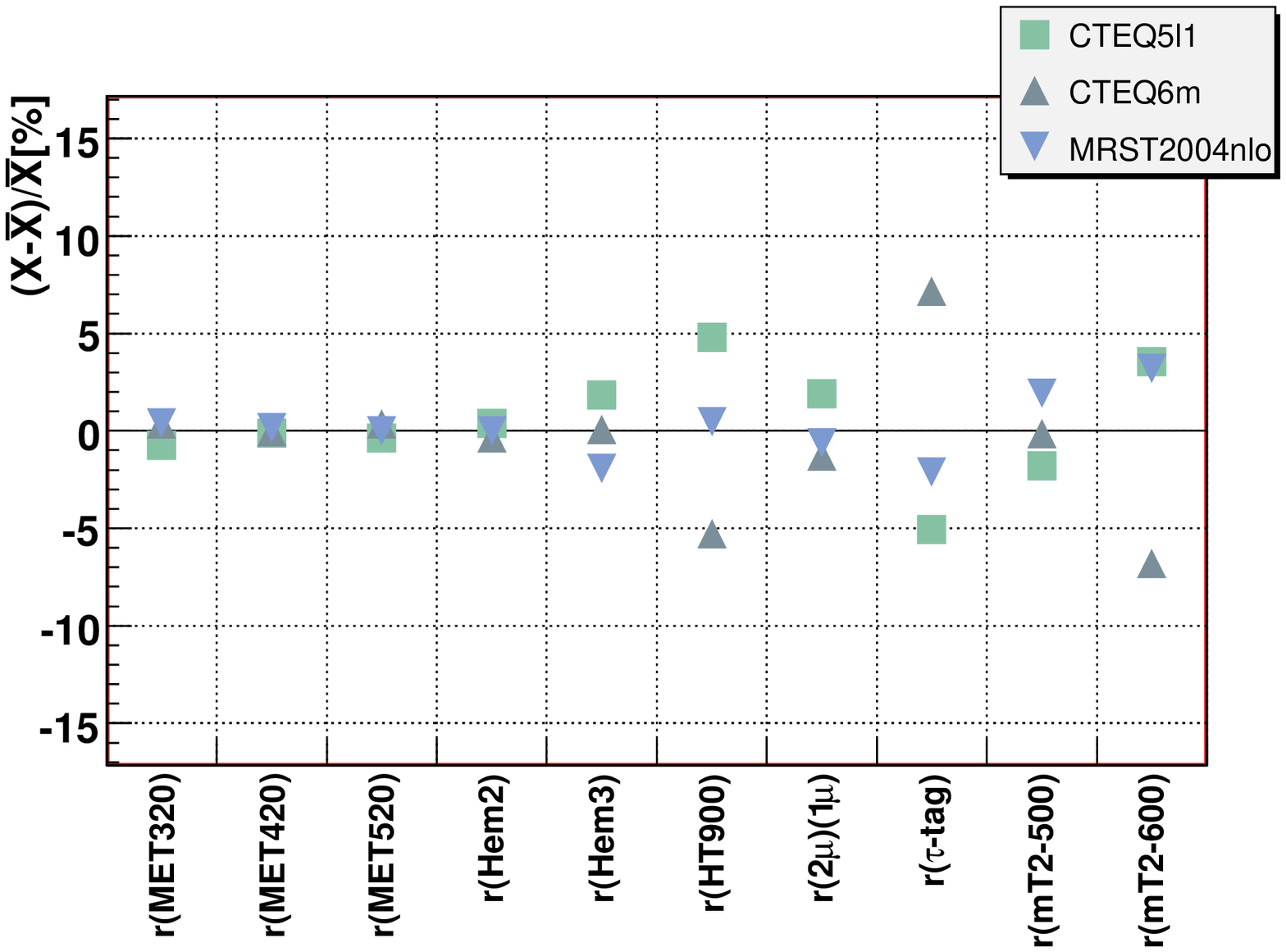}
\caption {\label{fig:CS61000pbCS4dPdf_MET_1} Pdf spreads for discriminating
ratios with 1000 pb$^{-1}$, in the comparison of look-alike models CS6 and CS4d, with
CS6 treated as the ``data''.}
\end{figure}



\begin{figure}[htp]
\includegraphics[width=0.48\textwidth]{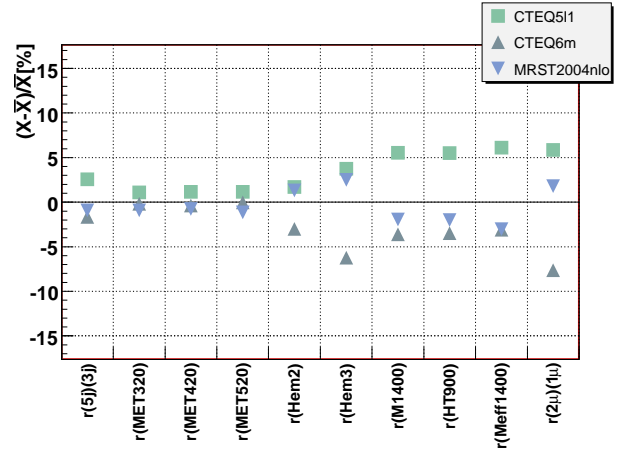}
\caption {\label{fig:CS61000pbLM8Pdf_MET_1} Pdf spreads for discriminating
ratios with 1000 pb$^{-1}$, in the comparison of look-alike models CS6 and LM8, with
CS6 treated as the ``data''.}
\end{figure}


\subsection{CS6 vs LM2p, LM5, LM8 and CS4d}\label{sec:7.4}

This is a complete example of how we could deduce the correct model, under
the assumption of SUSY, based
on how it fails to match with four incorrect SUSY models. 
We take CS6 as the ``data'' and compare it to LM2p, LM5, LM8 and CS4d. 

With 100 pb$^{-1}$, the ratio r(mT2-500/300)
is much smaller for the ``data'' than for LM2p, LM5, LM8 and CS4d,
with significance 5.1, 4.6, 3.3 and 2.9$\sigma$ respectively.
This indicates that the parent particle mass in the
data model is quite a bit lighter than in model CS4d, which has the lightest
squarks and gluinos of the four theory models.

To keep the overall data yield constant, we can contemplate two
possible modifications of the spectrum in CS4d: either make the
gluino lighter and the squarks heavier, or vice-versa.

The other striking result with 100 pb$^{-1}$ is that the ``data''
has only 1/4 as many events in the Muon20 box as the theory model LM8,
and 1/3 as many as the theory model CS4d. At the same time, the ``data''
has nearly three times \textit{more} $b$ tags in the Muon20 box
than does LM8, a 6.5$\sigma$ discrepancy. With a little more data
the same puzzling discrepancy turns up in the comparison with CS4d.

How to explain this? Recall that both models LM8 and CS4d produce
a large number of $b$'s and $W$'s, and LM8 also has large numbers of $Z$'s.
Events ending up in the Muon20 box have hard muons from either
$W$/$Z$ decays or from energetic cascades with semileptonic $B$ decays
in jets. If we removed all of the $W$'s and $Z$'s we would have many
fewer events in the Muon20 box. However precisely in this case
all of the events in the Muon20 box will have hard muons inside
$b$ jets. This would then explain a high $b$ tagging rate in this
box combined with a smaller overall count.

Already with 100 pb$^{-1}$ we have strong clues that the data
model has a light gluino, heavy squarks, and that the gluino
decays do not involve $W$'s, $Z$'s or sleptons. Table~\ref{tab:tautagging}
shows that the ``data'' has only 59 $\tau$ tags per 1000 pb$^{-1}$;
this gives a 6$\sigma$ deficiency in $\tau$ tags for the data relative
to model LM2p after 1000 pb$^{-1}$. The lack of $\tau$ tags 
indicates we are not making a
lot of $\tau$'s, but if we are not making either $W$'s or $\tau$'s then
we are probably not making charginos from gluino decays.
This suggests that the mass splitting between the gluino and the chargino
is relatively small, and that the gluino has a three-body decay.
A three-body decay mediated by a virtual chargino would imply
more muons, so we a led to the three-body decay $\tilde{g}$$\to$
$q\bar{q}\tilde{\chi}_1^0$ mediated by a virtual squark.

Putting it all
together (within the hypothesis of SUSY) leads to a model like CS6, 
with heavy squarks, a light gluino, and
a compressed gaugino spectrum. Production is dominated by gluino pairs,
and gluino decays are dominated by the three-body mode 
to $q\bar{q}\tilde{\chi}_1^0$. The only muons are from semileptonic $B$ decays. 

This scenario also makes
predictions that can be tested with more data. For example, the hemisphere
counts should be quite symmetrical, since we are almost always producing a pair
of the same particles with the same decays. Indeed this prediction is
borne out with 1000 pb$^{-1}$, where both r(Hem2) and r(Hem3) are $>6 \sigma$
smaller for CS6 than for LM8.

These hemisphere ratios also demonstrate the importance of the pdf
 uncertainties.
Figures~\ref{fig:CS61000pbCS4dErr_MET_1} and \ref{fig:CS61000pbLM8Err_MET_1}
show the breakdown of the uncertainties for some of the discriminating ratios
with 1000 pb$^{-1}$ in the comparisons of CS6 ``data'' with models CS4d and LM8.
For CS4d the r(Hem2) ratio, which discriminates at 4.6$\sigma$,
is systematics limited, while the r(Hem3) ratio still has a rather large
statistical uncertainty. For LM8 we notice that the theory systematic is the
largest uncertainty for both r(Hem2) and r(Hem3).

These differences are explained by Figures~\ref{fig:CS61000pbCS4dPdf_MET_1}
and \ref{fig:CS61000pbLM8Pdf_MET_1}, which show the spread in the values
of the ratios as we vary the parton distribution functions used in the simulation.
Note that the pdf spreads are model dependent: the 
spreads for the hemisphere ratios r(Hem2) and r(Hem3) are twice as large
for LM8 as they are for CS4d. This explains why the total theory systematic
for these ratios is larger for LM8 than for CS4d. However because of better
statistics the hemisphere ratios appear to discriminate better for CS6 vs LM8
than they do for CS6 vs CS4d. The caveat is that the validity of this statement 
depends crucially on whether our estimates of the pdf uncertainties are
at least roughly accurate.


\section{Summary of Group 2 results}\label{sec:8}

Table~\ref{tab:g2resultssummary} summarizes the results from the models of Group 2.
There are 6 pairwise model comparisons. One model is taken as the simulated data, with
the number of signal events scaled to either the 100 pb$^{-1}$ discovery data set or
a 1000 pb$^{-1}$ follow-up. In our analysis we include the correct statistical
uncertainty arising from the assumed 100 pb$^{-1}$ or 1000 pb$^{-1}$ of integrated
luminosity in the ``data'' sample, as described in Section \ref{sec:6}.B.

The most remarkable feature of these results is that we achieve greater than
4$\sigma$ discrimination of non-SUSY model LH2 from its SUSY look-alikes
NM4 and CS7, \textit{already at the moment of discovery}. With the larger
1000 pb$^{-1}$ data set, we achieve $> 5 \sigma$ discrimination in every
case, for more than five distinct ratios per comparison. Thus even with small
data sets we can both distinguish SUSY and non-SUSY explanations of
the same excess and have multiple handles to inform us about key properties
of the true model behind the data.

To see how this works in detail, let us take LH2 as our ``data''.
Suppose that we lived in a world where particle theorists believed
that any missing energy signal has to be explained by SUSY (until
recently we did in fact live in such a world). Then clever theorists
might construct the SUSY model NM4 shown in
Figure~\ref{fig:group2spectra}
as an explanation of the missing
energy discovery.

Applying our look-alike analysis, however, one detects a problem
already with 100 pb$^{-1}$. The $m_{T2}$ ratio r(mT2-500) (computed using
the NM4 LSP mass) is three times
larger for the ``data'' than for the SUSY model NM4, a nearly 5$\sigma$
discrepancy. As seen in Table~\ref{tab:LH2vsNM4_100MET}, this is 
positively correlated with the $m_{T2}$ ratio
r(mT2-500/300), which is more than twice as large for the ``data''
as for NM4, a 3$\sigma$ discrepancy, and uncorrelated with the jet
multiplicity, which shows no significant difference.

These results strongly suggest that the true model underlying
the ``data'' has heavier parent particles than does SUSY model NM4.
However the SUSY enthusiast will be quite confused by this conclusion,
since the kinematic ratio r(Meff1400) is more than two times \textit{greater}
for NM4 than it is for the ``data'', a 3$\sigma$ discrepancy that seems
to directly contradict our previous conclusion. A conservative
SUSY enthusiast might choose to wait for more data, but this will only
reinforce the confusion; with 1000 pb$^{-1}$ the ratio r(mT2-500/300)
indicates with 8.5$\sigma$ significance that the NM4 squarks are too light, while
r(Meff1400), r(M1400) and r(HT900) all indicate at $>5 \sigma$ that
SUSY NM4 events are too energetic!

At some point our SUSY enthusiast may decide to replace
SUSY model NM4 with SUSY model CS7.
This is a bright idea since the parent production in CS7 is
all gluinos, instead of all squarks as in NM4,
and the gluino kinematics are naturally softer then squark
kinematics, for comparable masses. Thus, as seen in
Table~\ref{tab:LH2vsCS7_100MET}, CS7 matches both the kinematics and the
overall yield of the ``data'' much better than NM4: even with 1000 pb$^{-1}$
r(Meff1400), r(M1400) and r(HT900) have no discrepancies as large as 3$\sigma$.

However our SUSY enthusiast still has serious problems. 
Table~\ref{tab:LH2vsCS7_100MET} shows that now the \met~distributions 
are way off: 
even with 100 pb$^{-1}$
the ratio r(MET420) is more than twice as big for the ``data'' as for CS7,
while r(MET520) is three times as big; these are both $>5 \sigma$ discrepancies.
Furthermore the jet multiplicities don't match: the ratio r(4j)(3j) is almost
twice as large for CS7 as for the ``data'', a 4$\sigma$ discrepancy with 100 pb$^{-1}$.

\begin{figure}[htp]
\includegraphics[width=0.48\textwidth]{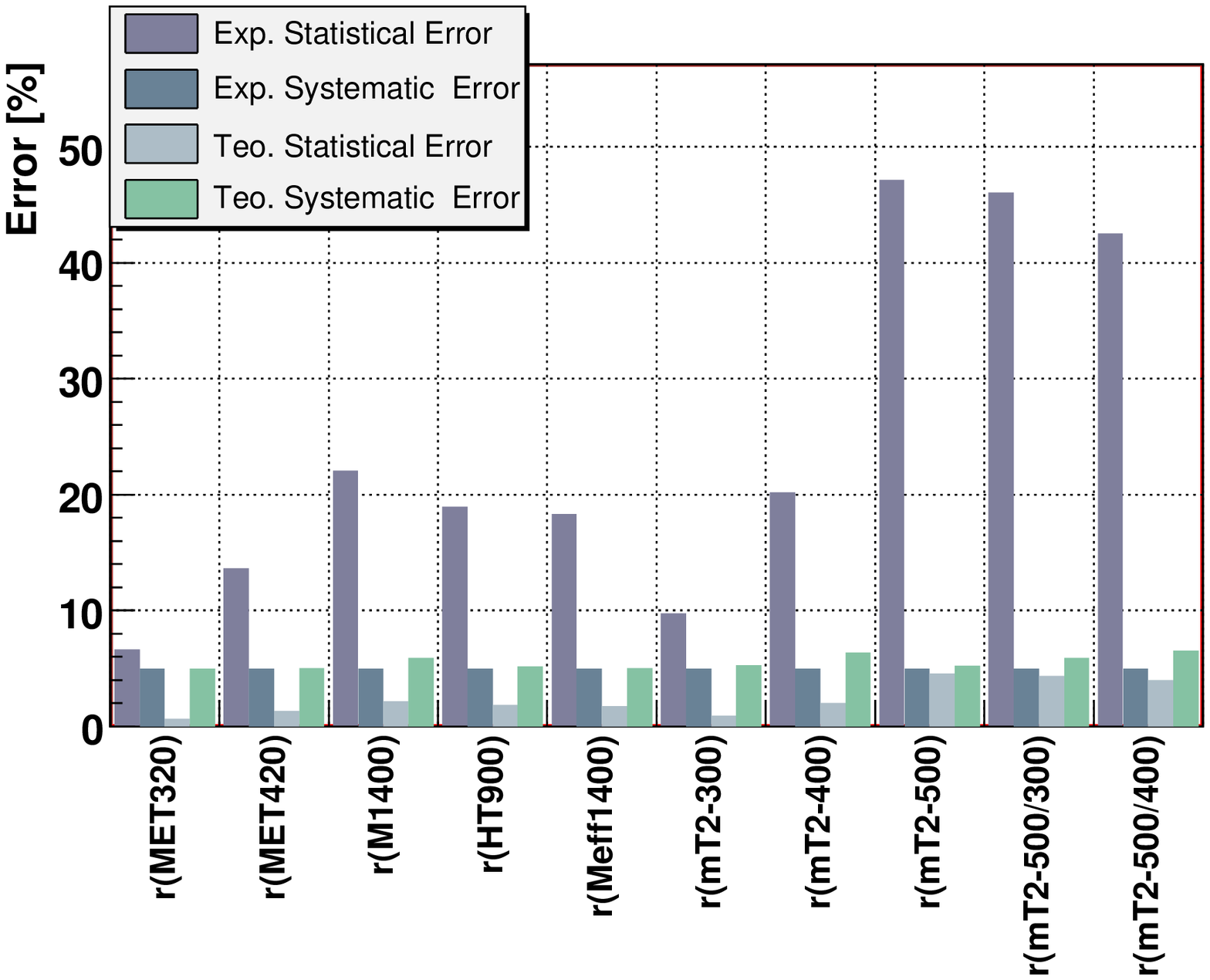}
\caption {\label{fig:LH2100pbNM4Err_MET_1} Breakdown of estimated uncertainties for discriminating
ratios with 100 pb$^{-1}$, in the comparison of look-alike models LH2 and NM4, with
LH2 treated as the ``data''.}
\end{figure}

\begin{figure}[htp]
\includegraphics[width=0.48\textwidth]{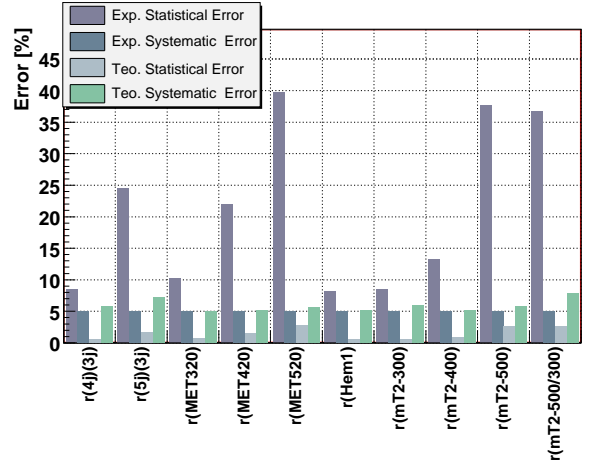}
\caption {\label{fig:LH2100pbCS7Err_MET_1} Breakdown of estimated uncertainties for discriminating
ratios with 100 pb$^{-1}$, in the comparison of look-alike models LH2 and CS7, with
LH2 treated as the ``data''.}
\end{figure}

Figures~\ref{fig:LH2100pbNM4Err_MET_1} and \ref{fig:LH2100pbCS7Err_MET_1}
demonstrate the robustness of these results, by showing the breakdown of the
experimental and theoretical uncertainties for the relevant ratios.
With the exception of r(4j)(3j), the uncertainties on all of the ratios
that we have been discussing are completely dominated by the low statistics
of our small ``data'' sample. Thus, for example, doubling the pdf uncertainties
would not alter any of the conclusions reached above. 

It is not obvious that 
our SUSY diehard can fix up a SUSY candidate to falsely explain 
the non-SUSY ``data'',
while surviving the scrutiny of
our look-alike analysis. This applies even for small data sets on the order of a few
hundred inverse picobarns. The key observation is that although SUSY models 
have many adjustable parameters,
the number of adjustable parameters \textit{relevant} to this look-alike analysis is small compared
to the number of robust discriminators.

\begin{table}[htp]
\begin{center}
\begin{tabular}{lrrr}
\toprule
\multicolumn{4}{c}{LH2 vs. NM4 [100 pb$^{-1}$]}\\
\midrule
 Variable & LH2 & NM4 & Separation \\
\midrule
\multicolumn{4}{c}{MET}\\
\midrule
r(mT2-500)  & 0.16 & 0.05 & 4.87 \\
r(mT2-400)  & 0.44 & 0.21 & 4.84 \\
r(mT2-300)  & 0.75 & 0.54 & 3.49 \\
r(Meff1400)  & 0.11 & 0.25 & 2.99 \\
r(mT2-500/300)  & 0.21 & 0.09 & 2.98 \\
r(M1400)  & 0.07 & 0.19 & 2.69 \\
r(mT2-400/300)  & 0.58 & 0.40 & 2.48 \\
r(HT900)  & 0.13 & 0.24 & 2.34 \\
r(MET420)  & 0.48 & 0.37 & 2.00 \\
r(mT2-500/400)  & 0.36 & 0.22 & 1.47 \\
\bottomrule
\end{tabular}
\end{center}
\caption {Best discriminating ratios in the MET box, with separations in units of $\sigma$,
for the comparison of LH2 vs. NM4, taking LH2 as the ``data'', 
assuming an integrated luminosity of 100 pb$^{-1}$.\label{tab:LH2vsNM4_100MET}}
\end{table}

\begin{table}[htp]
\begin{center}
\begin{tabular}{lrrr}
\toprule
\multicolumn{4}{c}{LH2 vs. CS7 [100 pb$^{-1}$]}\\
\midrule
 Variable & LH2 & CS7 & Separation \\
\midrule
\multicolumn{4}{c}{MET}\\
\midrule
r(mT2-500)  & 0.27 & 0.08 & 6.68 \\
r(MET420)  & 0.48 & 0.20 & 6.49 \\
r(MET520)  & 0.21 & 0.07 & 5.06 \\
r(MET320)  & 0.78 & 0.53 & 4.29 \\
r(mT2-500/300)  & 0.32 & 0.12 & 4.24 \\
r(4j)(3j)  & 0.36 & 0.61 & 4.04 \\
r(mT2-400)  & 0.63 & 0.40 & 4.00 \\
r(mT2-300)  & 0.85 & 0.62 & 3.55 \\
r(mT2-500/400)  & 0.43 & 0.19 & 3.52 \\
r(Hem1)  & 0.79 & 0.63 & 2.59 \\
\bottomrule
\end{tabular}
\end{center}
\caption {Best discriminating ratios in the MET box, with separations in units of $\sigma$,
for the comparison of LH2 vs. CS7, taking LH2 as the ``data'', 
assuming an integrated luminosity of 100 pb$^{-1}$.\label{tab:LH2vsCS7_100MET}}
\end{table}


\section{Discussion and outlook}\label{sec:9}

We have presented a concrete strategy for determining the underlying
theory model of an early missing energy discovery at the LHC.
Applying this look-alike analysis to a realistic simulation,
we were able to distinguish a non-SUSY
model from its SUSY look-alikes essentially at the moment of discovery,
with little more than 100 pb$^{-1}$ of integrated luminosity.
In 23 of 26 pairwise comparisons, mostly SUSY with SUSY, 
we were able to discriminate look-alikes
at better than 5$\sigma$ significance with at least one robust observable
and 1000 pb$^{-1}$ or less of integrated luminosity. Even in the three
cases with the worst discrimination we found strong hints of the
key properties of the underlying model; these would be confirmed
with more data and/or by our improving the look-alike analysis.

One surprise of our study (at least to us) was the sensitivity
and robustness of the ratios based on the stransverse mass $m_{T2}$.
Keep in mind that we did not apply the $m_{T2}$ distributions to
their originally intended use i.e. extracting masses from endpoints
and kinks, and we applied our $m_{T2}$ ratios to data sets 100 times
smaller than used in previous studies. Nevertheless we found that the
$m_{T2}$ ratios are among our best discriminators. One of the
most important features of the $m_{T2}$ ratios is that to
first approximation they do not depend on the spins of the parent
particles. Since ratios based on more traditional kinematic
distributions like $H_T$ and $M_{\rm eff}$ have a large
dependence on the spins of the parent particles, comparing
$m_{T2}$ ratios to these ratios is a powerful discriminator
for spin.

Our main goal in this study was to develop a look-alike analysis
for missing energy that can be successfully applied to the LHC data
in the first year of physics running. The crucial properties
of such an analysis are realism, robustness, validation and
sensitivity. We briefly summarize where we stand with respect
to establishing these properties.

\subsection*{Realism} 
We have employed state-of-the-art event generation
for the missing energy signals, but only at leading order in each
subprocess. In the next phase we include the possibility of extra
high $E_T$ jets at the matrix element level. We are performing this study
within the CMS collaboration and hence replace our 
fast simulation with the full CMS detector simulation. 
This also allows us to include the Standard Model backgrounds
in the full analysis, replacing the background subtraction assumed here.

\subsection*{Robustness} Our analysis is already quite robust against
disappointments in the performance of the LHC detectors during the
first physics run. We have assumed a minimal palette of reconstructed
objects and triggers. Because our analysis uses only ratios of correlated
inclusive counts, there is a large cancellation of theoretical and experimental
uncertainties, and we can make simple apples--with--apples comparisons between
different observables. 

Despite the fact that we are considering small
data sets, we have not employed any multivariate statistical methods.
As mentioned earlier such methods are left for the era of
demonstrated understanding of the correlations between observables.
By dispensing with these methods we lose sensitivity but gain a
cleaner more robust analysis.

We gain additional robustness from the physics redundancy
built into our choice of correlated observables. For example, jet multiplicity
is correlated with track counts in the cones, and sometimes with
the ratios of $m_{T2}$ counts to total number of events in the box.
The hemisphere ratios are correlated with the difference counts in the
cones. Muon counting is correlated with $b$ tagging. The four
trigger boxes provide us with four complete sets of ratios, allowing
further comparisons and cross-checks.

Our main deficiency in robustness is the limited number of theory
models simulated for this study. In the next phase we are including
a much larger number and variety of models.

A possible approach to expanding this analysis is to apply
the idea of OSETs \cite{ArkaniHamed:2007fw}, as a strategy
for effectively sampling the entire theory space.
A disadvantage of this approach is that, by definition, we give up our
spin sensitivity and more generally any discrimination based
on details of the matrix elements. 
 
\subsection*{Validation} 
No studies performed to date of the LHC phenomenology
of any BSM theory have adequately validated uncertainties.
The experimental uncertainties cannot be sufficiently validated 
until we have LHC data. The theoretical uncertainties could be
validated sooner, but this will require many more detailed
studies adhering to at least the degree of realism attempted here.

In the next phase of this analysis (currently being performed
within the CMS  collaboration) we are validating
the Standard Model backgrounds of the
inclusive missing energy signature using the CMS full simulation
framework. This, among other things, is allowing us to compute the
backgrounds in each of our trigger boxes and study the effect
of varying the selection criteria, e.g. relaxing or
modifying the ILV.

The full detector simulation itself needs to be validated against
real LHC data, using these same Standard Model processes as benchmarks.
Similar comments apply to the parton distribution functions.
In both cases we need to develop a more sophisticated parametrization
of the uncertainties.

The event generation chains used in our study, while state-of-the-art,
have not been adequately validated; we have performed several cross-validation 
checks and obtained significant discrepancies. This is an important task 
for the entire LHC theory community. 

\subsection*{Sensitivity} 

We have demonstrated the importance of being able to
obtain subsamples of the discovery data set enriched in $\tau$'s and
$b$'s. To do this successfully will require a different approach to flavor
tagging in the early LHC running, an approach that emphasizes robustness
and fast validation over efficiency and purity. It seems likely that a dedicated
effort could achieve results as good as our preliminary study, and
possibly much better.

We have seen that one of the virtues of the $m_{T2}$ ratios is
sensitivity to the number of weakly interacting particles in the
final state. It is important to study this further, along with
the potential to extract estimates of the masses of parent
particles and the LSP from small data sets.

\subsection*{Dark matter at 100 pb$^{-1}$} 

We have demonstrated a concrete strategic solution to the LHC Inverse Problem
applicable under realistic conditions to early physics running at the LHC.
Since the missing energy signature is motivated by the existence of
dark matter, we should also address what might be called the Dark Matter
Inverse Problem: given a missing energy discovery at the LHC, what can
we learn about dark matter and the cosmological events that produced it?
This problem has not been addressed at all for the 100 pb$^{-1}$ era 
of LHC running. Given an early missing energy discovery at the LHC,
this problem will become one of the most interesting questions in
particle physics, especially tied in to results from the ongoing direct and
indirect dark matter searches.

\begin{table}[htp]
\begin{center}
\setlength{\tabcolsep}{1.6mm}
\begin{tabular}{lrrr}
\toprule
\multicolumn{4}{c}{LM5 vs. CS4d [100 pb$^{-1}$]}\\
\midrule
 Variable & LM5 & CS4d & Separation \\
\midrule
\multicolumn{4}{c}{MET}\\
\midrule
r(HT900)  & 0.38 & 0.24 & 3.59 \\
r(Meff1400)  & 0.32 & 0.21 & 3.15 \\
r(MET520)  & 0.28 & 0.19 & 3.12 \\
\bottomrule
\multicolumn{4}{c}{DiJet}\\
\midrule
r(DiJet)  & 0.22 & 0.15 & 2.37 \\
r(MET520)  & 0.28 & 0.23 & 0.56 \\
r(Hem2)  & 0.27 & 0.32 & 0.51 \\
\bottomrule
\multicolumn{4}{c}{TriJet}\\
\midrule
r(TriJet)  & 0.22 & 0.17 & 1.89 \\
r(Meff1400)  & 0.65 & 0.59 & 0.62 \\
r(HT900)  & 0.73 & 0.67 & 0.59 \\
\bottomrule
\multicolumn{4}{c}{Muon20}\\
\midrule
r(HT900)  & 0.35 & 0.25 & 1.17 \\
r(mT2-300)  & 0.27 & 0.36 & 1.00 \\
r(mT2-400)  & 0.26 & 0.34 & 0.93 \\
\bottomrule
\end{tabular}
\end{center}
\caption {Largest separation (in units of $\sigma$)
for the comparison of LM5 vs. CS4d (error on CS4d) assuming an integrated luminosity of 100 pb$^{-1}$.\label{tab:LM5vsCS4d_100}}
\end{table}

\begin{table}[htp]
\begin{center}
\setlength{\tabcolsep}{1.6mm}
\begin{tabular}{lrrr}
\toprule
\multicolumn{4}{c}{LM5 vs. CS4d [100 pb$^{-1}$]}\\
\midrule
 Variable & LM5 & CS4d & Separation \\
\midrule
\multicolumn{4}{c}{MET}\\
\midrule
r(HT900)  & 0.38 & 0.24 & 3.59 \\
r(Meff1400)  & 0.32 & 0.21 & 3.15 \\
r(MET520)  & 0.28 & 0.19 & 3.12 \\
r(mT2-600/300)  & 0.29 & 0.17 & 2.83 \\
r(mT2-600/400)  & 0.31 & 0.18 & 2.82 \\
r(mT2-600/500)  & 0.46 & 0.30 & 2.22 \\
r(MET420)  & 0.49 & 0.40 & 2.14 \\
r(mT2-300)  & 0.36 & 0.45 & 1.98 \\
r(mT2-400)  & 0.34 & 0.43 & 1.87 \\
r(mT2-600)  & 0.11 & 0.08 & 1.47 \\
\bottomrule
\end{tabular}
\end{center}
\caption {Largest separation (in units of $\sigma$)
for the comparison of LM5 vs. CS4d (error on CS4d) assuming an integrated luminosity of 100 pb$^{-1}$.\label{tab:LM5vsCS4d_100MET}}
\end{table}

\begin{table}[htp]
\begin{center}
\setlength{\tabcolsep}{1.6mm}
\begin{tabular}{lrrr}
\toprule
\multicolumn{4}{c}{LM5 vs. CS4d [1000 pb$^{-1}$]}\\
\midrule
 Variable & LM5 & CS4d & Separation \\
\midrule
\multicolumn{4}{c}{MET}\\
\midrule
r(MET520)  & 0.28 & 0.19 & 7.07 \\
r(HT900)  & 0.38 & 0.24 & 6.41 \\
r(mT2-600/400)  & 0.31 & 0.18 & 6.10 \\
\bottomrule
\multicolumn{4}{c}{DiJet}\\
\midrule
r(DiJet)  & 0.22 & 0.15 & 4.80 \\
r(mT2-600/300)  & 0.29 & 0.15 & 2.69 \\
r(mT2-600/400)  & 0.30 & 0.16 & 2.69 \\
\bottomrule
\multicolumn{4}{c}{TriJet}\\
\midrule
r(TriJet)  & 0.22 & 0.17 & 3.50 \\
r(MET520)  & 0.28 & 0.24 & 1.31 \\
r(Meff1400)  & 0.65 & 0.59 & 1.29 \\
\bottomrule
\multicolumn{4}{c}{Muon20}\\
\midrule
r(mT2-600/300) & 0.23 & 0.13 & 3.04 \\
r(mT2-600/400) & 0.25 & 0.14 & 2.98 \\
r(HT900)  & 0.35 & 0.25 & 2.63 \\
\bottomrule
\end{tabular}
\end{center}
\caption {Best discriminating ratios in each trigger box, with separations in units of $\sigma$,
for the comparison of LM5 vs. CS4d, taking LM5 as the ``data'',
assuming an integrated luminosity of 1000 pb$^{-1}$.\label{tab:LM5vsCS4d_1000}}
\end{table}

\begin{table}[htp]
\begin{center}
\setlength{\tabcolsep}{1.6mm}
\begin{tabular}{lrrr}
\toprule
\multicolumn{4}{c}{LM5 vs. CS4d [1000 pb$^{-1}$]}\\
\midrule
 Variable & LM5 & CS4d & Separation \\
\midrule
\multicolumn{4}{c}{MET}\\
\midrule
r(MET520)  & 0.28 & 0.19 & 7.07 \\
r(HT900)  & 0.38 & 0.24 & 6.41 \\
r(mT2-600/400) & 0.31 & 0.18 & 6.10 \\
r(mT2-600/300) & 0.29 & 0.17 & 6.04 \\
r(Meff1400)  & 0.32 & 0.21 & 5.94 \\
r(mT2-600/500) & 0.46 & 0.30 & 4.88 \\
r(MET420)  & 0.49 & 0.40 & 3.82 \\
r(mT2-600)  & 0.11 & 0.08 & 3.52 \\
r(mT2-300)  & 0.36 & 0.45 & 3.12 \\
r(mT2-400)  & 0.34 & 0.43 & 3.02 \\
\bottomrule
\end{tabular}
\end{center}
\caption {Best discriminating ratios in the MET box, with separations in units of $\sigma$,
for the comparison of LM5 vs. CS4d, taking LM5 as the ``data'',
assuming an integrated luminosity of 1000 pb$^{-1}$.\label{tab:LM5vsCS4d_1000MET}}
\end{table}


\begin{table}[htp]
\begin{center}
\begin{tabular}{lrrr}
\toprule
\multicolumn{4}{c}{LM2p vs. LM5 [100 pb$^{-1}$]}\\
\midrule
 Variable & LM2p & LM5 & Separation \\
\midrule
\multicolumn{4}{c}{MET}\\
\midrule
r(5j)(3j)  & 0.33 & 0.40 & 1.64 \\
r(mT2-300)  & 0.41 & 0.34 & 1.44 \\
r(mT2-400)  & 0.30 & 0.25 & 1.34 \\
r(4j)(3j)  & 0.64 & 0.69 & 1.18 \\
r($\tau$-tag)  & 0.07 & 0.05 & 1.17 \\
r(10t-c45)  & 0.30 & 0.36 & 1.14 \\
r(10t-c30)  & 0.16 & 0.20 & 1.13 \\
r(10t-c75)  & 0.48 & 0.53 & 1.07 \\
r(10t-c60)  & 0.41 & 0.47 & 1.07 \\
r(mT2-500)  & 0.16 & 0.13 & 1.05 \\
\bottomrule
\end{tabular}
\end{center}
\caption {Best discriminating ratios in the MET box, with separations in units of $\sigma$,
for the comparison of LM2p vs. LM5, taking LM2p as the ``data'',
assuming an integrated luminosity of 100 pb$^{-1}$.\label{tab:LM2pvsLM5_100MET}}
\end{table}


\begin{figure}[htp]
\includegraphics[width=0.48\textwidth]{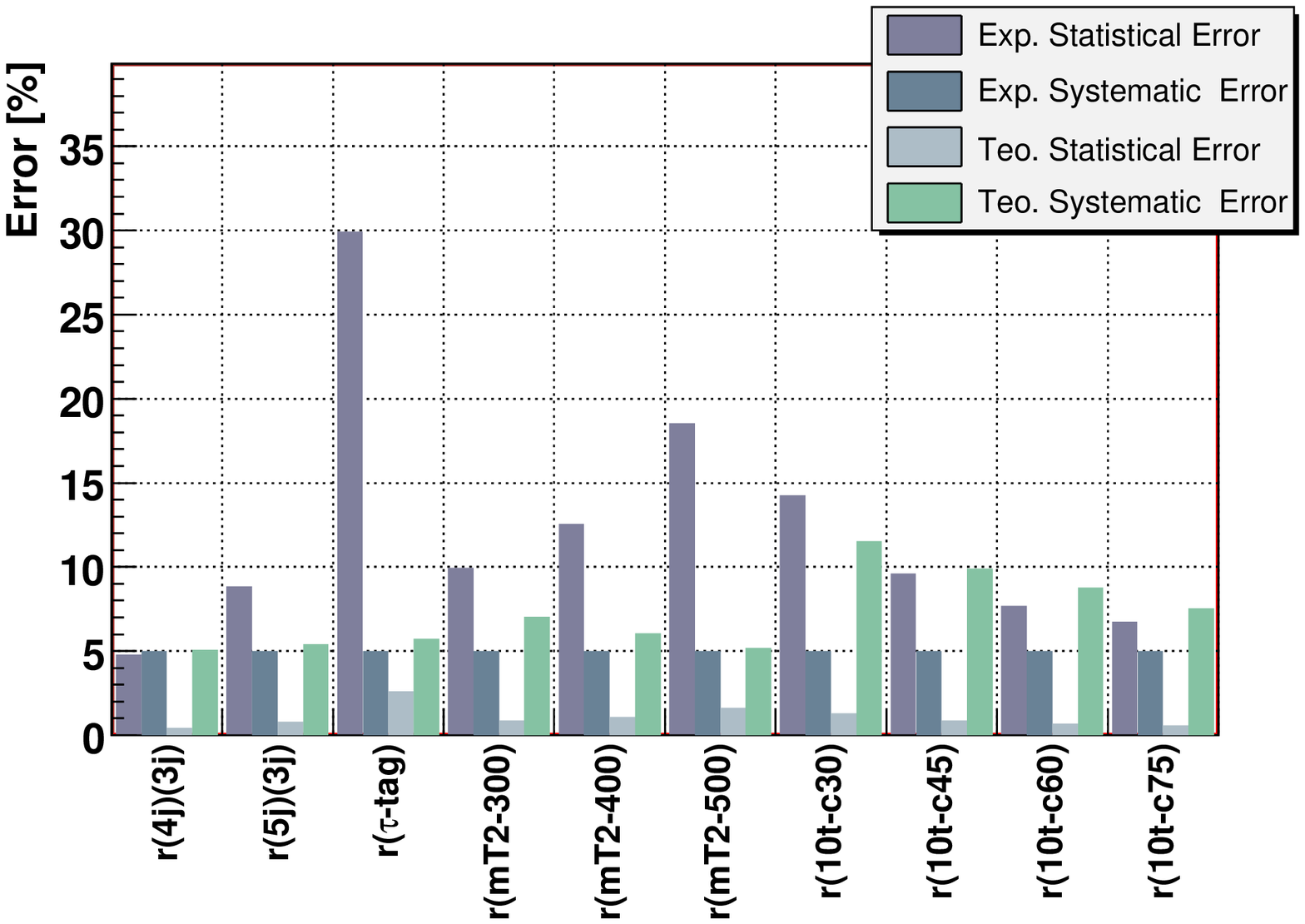}
\caption {\label{fig:LM2p100pbLM5Err_MET_1} Breakdown of estimated uncertainties for discriminating
ratios with 100 pb$^{-1}$, in the comparison of look-alike models LM2p and LM5, with
LM2p treated as the ``data''.}
\end{figure}



\begin{figure}[htp]
\includegraphics[width=0.48\textwidth]{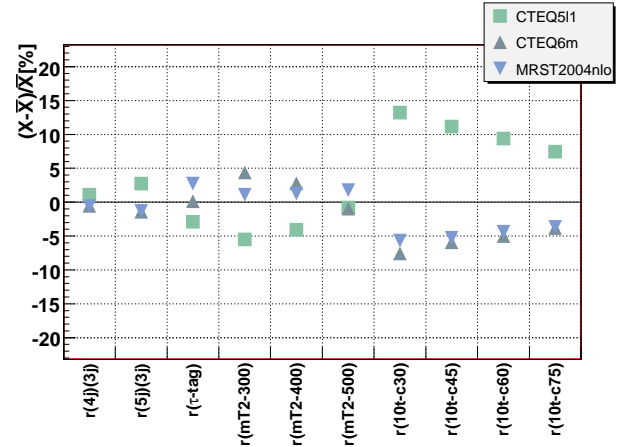}
\caption {\label{fig:LM2p100pbLM5Pdf_MET_1} Pdf spreads for discriminating
ratios with 100 pb$^{-1}$, in the comparison of look-alike models LM2p and LM5, with
LM2p treated as the ``data''.}
\end{figure}




\subsection*{Acknowledgments}
The authors are grateful to Keith Ellis, Gordon Kane, Filip Moortgat, 
Stephen Mrenna, Luc Pape, Tilman Plehn, Chris Quigg, Sezen Sekmen, Phillip Schuster,
Peter Skands and Natalia Toro for useful discussions.
JL and MS acknowledge the Michigan Center for Theoretical Physics
for hospitality during the concluding phase of this work;
JH similarly acknowledges the Kavli Institute for Theoretical Physics.
Fermilab is operated by the Fermi Research Alliance LLC under contract
DE-AC02-07CH11359 with the U.S. Dept. of Energy.
Research at Argonne National Laboratory Division is supported by
the U.S. Dept. of Energy under contract DE-AC02-06CH11357.


\appendix


\begin{figure}[htp]
\includegraphics[width=0.48\textwidth]{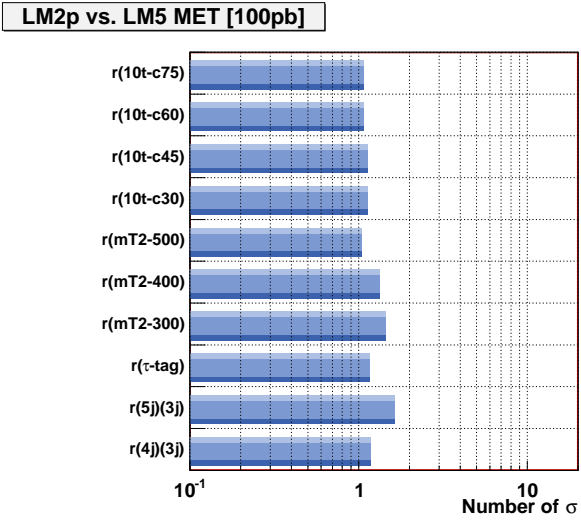}
\caption {\label{fig:LM2p100pbLM5Pull_MET_1} Pulls of the best discriminating
ratios with 100 pb$^{-1}$, in the comparison of look-alike models LM2p and LM5, with
LM2p treated as the ``data''.}
\end{figure}

\begin{table}[htp]
\begin{center}
\begin{tabular}{lrrr}
\toprule
\multicolumn{4}{c}{LM2p vs. LM5 [1000 pb$^{-1}$]}\\
\midrule
 Variable & LM2p & LM5 & Separation \\
\midrule
\multicolumn{4}{c}{MET}\\
\midrule
r($\tau$-tag)  & 0.07 & 0.05 & 3.14 \\
r(5j)(3j)  & 0.33 & 0.40 & 2.77 \\
r(mT2-400)  & 0.30 & 0.25 & 2.56 \\
r(mT2-500)  & 0.16 & 0.13 & 2.53 \\
r(mT2-300)  & 0.41 & 0.34 & 2.27 \\
r(10t-c30)  & 0.16 & 0.20 & 1.67 \\
r(MET520)  & 0.31 & 0.28 & 1.67 \\
r(20t-c45)  & 0.07 & 0.09 & 1.61 \\
r(10tdiff-c30)  & 0.14 & 0.16 & 1.61 \\
r(4j)(3j)  & 0.64 & 0.69 & 1.56 \\
\bottomrule
\end{tabular}
\end{center}
\caption {Best discriminating ratios in the MET box, with separations in units of $\sigma$,
for the comparison of LM2p vs. LM5, taking LM2p as the ``data'', 
assuming an integrated luminosity of 1000 pb$^{-1}$.\label{tab:LM2pvsLM5_1000MET}}
\end{table}

\begin{table}[htp]
\begin{center}
\begin{tabular}{lrrr}
\toprule
\multicolumn{4}{c}{CS4d vs. LM8 [100 pb$^{-1}$]}\\
\midrule
 Variable & CS4d & LM8 & Separation \\
\midrule
\multicolumn{4}{c}{MET}\\
\midrule
r(5j)(3j)  & 0.40 & 0.54 & 2.83 \\
r(mT2-300)  & 0.41 & 0.32 & 2.14 \\
r(4j)(3j)  & 0.70 & 0.81 & 2.11 \\
r(10t-c30)  & 0.20 & 0.28 & 1.93 \\
r(20t-c60)  & 0.16 & 0.23 & 1.78 \\
r(20t-c45)  & 0.08 & 0.13 & 1.73 \\
r(10t-c30)(10t-c90)  & 0.34 & 0.43 & 1.72 \\
r(10t-c45)  & 0.36 & 0.46 & 1.70 \\
r(30t-c90)  & 0.11 & 0.17 & 1.68 \\
r(Meff1400)  & 0.21 & 0.28 & 1.65 \\
\bottomrule
\end{tabular}
\end{center}
\caption {Best discriminating ratios in the MET box, with separations in units of $\sigma$,
for the comparison of CS4d vs. LM8, taking CS4d as the ``data'',
assuming an integrated luminosity of 100 pb$^{-1}$.\label{tab:CS4dvsLM8_100MET}}
\end{table}

\begin{table}[htp]
\begin{center}
\begin{tabular}{lrrr}
\toprule
\multicolumn{4}{c}{CS4d vs. LM8 [1000 pb$^{-1}$]}\\
\midrule
 Variable & CS4d & LM8 & Separation \\
\midrule
\multicolumn{4}{c}{MET}\\
\midrule
r(5j)(3j)  & 0.40 & 0.54 & 4.21 \\
r(10tdiff-c30)  & 0.15 & 0.20 & 3.63 \\
r(Hem3)  & 0.07 & 0.11 & 3.58 \\
r(mT2-300) & 0.41 & 0.32 & 3.43 \\
r(mT2-400)  & 0.25 & 0.20 & 3.24 \\
r(20t-c30)  & 0.02 & 0.04 & 3.09 \\
r(20tdiff-c45)  & 0.06 & 0.08 & 3.05 \\
r(20tdiff-c60)  & 0.12 & 0.16 & 3.04 \\
r(20t-c45)  & 0.08 & 0.13 & 3.03 \\
r(10t-c30)(10t-c90)  & 0.34 & 0.43 & 2.95 \\
\bottomrule
\end{tabular}
\end{center}
\caption {Best discriminating ratios in the MET box, with separations in units of $\sigma$,
for the comparison of CS4d vs. LM8, taking CS4d as the ``data'',
assuming an integrated luminosity of 1000 pb$^{-1}$.\label{tab:CS4dvsLM8_1000MET}}
\end{table}


\begin{figure}[htp]
\includegraphics[width=0.48\textwidth]{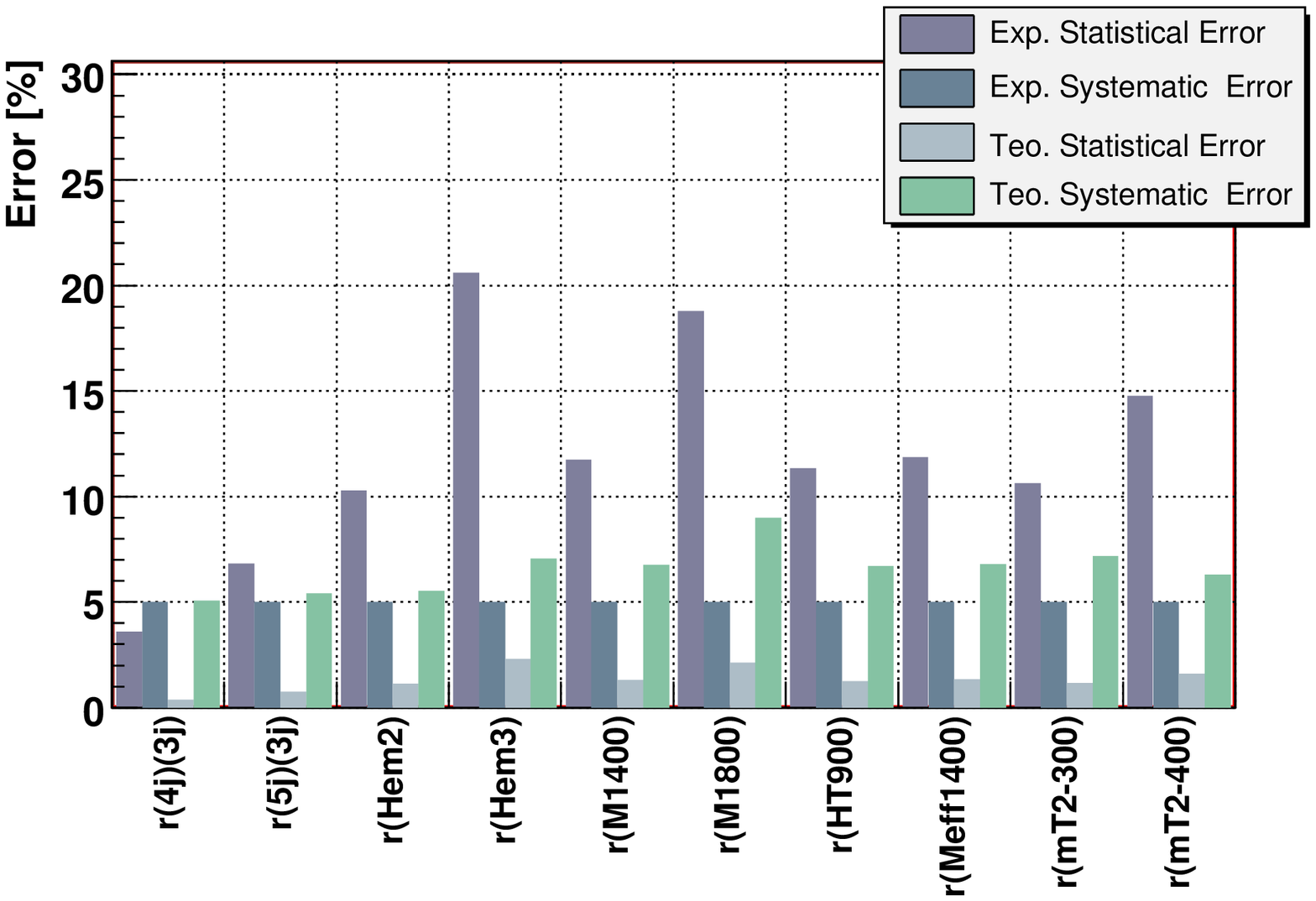}
\caption {\label{fig:CS4d100pbLM8Err_MET_1} Breakdown of estimated uncertainties for discriminating
ratios with 100 pb$^{-1}$, in the comparison of look-alike models CS4d and LM8, with
CS4d treated as the ``data''.}
\end{figure}



\begin{figure}[htp]
\includegraphics[width=0.48\textwidth]{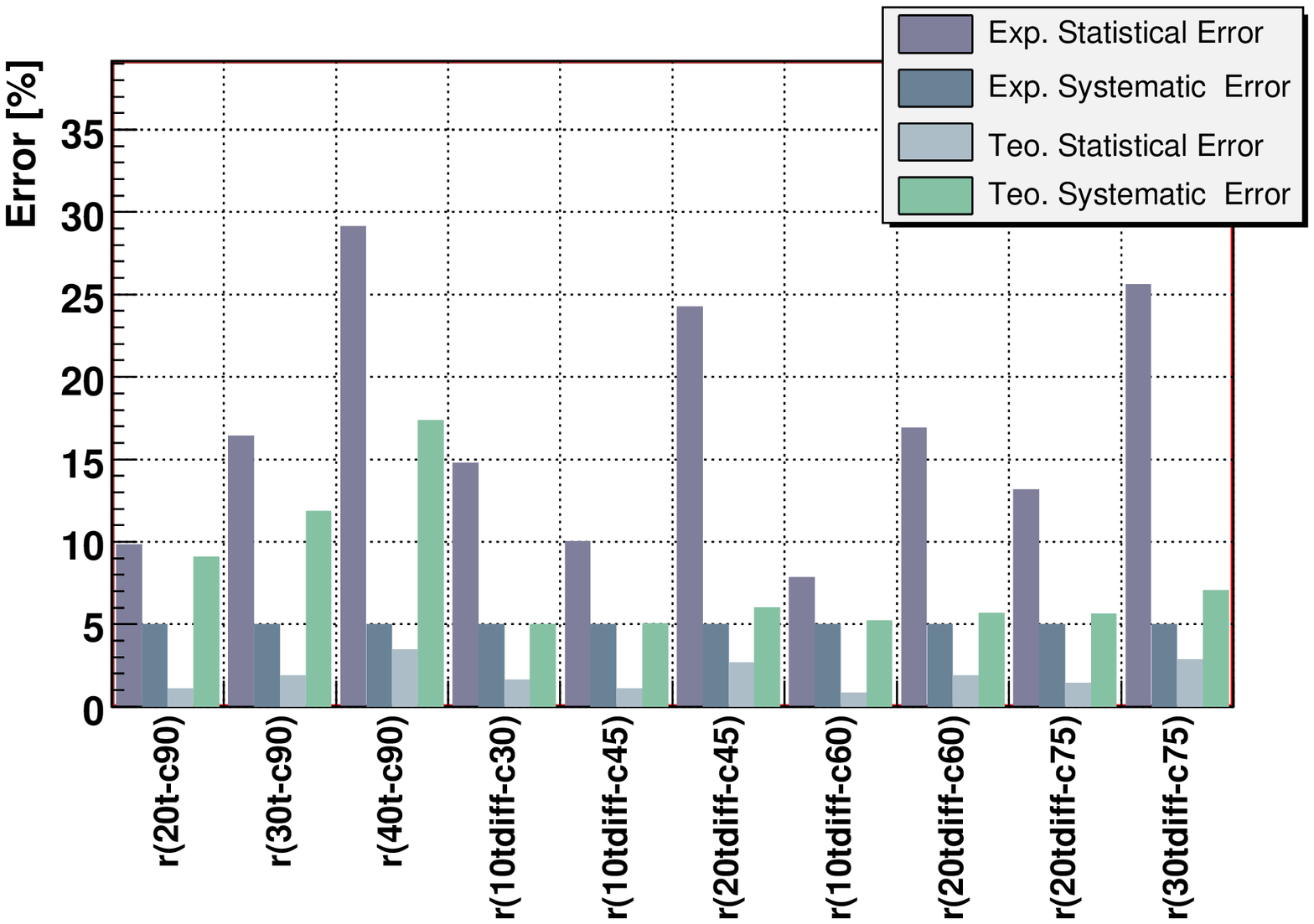}
\caption {\label{fig:CS4d100pbLM8Err_MET_3} Breakdown of estimated uncertainties for discriminating
ratios with 100 pb$^{-1}$, in the comparison of look-alike models CS4d and LM8, with
CS4d treated as the ``data''.}
\end{figure}


\begin{table}[htp]
\begin{center}
\setlength{\tabcolsep}{1.5mm}
\begin{tabular}{lrrr}
\toprule
\multicolumn{4}{c}{CS6 vs. LM2p [100 pb$^{-1}$]}\\
\midrule
 Variable & CS6 & LM2p & Separation \\
\midrule
\multicolumn{4}{c}{MET}\\
\midrule
r(MET420)  & 0.22 & 0.53 & 6.99 \\
r(MET520)  & 0.09 & 0.31 & 6.10 \\
r(mT2-500/300) & 0.08 & 0.39 & 5.14 \\
\bottomrule
\multicolumn{4}{c}{DiJet}\\
\midrule
r(DiJet)  & 0.11 & 0.21 & 3.04 \\
r(5j)(3j)  & 0.54 & 0.32 & 3.00 \\
r(4j)(3j)  & 0.83 & 0.62 & 2.56 \\
\bottomrule
\multicolumn{4}{c}{TriJet}\\
\midrule
r(MET420)  & 0.29 & 0.53 & 2.96 \\
r(MET320)  & 0.55 & 0.77 & 2.84 \\
r(HT900)  & 0.57 & 0.75 & 2.46 \\
\bottomrule
\multicolumn{4}{c}{Muon20}\\
\midrule
r($b$-tag)  & 0.74 & 0.36 & 4.25 \\
r(Muon20)  & 0.06 & 0.14 & 3.14 \\
r(MET420)  & 0.23 & 0.48 & 2.55 \\
\bottomrule
\end{tabular}
\end{center}
\caption {Best discriminating ratios in each trigger box, with separations in units of $\sigma$,
for the comparison of CS6 vs. LM2p, taking CS6 as the ``data'',
assuming an integrated luminosity of 100 pb$^{-1}$.\label{tab:CS6vsLM2p_100}}
\end{table}

\begin{table}[htp]
\begin{center}
\setlength{\tabcolsep}{1.5mm}
\begin{tabular}{lrrr}
\toprule
\multicolumn{4}{c}{CS6 vs. LM2p [100 pb$^{-1}$]}\\
\midrule
 Variable & CS6 & LM2p & Separation \\
\midrule
\multicolumn{4}{c}{MET}\\
\midrule
r(MET420)  & 0.22 & 0.53 & 6.99 \\
r(MET520)  & 0.09 & 0.31 & 6.10 \\
r(mT2-500/300)  & 0.08 & 0.39 & 5.14 \\
r(mT2-500/400)  & 0.17 & 0.54 & 5.11 \\
r(MET320)  & 0.54 & 0.78 & 4.93 \\
r(HT900)  & 0.18 & 0.38 & 4.80 \\
r(mT2-500)  & 0.03 & 0.16 & 4.56 \\
r(Meff1400)  & 0.16 & 0.32 & 3.91 \\
r(mT2-400/300) & 0.49 & 0.73 & 3.88 \\
r(mT2-400/300)  & 0.49 & 0.73 & 3.88 \\
\bottomrule
\end{tabular}
\end{center}
\caption {Best discriminating ratios in the MET box, with separations in units of $\sigma$,
for the comparison of CS6 vs. LM2p, taking CS6 as the ``data'',
assuming an integrated luminosity of 100 pb$^{-1}$.\label{tab:CS6vsLM2p_100MET}}
\end{table}


\begin{figure}[htp]
\includegraphics[width=0.48\textwidth]{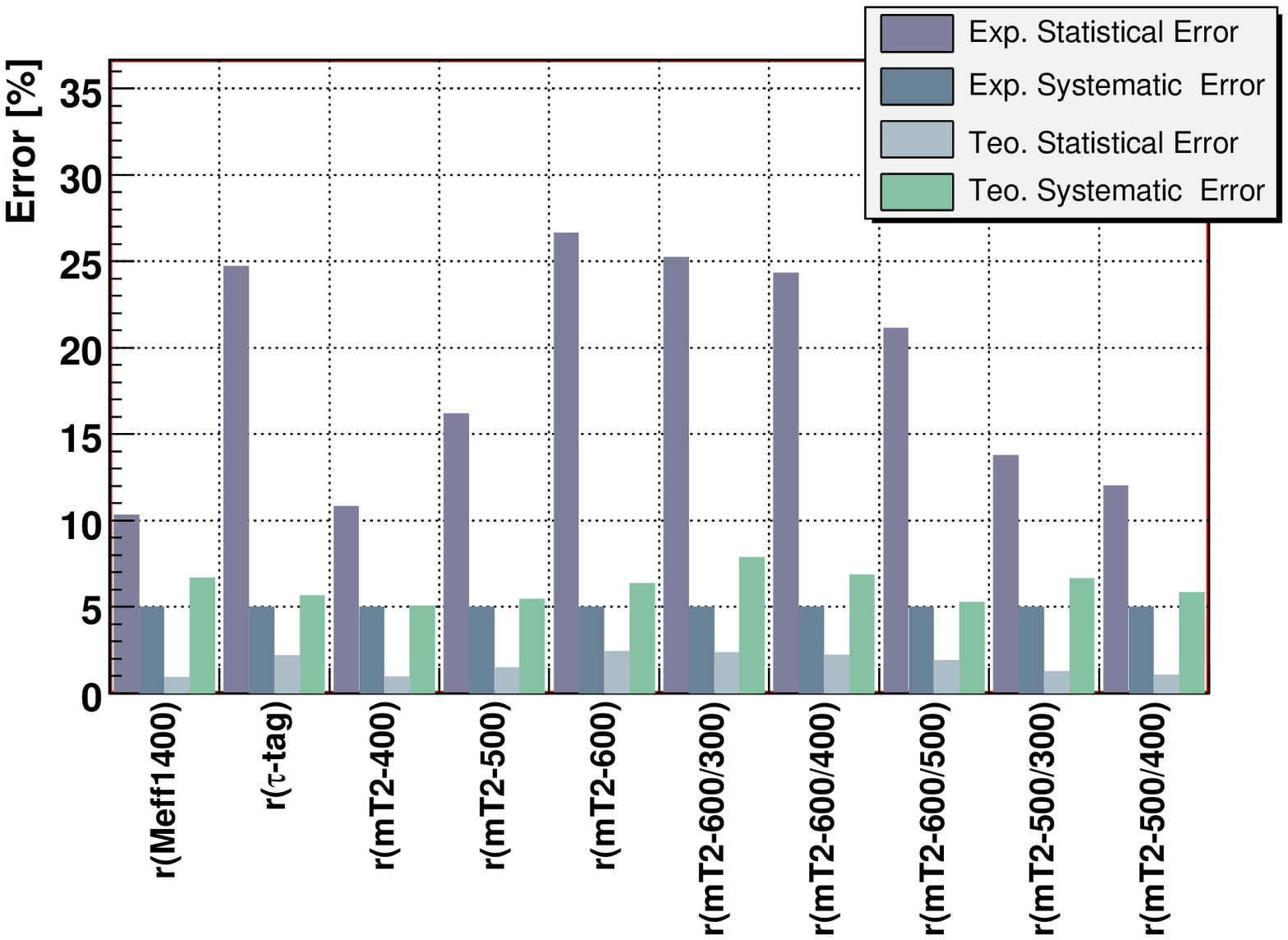}
\caption {\label{fig:CS6100pbLM2pErr_MET_3} Breakdown of estimated uncertainties for discriminating
ratios with 100 pb$^{-1}$, in the comparison of look-alike models CS6 and LM2p, with
CS6 treated as the ``data''.}
\end{figure}



\begin{figure}[htp]
\includegraphics[width=0.48\textwidth]{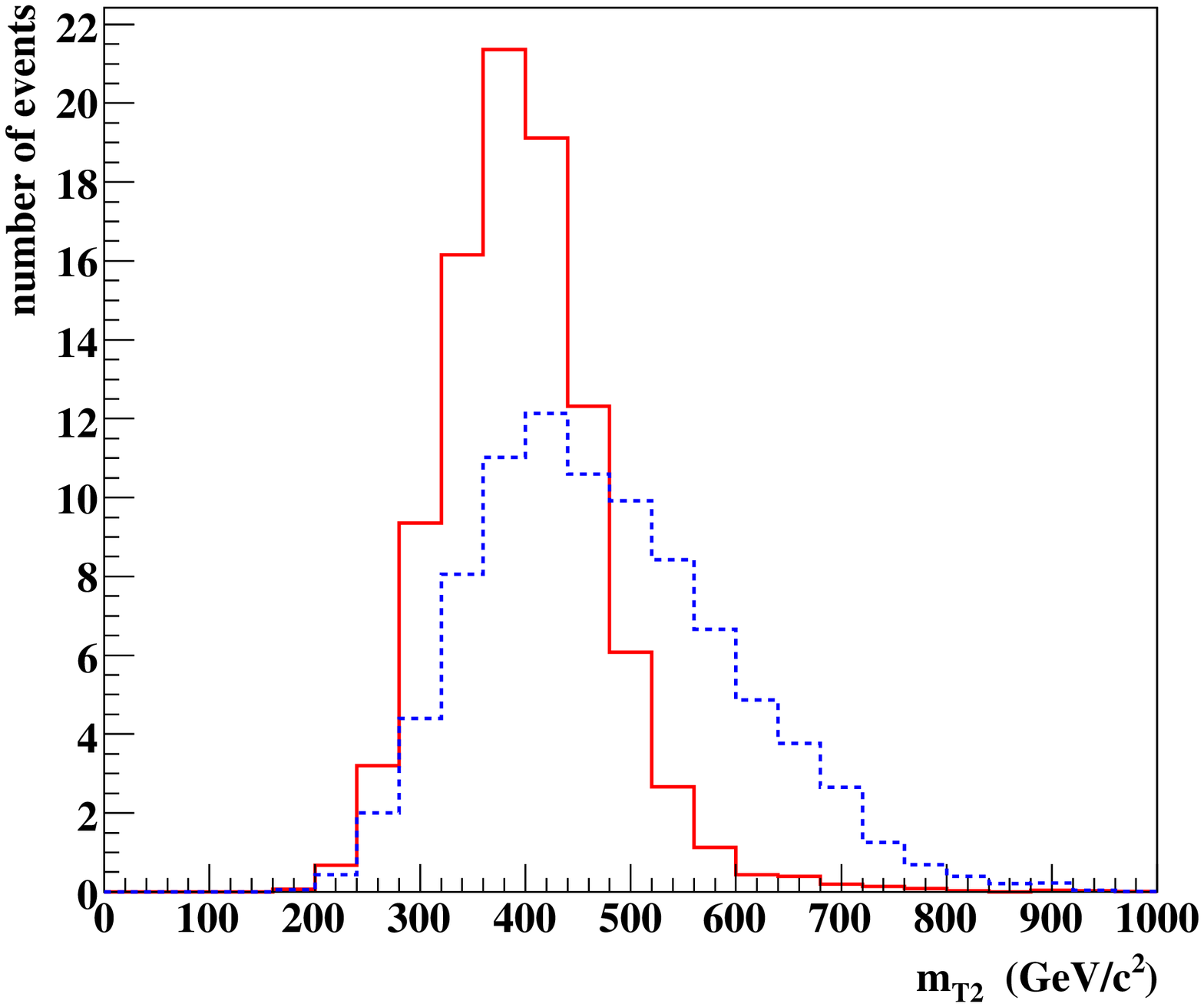}
\caption {\label{fig:mt2CS6LM2p} Comparison of the $m_{T2}$ distribution
of the CS6 ``data'' (solid red line) to that of the theory model LM2p 
(dashed blue line) for 100 pb$^{-1}$. Here
$m_{T2}$ is computed using the LSP mass of the theory model LM2p.}
\end{figure}


\section{Detailed results for Group 1}

\subsection{LM5 vs CS4d}

This comparison is described in section~\ref{sec:7.1}. We treat model LM5
as the data and CS4d as the theory model. For 100 pb$^{-1}$
of integrated luminosity, Table~\ref{tab:LM5vsCS4d_100}
shows the three best discriminating ratios as defined in each
of the trigger boxes: MET, DiJet, TriJet and Muon20.
Table~\ref{tab:LM5vsCS4d_100MET}
shows the ten best discriminating ratios in the MET box.
The separation in units of $\sigma$ is computed from the total
estimated theoretical and experimental uncertainties as
described in section~\ref{sec:6}. 

The same information for 1000 pb$^{-1}$ is shown in
Tables~\ref{tab:LM5vsCS4d_1000} and \ref{tab:LM5vsCS4d_1000MET}.
The notation r(DiJet), r(TriJet) and r(Muon20)
denotes the ratio of the number of events in that box
with the number of events in the MET box. Since (plus or minus
about one event in 1000 pb$^{-1}$) the DiJet, TriJet and Muon20
boxes are subsamples of the MET box, these are ratios
of inclusive counts.

\subsection{LM2p vs LM5}

This comparison is described in section~\ref{sec:7.2}. We treat model LM2p
as the data and LM5 as the theory model.
For 100 pb$^{-1}$, the ten best discriminating ratios are listed in
Table~\ref{tab:LM2pvsLM5_100MET} and the pulls are displayed
in Figure~\ref{fig:LM2p100pbLM5Pull_MET_1}. While the $\tau$ tag
ratio is not discriminating well, Figure~\ref{fig:LM2p100pbLM5Err_MET_1}
shows that this is due entirely to poor statistics in this small simulated
data sample: the experimental statistical uncertainty is 30\%, compared
to experimental and theory systematics both estimated at around 5\%.
The small theory statistical uncertainty shown is the error
from the finite Monte Carlo statistics in simulating the theory model.
Table~\ref{tab:LM2pvsLM5_1000MET} shows the improvement in the
discriminating power of r($\tau$-tag) with 1000 pb$^{-1}$, due to the
reduction in the experimental statistical uncertainty. Note that
in this difficult case no other ratio discriminates with better
than 3$\sigma$ significance.

In Figure~\ref{fig:LM2p100pbLM5Err_MET_1} one also notices large differences
in the relative size of the theory systematics for the different ratios.
These are due to large differences in the spread of values when we vary the
parton distribution functions used in the simulation, as shown in
Figure~\ref{fig:LM2p100pbLM5Pdf_MET_1}. Notice that the pdf spreads
vary from less than 5\% for the jet multiplicity ratio r(4j)(3j) to greater than 20\% for
the cone track count ratio r(10t-c30).

This is an important generic feature of our results. We find the pdf uncertainties to
be process dependent and thus model dependent. We find also that the
relative pdf uncertainties for different ratios in the same model vary
by factors as large as 4 or 5.

\subsection{CS4d vs LM8}

This comparison is described in section~\ref{sec:7.3}. We treat model CS4d
as the data and LM8 as the theory model.
Tables~\ref{tab:CS4dvsLM8_100MET} and \ref{tab:CS4dvsLM8_1000MET} show the ten
best discriminating ratios for 100 pb$^{-1}$ and 1000 pb$^{-1}$ respectively.
We observe that the best discriminator with 100 pb$^{-1}$, the jet multiplicity
ratio r(5j)(3j), remains the best with 1000 pb$^{-1}$. However the second
and third best dsicriminators with 1000 pb$^{-1}$, r(10tdiff-c30) and r(Hem3),
are not even among the top ten best with 100 pb$^{-1}$.

These again are generic features of our results. 
Figures~\ref{fig:CS4d100pbLM8Err_MET_1} and \ref{fig:CS4d100pbLM8Err_MET_3}
show that with 100 pb$^{-1}$ r(5j)(3j) already has a statistical uncertainty
comparable to the estimated systematics, while many other ratios still have
large 15 to 20\% statistical uncertainties. This includes r(10tdiff-c30) and
r(Hem3), ratios that are sensitive to what fraction of events have
large hemisphere differences in object counts or track counts.
Thus qualitatively new features of the events emerge automatically as good discriminators
as the integrated luminosity goes up, without changing the design of the
look-alike analysis.

\subsection{CS6 vs LM2p, LM5, LM8 and CS4d}

This comparison is described in section~\ref{sec:7.4}. We treat model CS6
as the data and compare to theory models LM2p, LM5, LM8 and CS4d.
Tables~\ref{tab:CS6vsLM2p_100} and \ref{tab:CS6vsLM2p_100MET} show the
best discriminating ratios for the comparison of LM2p to the CS6 ``data''
with 100 pb$^{-1}$. We see that two \met~ratios and two $m_{T2}$ ratios
already discriminate at better than 5$\sigma$.

Figure~\ref{fig:CS6100pbLM2pErr_MET_3} shows the breakdown of the
uncertainites for some of the ratios in the this comparison.
Observe that the kinematic ratio r(Meff1400) has a smaller
experimental statistical uncertainty than do the $m_{T2}$ ratios
r(mT2-500/300) and r(mT2-500/400); nevertheless Table~\ref{tab:CS6vsLM2p_100MET}
shows that r(Meff1400) is a worse discriminator.

This is a generic feature of our results. The distributions in $M_{\rm eff}$
are rather broad, whereas the $m_{T2}$ distributions are steeply falling
as one approaches the endpoint region. As seen in Figure~\ref{fig:mt2CS6LM2p},
when the parent particle mass differences are large,
$m_{T2}$ is an intrinsically good discriminator even for quite small data samples
within very modest resolutions.

Tables~\ref{tab:CS6vsLM8_100} and \ref{tab:CS6vsCS4d_100} show the
best discriminating ratios for the comparison of LM8 ans CS4d to the CS6 ``data''
with 100 pb$^{-1}$. These results show the importance of the ratios in the
Muon20 box.

\begin{table}[htp]
\begin{center}
\begin{tabular}{lrrr}
\toprule
\multicolumn{4}{c}{CS6 vs. LM8 [100 pb$^{-1}$]}\\
\midrule
 Variable & CS6 & LM8 & Separation \\
\midrule
\multicolumn{4}{c}{MET}\\
\midrule
r(MET420)  & 0.22 & 0.39 & 4.03 \\
r(Hem2)  & 0.19 & 0.34 & 3.60 \\
r(MET520)  & 0.09 & 0.20 & 3.45 \\
\bottomrule
\multicolumn{4}{c}{DiJet}\\
\midrule
r(DiJet)  & 0.11 & 0.18 & 2.21 \\
r(Hem2)  & 0.28 & 0.35 & 0.87 \\
r(MET320)  & 0.59 & 0.66 & 0.86 \\
\bottomrule
\multicolumn{4}{c}{TriJet}\\
\midrule
r(Meff1400)  & 0.50 & 0.64 & 1.58 \\
r(HT900)  & 0.57 & 0.68 & 1.35 \\
r(MET320)  & 0.55 & 0.66 & 1.21 \\
\bottomrule
\multicolumn{4}{c}{Muon20}\\
\midrule
r($b$-tag)  & 0.74 & 0.29 & 6.45 \\
r(Muon20)  & 0.06 & 0.24 & 5.19 \\
r(MET420)  & 0.23 & 0.36 & 1.78 \\
\bottomrule
\end{tabular}
\end{center}
\caption {Best discriminating ratios in each trigger box, with separations in units of $\sigma$,
for the comparison of CS6 vs. LM8, taking CS6 as the ``data'',
assuming an integrated luminosity of 100 pb$^{-1}$.\label{tab:CS6vsLM8_100}}
\end{table}

\begin{table}[htp]
\begin{center}
\begin{tabular}{lrrr}
\toprule
\multicolumn{4}{c}{CS6 vs. CS4d [100 pb$^{-1}$]}\\
\midrule
 Variable & CS6 & CS4d & Separation \\
\midrule
\multicolumn{4}{c}{MET}\\
\midrule
r(MET420)  & 0.22 & 0.40 & 4.25 \\
r(MET320)  & 0.54 & 0.70 & 3.26 \\
r(MET520)  & 0.09 & 0.19 & 3.15 \\
\bottomrule
\multicolumn{4}{c}{DiJet}\\
\midrule
r(5j)(3j)  & 0.54 & 0.39 & 1.53 \\
r(DiJet)  & 0.11 & 0.15 & 1.42 \\
r(4j)(3j)  & 0.83 & 0.69 & 1.41 \\
\bottomrule
\multicolumn{4}{c}{TriJet}\\
\midrule
r(MET320)  & 0.55 & 0.70 & 1.63 \\
r(MET420)  & 0.29 & 0.43 & 1.53 \\
r(HT900)  & 0.57 & 0.67 & 1.17 \\
\bottomrule
\multicolumn{4}{c}{Muon20}\\
\midrule
r(Muon20)  & 0.06 & 0.18 & 3.95 \\
r(MET320)  & 0.54 & 0.69 & 1.73 \\
r(MET420)  & 0.23 & 0.37 & 1.59 \\
\bottomrule
\end{tabular}
\end{center}
\caption {Best discriminating ratios in each trigger box, with separations in units of $\sigma$,
for the comparison of CS6 vs. CS4d, taking CS6 as the ``data'', 
assuming an integrated luminosity of 100 pb$^{-1}$.\label{tab:CS6vsCS4d_100}}
\end{table}

\begin{table}[htp]
\begin{center}
\begin{tabular}{lrrr}
\toprule
\multicolumn{4}{c}{LH2 vs. NM4 [1000 pb$^{-1}$]}\\
\midrule
 Variable & LH2 & NM4 & Separation \\
\midrule
\multicolumn{4}{c}{MET}\\
\midrule
r(mT2-500)  & 0.16 & 0.05 & 14.11 \\
r(mT2-400)  & 0.44 & 0.21 & 11.13 \\
r(mT2-500/300)  & 0.21 & 0.09 & 8.52 \\
\bottomrule
\multicolumn{4}{c}{DiJet}\\
\midrule
r(mT2-400)  & 0.32 & 0.12 & 7.89 \\
r(mT2-300)  & 0.64 & 0.32 & 7.79 \\
r(DiJet)  & 0.11 & 0.22 & 5.94 \\
\bottomrule
\multicolumn{4}{c}{TriJet}\\
\midrule
r(mT2-300)  & 0.62 & 0.19 & 10.96 \\
r(mT2-400)  & 0.34 & 0.07 & 10.91 \\
r(TriJet)  & 0.06 & 0.15 & 5.94 \\
\bottomrule
\multicolumn{4}{c}{Muon20}\\
\midrule
r(mT2-400)  & 0.38 & 0.14 & 5.03 \\
r(mT2-300)  & 0.72 & 0.42 & 4.30 \\
r(Meff1400)  & 0.10 & 0.34 & 3.50 \\
\bottomrule
\end{tabular}
\end{center}
\caption {Best discriminating ratios in each trigger box, with separations in units of $\sigma$,
for the comparison of LH2 vs. NM4, taking LH2 as the ``data'',
assuming an integrated luminosity of 1000 pb$^{-1}$.\label{tab:LH2vsNM4_1000}}
\end{table}

\begin{table}[htp]
\begin{center}
\begin{tabular}{lrrr}
\toprule
\multicolumn{4}{c}{LH2 vs. NM4 [1000 pb$^{-1}$]}\\
\midrule
 Variable & LH2 & NM4 & Separation \\
\midrule
\multicolumn{4}{c}{MET}\\
\midrule
r(mT2-500)  & 0.16 & 0.05 & 14.11 \\
r(mT2-400)  & 0.44 & 0.21 & 11.13 \\
r(mT2-500/300) & 0.21 & 0.09 & 8.52 \\
r(Meff1400)  & 0.11 & 0.25 & 7.24 \\
r(M1400)  & 0.07 & 0.19 & 6.57 \\
r(mT2-300)  & 0.75 & 0.54 & 6.26 \\
r(mT2-400/300)  & 0.58 & 0.40 & 5.77 \\
r(HT900)  & 0.13 & 0.24 & 5.67 \\
r(M1800)  & 0.02 & 0.07 & 4.82 \\
r(MET420)  & 0.48 & 0.37 & 4.32 \\
\bottomrule
\end{tabular}
\end{center}
\caption {Best discriminating ratios in the MET box, with separations in units of $\sigma$,
for the comparison of LH2 vs. NM4, taking LH2 as the ``data'', 
assuming an integrated luminosity of 1000 pb$^{-1}$.\label{tab:LH2vsNM4_1000MET}}
\end{table}

\begin{table}[htp]
\begin{center}
\begin{tabular}{lrrr}
\toprule
\multicolumn{4}{c}{LH2 vs. CS7 [1000 pb$^{-1}$]}\\
\midrule
 Variable & LH2 & CS7 & Separation \\
\midrule
\multicolumn{4}{c}{MET}\\
\midrule
r(mT2-500)  & 0.27 & 0.08 & 18.87 \\
r(MET420)  & 0.48 & 0.20 & 16.73 \\
r(MET520)  & 0.21 & 0.07 & 14.49 \\
\bottomrule
\multicolumn{4}{c}{DiJet}\\
\midrule
r(4j)(3j)  & 0.20 & 0.67 & 7.30 \\
r(mT2-300)  & 0.72 & 0.31 & 6.73 \\
r(mT2-400)  & 0.53 & 0.22 & 6.26 \\
\bottomrule
\multicolumn{4}{c}{TriJet}\\
\midrule
r(mT2-500)  & 0.20 & 0.04 & 8.83 \\
r(mT2-300)  & 0.68 & 0.32 & 7.43 \\
r(mT2-400)  & 0.53 & 0.22 & 7.18 \\
\bottomrule
\multicolumn{4}{c}{Muon20}\\
\midrule
r(mT2-300)  & 0.84 & 0.35 & 1.57 \\
r(mT2-400)  & 0.60 & 0.24 & 1.32 \\
\bottomrule
\end{tabular}
\end{center}
\caption {Best discriminating ratios in each trigger box, with separations in units of $\sigma$,
for the comparison of LH2 vs. CS7, taking LH2 as the ``data'',
assuming an integrated luminosity of 1000 pb$^{-1}$.\label{tab:LH2vsCS7_1000}}
\end{table}

\begin{table}[htp]
\begin{center}
\begin{tabular}{lrrr}
\toprule
\multicolumn{4}{c}{LH2 vs. CS7 [1000 pb$^{-1}$]}\\
\midrule
 Variable & LH2 & CS7 & Separation \\
\midrule
\multicolumn{4}{c}{MET}\\
\midrule
r(mT2-500)  & 0.27 & 0.08 & 18.87 \\
r(MET420)  & 0.48 & 0.20 & 16.73 \\
r(MET520)  & 0.21 & 0.07 & 14.49 \\
r(mT2-600)  & 0.05 & 0.01 & 14.11 \\
r(mT2-500/300)  & 0.32 & 0.12 & 11.17 \\
r(mT2-500/400)  & 0.43 & 0.19 & 9.77 \\
r(mT2-600/300)  & 0.06 & 0.01 & 9.77 \\
r(mT2-400)  & 0.63 & 0.40 & 8.46 \\
r(MET320)  & 0.78 & 0.53 & 8.17 \\
\bottomrule
\end{tabular}
\end{center}
\caption {Best discriminating ratios in the MET box, with separations in units of $\sigma$,
for the comparison of LH2 vs. CS7, taking LH2 as the ``data'', 
assuming an integrated luminosity of 1000 pb$^{-1}$.\label{tab:LH2vsCS7_1000MET}}
\end{table}


\section{Detailed results for Group 2}

\subsection{LH2 vs NM4 and CS7}

This comparison is described in section~\ref{sec:8}. We treat the little
Higgs model LH2
as the data and compare to SUSY models NM4 and CS7.
Tables~\ref{tab:LH2vsNM4_1000}-\ref{tab:LH2vsCS7_1000MET} show the
best discriminating ratios for the comparison of NM4 and CS7 to the LH2 ``data''
with 1000 pb$^{-1}$.

With the better statistics of 1000 pb$^{-1}$,
Table~\ref{tab:LH2vsNM4_1000MET} illustrates even more clearly the conundrum
discussed in section ~\ref{sec:8}. The $m_{T2}$ ratio
r(mT2-500/300) shows unquestionably that the parent particle
masses in model NM4 are too small to fit the ``data''. The
\met~distribution of NM4 also appears to be too soft, with
4.3$\sigma$ significance. However the kinematic distributions for NM4
represented by r(Meff1400), r(M1400) and r(HT900) are all too hard,
with $>5 \sigma$ significance. 

The other impressive feature of these tables is that with
1000 pb$^{-1}$ we acquire several highly discriminating ratios
in the DiJet and TriJet boxes. With real data this
would provide an impressive redundancy of cross-checks, still
within the original design of our look-alike analysis.

The large number of independent highly discriminating robust ratios
seen here provide a powerful tool to resolve SUSY look-alikes from
non-SUSY look-alikes.


\section{Comparison of squark production with heavy quark production}

\subsection{smuon production versus muon production}

Let's compare the QED processes $e^+e^- \to \mu^+\mu^-$ and $e^+e^-
\to \smur\smurbar$. We will use the conventions and notation of Peskin
and Schroeder (PS) \cite{Peskin:1995ev}, and work in the approximation
that the electron and positron are massless. In this notation $p$ and
$p'$ denote the incoming 4-momenta of the electron and positron, while
$k$ and $k'$ denote the outgoing 4-momenta of the muons or smuons. The
photon 4-momentum is denoted by $q = p+p'$. We will use $m$
interchangably to denote the mass of the muon or smuon, assuming them
(in this pedagogical example) to be degenerate.

The leading order QED matrix element for $e^+e^- \to \mu^+\mu^-$ is
\be
\label{eq:mumatrixelement}
\bar{v}^{s'}(p')(-ie\gamma^{\mu})u^s(p)\left( \frac{-i}{q^2} \right) 
\bar{u}^r(k)(-ie\gamma_{\mu})v^{r'}(k')
\ee
The corresponding matrix element for $e^+e^- \to \smur\smurbar$ is
\be
\label{eq:smumatrixelement}
\bar{v}^{s'}(p')(-ie\gamma^{\mu})u^s(p)\left( \frac{-i}{q^2} \right) 
(-ie(k_{\mu}-k'_{\mu}))
\ee
In each case, we compute the squared matrix element, averaging over the
spins of the electrons. For $e^+e^- \to \mu^+\mu^-$ we also sum over the
spins of the muons, thus
\be
\label{eq:sqmumatrixelement}
\textstyle{
\frac{1}{4}\sum_{s,s',r,r'}\vert {\cal M}(s,s',r,r') \vert^2 =
\frac{e^4}{4q^4}{\rm tr}\left[ \ppslash\gamma^{\mu}\pslash\gamma^{\nu} \right] 
}
\nn\\
\textstyle{
\times {\rm tr}
\left[ (\kslash + m)\gamma_{\mu}(\kpslash - m)\gamma_{\nu} \right]
} 
\; ,
\ee
while for $e^+e^- \to \smur\smurbar$ we have
\be
\label{eq:sqsmumatrixelement}
&\hspace{-120pt}
\textstyle{
\frac{1}{4}\sum_{s,s'}\vert {\cal M}(s,s') \vert^2 = 
}
\nn \\
&\hspace{20pt}
\textstyle{
\frac{e^4}{4q^4}{\rm tr}\left[ \ppslash\gamma^{\mu}
(k_{\mu}-k'_{\mu})\pslash\gamma^{\nu}(k_{\nu}-k'_{\nu}) \right] 
}
\; .
\ee
From now on we will follow the convention of 
Ellis, Stirling and Webber (ESW) \cite{Ellis:1991qj} and use a barred
summation to denote the average over initial spins and sum 
over final spins (if any). Thus performing the traces 
(\ref{eq:sqmumatrixelement}) becomes
\be
\label{eq:trsqmumatrixelement}
\hspace*{-9pt}\textstyle{
\overline{\sum}\vert {\cal M} \vert^2 =
\frac{8e^4}{q^4} \hspace*{-3pt} \left[
(p\tdot k)(p'\tdot k')\hspace{-2pt}+\hspace{-2pt}(p\tdot k')(p'\tdot k) 
\hspace{-2pt}+\hspace{-2pt}m^2(p\tdot p') \right]
}
\ee
while (\ref{eq:sqsmumatrixelement}) becomes
\be
\label{eq:trsqsmumatrixelement}
&\hspace{-7pt}
\textstyle{
\overline{\sum}\vert {\cal M} \vert^2 =
\frac{2e^4}{q^4}\bigl[
(p\tdot p')(k\tdot k')-(p\tdot k)(p'\tdot k')-(p\tdot k')(p'\tdot k)
}
\nn\\
&\hspace{10pt}
\textstyle{
+(p\tdot k)(p'\tdot k)+(p\tdot k')(p'\tdot k')-m^2(p\tdot p')
\bigr]
}
\; .
\ee
The kinematics of both cases are identical; in the center of mass frame:
\be
p\tdot p' &= \displaystyle{\frac{s}{2},} \quad k\tdot k' 
&= \quad\frac{s}{4}(1+\beta^2) 
\, ;\nn\\
p\tdot k &= \quad p'\tdot k' &= \quad\frac{s}{4}(1-\beta\,{\rm cos}\,\theta)
\,;\\
p\tdot k' &= \quad p'\tdot k &= \quad\frac{s}{4}(1+\beta\,{\rm cos}\,\theta) 
\,,\nn
\ee
where $\theta$ is the polar angle of the final state muon or smuon, and
\be
\label{eq:betadef}
\textstyle{
\beta \equiv \sqrt{1-\frac{4
m^2}{s}}
} \; .
\ee 
Substituting, (\ref{eq:trsqmumatrixelement}) becomes
\be
\textstyle{
\overline{\sum}\vert {\cal M} \vert^2 =
e^4\left[ 2-\beta^2(1-{\rm cos}^2\theta)\right]
}
\, ,
\ee
and (\ref{eq:trsqsmumatrixelement}) becomes
\be
\textstyle{
\overline{\sum}\vert {\cal M} \vert^2 =
\frac{e^4}{2}\beta^2(1-{\rm cos}^2\theta)
}
\,.
\ee
Thus the differential cross section for
$e^+e^- \to \mu^+\mu^-$ at leading order in QED is
\be
\label{eq:mudiffcross}
\frac{d\sigma}{d({\rm cos}\,\theta )} = 
\frac{\pi\alpha^2}{2s}\beta\,\left[ 2-\beta^2(1-{\rm cos}^2\theta)\right]
\, ,
\ee
while for
$e^+e^- \to \smur\smurbar$ we get
\be
\label{eq:smudiffcross}
\frac{d\sigma}{d({\rm cos}\,\theta )} = 
\frac{\pi\alpha^2}{4s}\beta^3\,(1-{\rm cos}^2\theta)
\, ,
\ee
agreeing with Farrar and Fayet \cite{Farrar:1980uy}.

Since a Dirac muon has two complex scalar superpartners, $\smur$ and
$\smul$, from now on we will write the combined cross section for
$e^+e^- \to \smur\smurbar + \smul\smulbar$, which is just twice the
expression in (\ref{eq:smudiffcross}).  Thus we have the total cross
sections:

\be
\sigma (e^+e^- \to \mu^+\mu^-) &=& \frac{2\pi\alpha^2}{3s}\beta(3-\beta^2)
\,;\nn\\
\sigma (e^+e^- \to \smur\smurbar + \smul\smulbar) &=&
\frac{2\pi\alpha^2}{3s}\beta^3 \,.
\ee

In the high energy limit $\beta \to 1$, the leading order muon pair
cross section is exactly twice the smuon pair cross section, as noted
for example in \cite{Choi:2006mr}.

\subsection{$\mathbf{q\bar{q} \to Q\bar{Q}}$ versus $\mathbf{q\bar{q} \to \sq\sqbar}$}

From these formulae it is easy to obtain the leading order cross
sections for hadroproduction of heavy quarks/squarks from light
$q\bar{q}$ initial parton states. First we introduce a kinematic
notation more suitable for hadroproduction. The subprocess Mandelstam
invariants are given by
\be
\hat{t} &=& m^2 -\frac{\hat{s}}{2}(1-\beta\,{\rm cos}\,\theta ) \; ;\nn\\
\hat{u} &=& m^2 -\frac{\hat{s}}{2}(1+\beta\,{\rm cos}\,\theta ) \; ;\\
2m^2 &=& \hat{s} +\hat{t} +\hat{u} \; .\nn
\ee
It is convenient to use the dimensionless variables defined by ESW:
\be
\tau_1 &=& \frac{m^2-\hat{t}}{\hat{s}} \; ; \nn\\
\tau_2 &=& \frac{m^2-\hat{u}}{\hat{s}} \; ; \\
\rho &=& 1-\beta^2 \; ; \nn\\
1 &=& \tau_1 + \tau_2 \; .\nn
\ee
We change variables using
\be
\frac{d({\rm cos}\,\theta)}{d\hat{t}} = \frac{2}{\beta\hat{s}} \; .
\ee
Thus (\ref{eq:mudiffcross}), the leading order QED differential cross
section for $e^+e^- \to \mu^+\mu^-$, becomes
\be\label{eq:munice}
\frac{d\sigma}{d\hat{t}} =
\frac{2\pi\alpha^2}{\hat{s}^2}
\left[ \tau_1^2 +\tau_2^2 + \frac{\rho}{2} \right]
\; ,
\ee
while the $e^+e^- \to \smur\smurbar + \smul\smulbar$ cross section is
\be\label{eq:smunice}
\frac{d\sigma}{d\hat{t}} =
\frac{2\pi\alpha^2}{\hat{s}^2}
\left[ 1- \tau_1^2 -\tau_2^2 - \frac{\rho}{2} \right]
\; .
\ee

To convert these formulae into cross sections for the leading order
QCD subprocesses $q\bar{q} \to Q\bar{Q}$ (heavy quark production) and
$q\bar{q} \to \sqr\sqrbar + \sql\sqlbar$ (squark production), we
replace $\alpha$ by $\alpha_s$, and insert a factor of $2/9$ to
account for the color factor, averaging over initial colors, and
summing over final colors.  To make an apples--with--apples comparison
we assume that the squarks are of a different flavor than the initial
state partons; then both leading order processes arise from a single
$s$-channel diagram.

Thus we get the fully differential cross section
for $q\bar{q} \to Q\bar{Q}$:
\be
\label{eq:Qdiffcross}
&&\hspace{-24pt}
\frac{d^3\sigma}{dx_1dx_2d\hat{t}} =
\frac{4\pi\alpha_s^2}{9\hat{s}^2}
f_1(x_1)f_2(x_2)
\left[ \tau_1^2 +\tau_2^2 + \frac{\rho}{2} \right]
,
\ee
and for $q\bar{q} \to \sqr\sqrbar + \sql\sqlbar$ we have
\be
\label{eq:sqdiffcross}
&&\hspace{-31pt}
\frac{d^3\sigma}{dx_1dx_2d\hat{t}} =
\frac{4\pi\alpha_s^2}{9\hat{s}^2}
f_1(x_1)f_2(x_2)
\left[ 1 - \tau_1^2 - \tau_2^2 - \frac{\rho}{2}\right]
\ee
where $f_1(x_1)$ and $f_2(x_2)$ are the pdfs for the initial
state quark and antiquark.

We can compare the heavy quark cross section (\ref{eq:Qdiffcross})
to ESW by making the change of variables
\be
\frac{d^4\sigma}{dy_3dy_4d^2p_T} =
\frac{x_1x_2}{\pi}\times
\frac{d^3\sigma}{dx_1dx_2d\hat{t}} 
\; ,
\ee
where $y_3$, $y_4$ are the lab frame rapidities of the 
heavy quarks, and $p_T$ is the transverse momentum.
Thus
\be
\label{eq:finalQdiffcross}
&&\hspace{-10pt}
\frac{d^4\sigma}{dy_3dy_4d^2p_T} =
\frac{4\alpha_s^2}{9\hat{s}^2}
x_1f_1x_2f_2
\left[ \tau_1^2 +\tau_2^2 + \frac{\rho}{2} \right]
,
\ee
agreeing with eqn. 10.51 of ESW.

Similarly, we can compare the differential cross section
for squark production  (\ref{eq:sqdiffcross}) to
the literature. The comparison requires a couple of kinematic
identities:
\be
\label{eq:firstMident}
\hat{u}\hat{t} - m^4 &=& \frac{\hat{s}^2}{8}
\left[ 1 - \tau_1^2 - \tau_2^2 - \frac{\rho}{2} \right]
\; ;\\
\label{eq:secondMident}
\hat{u}\hat{t} - m^4 &=&
\frac{1}{4}
\left[ \hat{s}(\hat{s}-4m^2)-(\hat{u}-\hat{t})^2 \right]
\; .
\ee
Using (\ref{eq:firstMident}), we see that
the differential cross section
for squark production  (\ref{eq:sqdiffcross}) agrees
with Dawson, Eichten and Quigg (DEQ) \cite{Dawson:1983fw},
and (\ref{eq:secondMident}) shows that we agree with
Harrison and Llewellyn Smith \cite{Harrison:1982yi}.

The total cross sections are obtained by integration:
\be
&&
\sigma (q\bar{q}\to Q\bar{Q}) =
\int_0^1 dx_1dx_2 \int_{\hat{t}^{min}}^{\hat{t}^{max}}
\hspace{-10pt} d\hat{t}
\frac{d^3\sigma}{dx_1dx_2d\hat{t}}  \\
&& = \frac{4\pi\alpha_s^2}{9}\int dx_1dx_2
\frac{f_1f_2}{\hat{s}}\int_{\tau_1^{min}}^{\tau_1^{max}}
\hspace{-15pt} d\tau_1
(1-2\tau_1+2\tau_1^2+\frac{\rho}{2})
\,. \nn
\ee
Using $\tau_1^{max} = \frac{1}{2}(1+\beta )$, 
$\tau_1^{min} = \frac{1}{2}(1-\beta )$, this becomes
\be
\label{eq:finalQcross}
&& \hspace{-25pt}
\sigma (q\bar{q}\to Q\bar{Q}) = \frac{8\pi\alpha_s^2}{27}
\hspace{-4pt} \int \hspace{-2pt}
dx_1dx_2f_1f_2 \frac{(\hat{s}+2m^2)}{\hat{s}^2}\,
\beta
\ee
which agrees with the result of Combridge \cite{Combridge:1978kx}. 
The analogous total cross section for squark production is
\be
\label{eq:finalsqcross}
&& \hspace{-20pt}
\sigma (q\bar{q}\to \sqr\sqrbar + \sql\sqlbar ) = \nn\\
&&\frac{4\pi\alpha_s^2}{27}
\hspace{-4pt} \int \hspace{-2pt}
dx_1dx_2f_1f_2 \frac{(\hat{s}-4m^2)}{\hat{s}^2}\,
\beta \\
&&=\frac{4\pi\alpha_s^2}{27}
\hspace{-4pt} \int \hspace{-2pt}
dx_1dx_2f_1f_2 \frac{1}{\hat{s}}\,
\beta^3 \; ,\nn
\ee
which agrees with 
Harrison and Llewellyn Smith \cite{Harrison:1982yi}.

\subsection{$\mathbf{gg \to Q\bar{Q}}$ versus $\mathbf{gg \to \sq\sqbar}$}

For gluon fusion, we start with the leading order
matrix elements as given by ESW and DEQ. At leading order
there are three diagrams for each process, corresponding
to the $s$, $t$ and $u$ channels, and an additional
gluon seagull diagram for the squark case. In the $t$ and $u$ channels
we are not only producing particles of different spins but
also exchanging particles of different spins.  
For $gg \to Q\bar{Q}$ 
we have (see Table 10.2 of \cite{Ellis:1991qj}):
\be
\label{eq:ellisggme}
&&\hspace{-32pt}
\textstyle{
\overline{\sum}\vert {\cal M} \vert^2 =
g_s^4\left( \frac{1}{6\tau_1\tau_2}-\frac{3}{8}\right)
\left( \tau_1^2 +\tau_2^2 +\rho -\frac{\rho^2}{4\tau_1\tau_2} \right)
,}
\ee
while for $gg \to \sqr\sqrbar + \sql\sqlbar$ we have 
(see eqn. 3.26 of \cite{Dawson:1983fw}):
\be
\label{eq:quiggggme}
&&\hspace{-32pt}
\textstyle{
\overline{\sum}\vert {\cal M} \vert^2 =
g_s^4\left( \frac{7}{48} + \frac{3(\hat{u}-\hat{t})^2}{16\hat{s}^2} \right)
}
\nn\\
&&\hspace{-15pt}
\times\textstyle{
\left[ 1 + \frac{2m^2\hat{t}}{(\hat{t}-m^2)^2}
+\frac{2m^2\hat{u}}{(\hat{u}-m^2)^2}
+\frac{4m^4}{(\hat{t}-m^2)(\hat{u}-m^2)} \right]
.}
\ee
Converting to the ESW kinematic variables and
expanding, (\ref{eq:quiggggme}) becomes
\be
\label{eq:quiggggmeb}
&&\hspace{-8pt}
\textstyle{
\overline{\sum}\vert {\cal M} \vert^2 =
}
\\
&&\hspace{0pt}
\textstyle{
g_s^4\left[ \frac{1}{3} + \frac{3}{8}\rho - \frac{3}{4}\tau_1\tau_2
-\frac{\rho}{6\tau_1\tau_2}-\frac{3\rho^2}{32\tau_1\tau_2}
+\frac{\rho^3}{24\tau_1^2\tau_2^2} \right]
.\nn
}
\ee
This can be refactored into
\be
\label{eq:quiggggmec}
&&\hspace{-15pt}
\textstyle{
\overline{\sum}\vert {\cal M} \vert^2 =
}
\\
&&\hspace{5pt}
\textstyle{
g_s^4\left( \frac{1}{6\tau_1\tau_2}-\frac{3}{8}\right)
\left( 1 - \tau_1^2 - \tau_2^2 - \rho + \frac{\rho^2}{4\tau_1\tau_2} \right)
.\nn
}
\ee
Notice that the sum of the squared matrix elements for heavy quark
and squark production with the same mass has a very simple form:
\be
\label{eq:simplesusy}
&&\hspace{-55pt}
\textstyle{
\overline{\sum}\vert {\cal M} \vert^2 
(gg\to Q\bar{Q} + \sqr\sqrbar + \sql\sqlbar)
=
}
\\
&&\hspace{45pt}
\textstyle{
g_s^4\left( \frac{1}{6\tau_1\tau_2}-\frac{3}{8}\right)
.\nn
}
\ee
An analogous simplification also occurs in the $q\bar{q}$
initiated production, as is obvious from comparing
(\ref{eq:munice}) to (\ref{eq:smunice}).

It does not appear that these elegant SUSY relations have
ever been noticed in the literature! This may be because these
are not, strictly speaking, MSSM relations. In the MSSM,
electroweak symmetry breaking does not occur in the SUSY limit
where the soft breaking terms are turned off. Since the MSSM
fermions are massless in the absence of EWSB, there are no
MSSM cross section relations between degenerate heavy quarks and squarks.
In the simple processes that we considered above, the
SUSY limit actually corresponds to some more generic
vectorlike SUSY theory.

The corresponding fully differential cross section
for $gg \to Q\bar{Q}$ is
\be
\label{eq:ggQdiffcross}
&&\hspace{-15pt}
\frac{d^3\sigma}{dx_1dx_2d\hat{t}} =
\frac{\pi\alpha_s^2}{\hat{s}^2}
f_1(x_1)f_2(x_2)
\\
&&
\times\left( \frac{1}{6\tau_1\tau_2}-\frac{3}{8}\right)
\left( \tau_1^2 +\tau_2^2 +\rho -\frac{\rho^2}{4\tau_1\tau_2} \right)
,\nn
\ee
and for $gg \to \sqr\sqrbar + \sql\sqlbar$ we have
\be
\label{eq:ggsqdiffcross}
&&\hspace{-22pt}
\frac{d^3\sigma}{dx_1dx_2d\hat{t}} =
\frac{\pi\alpha_s^2}{\hat{s}^2}
f_1(x_1)f_2(x_2)
\\
&&
\times\left( \frac{1}{6\tau_1\tau_2}-\frac{3}{8}\right)
\left( 1 - \tau_1^2 - \tau_2^2 - \rho + \frac{\rho^2}{4\tau_1\tau_2} \right)
.\nn
\ee
The total cross sections are
\be
\label{eq:ggQtotalcross}
&&\hspace{-10pt}
\sigma (gg\to Q\bar{Q}) = \frac{\pi\alpha_s^2}{48}
\hspace{-1pt} \int \hspace{-1pt}
dx_1dx_2\frac{f_1f_2}{\hat{s}}
\Bigl[ \; -59\beta 
\nn\\
&&\hspace{0pt}
+31\beta^3 +[32 - 16\beta^2 + (1-\beta^2)^2]\,{\rm ln}\,\frac{1+\beta}{1-\beta}
\Bigr]
\,;
\\
\label{eq:ggsqtotalcross}
&&\hspace{-10pt}
\sigma (gg\to \sqr\sqrbar \hspace{-1pt}+\hspace{-1pt} \sql\sqlbar ) 
= \frac{\pi\alpha_s^2}{48}
\hspace{-2pt} \int \hspace{-2pt}
dx_1dx_2\frac{f_1f_2}{\hat{s}}
\Bigl[ \; 41\beta 
\nn\\
&&\hspace{0pt}
-31\beta^3 -[16(1-\beta^2) + (1-\beta^2)^2]\,{\rm ln}\,\frac{1+\beta}{1-\beta}
\Bigr]
\,.
\ee

\subsection{$\mathbf{gq \to QW}$ versus $\mathbf{gq \to \sq\tilde{W}}$}

The process $gq \to  QW$ contributes to single top production in the 
Standard Model via $qb \to tW^-$. At leading order there is both an 
$s$-channel and a $t$-channel diagram. This can be compared to the
SUSY process $gq \to \sq\tilde{W}$, i.e., the associated 
production of a squark with a Wino. Again in the corresponding $t$-channel
diagrams we are exchanging particles of different spins (a quark versus
a squark).

This process obviously cares about EWSB and the fact that we have chiral
fermions rather than vectorlike ones. Thus as already explained above we
do not expect an elegant SUSY limit relating the two processes.
Indeed one observes intrinsic differences 
already at the level of comparing the $W$ and Wino decay widths
into $Q\bar{q}$ and $\sq\bar{q}$, respectively. Assuming that these
decays were kinematically allowed, we can extract the leading order
expressions from eqn. 5.15 of \cite{Greiner:1993qp} and
eqn. B.88a of \cite{Baer:2006rs}, in the limit that we neglect the light quark
mass:
\be
&&\hspace{-25pt}
\Gamma (W\to Q\bar{q}) = \frac{3g_W^2}{48\pi}m_W
\hspace{-2pt}
\left[1+\frac{m^2}{2m_W^2}\right]
\hspace{-4pt}
\left[1-\frac{m^2}{m_W^2}\right]^2
\hspace{-4pt}
;
\\
&&\hspace{-25pt}
\Gamma (\tilde{W}\to \sq\bar{q}) = \frac{3g_W^2}{48\pi}m_{\tilde{W}}
\hspace{-2pt}
\left[\frac{3}{2}\right]
\hspace{-4pt}
\left[1-\frac{m^2}{m_{\tilde{W}}^2}\right]^2
\hspace{-4pt}
,
\ee
where as before $m$ denotes the heavy quark or squark mass.
These formulae coincide in the limit $m = m_W = m_{\tilde{W}}$,
but this is a kinematic limit, not a SUSY limit.

The fully differential cross section for $gq \to QW$ was computed at
leading order by Halzen and Kim \cite{Halzen:1987qm}, while the leading order
cross section for $gq \to \sq\tilde{W}$ is given in DEQ \cite{Dawson:1983fw}.
We convert these expressions to ESW notation, introducing a new
dimensionless variable $\delta$:
\be
\delta \equiv \frac{m_W^2 - m^2}{\hat{s}} \; ,
\ee
with $m_W$ replaced by $m_{\tilde{W}}$ in the SUSY case.
We also replace (\ref{eq:betadef}) by the definition of $\beta$
appropriate for two unequal mass final state particles:
\be
\beta \equiv \sqrt{\left[1-\frac{(m+m_W)^2}{\hat{s}}\right]
\left[1-\frac{(m-m_W)^2}{\hat{s}}\right]}
\, .
\ee

Thus for $gq \to QW$ we obtain
\be
\label{eq:gqQWdiffcross}
&&\hspace{-5pt}
\frac{d^3\sigma}{dx_1dx_2d\hat{t}} =
\frac{g_W^2\alpha_s}{48\hat{s}^2}
f_1(x_1)f_2(x_2)
\Biggl[
-1 + \frac{4\delta}{\rho + 4\delta}
\\
&&
+
\left[\frac{3}{2}-\frac{2\delta}{\rho+4\delta}\right]
\left[ 2\delta + \tau_1 +\frac{1-2\delta (1-\delta )}{\tau_1}
+\frac{\rho\delta}{2\tau_1^2} \right]
\Biggr]
,\nn
\ee
and for $gq \to \sq\tilde{W}$ we have
\be
\label{eq:gqsqWinodiffcross}
&&\hspace{-35pt}
\frac{d^3\sigma}{dx_1dx_2d\hat{t}} =
\frac{g_W^2\alpha_s}{48\hat{s}^2}
f_1(x_1)f_2(x_2)
\\
&&
\times\Biggl[ 1
-2\delta - \tau_1 +\frac{2\delta (1-\delta )}{\tau_1}
-\frac{\rho\delta}{2\tau_1^2}
\Biggr]
.\nn
\ee

The total cross sections are
\be
\label{eq:gqQWtotalcross}
&&\hspace{0pt}
\sigma (gq \to QW) = \frac{g_W^2\alpha_s}{48}
\hspace{-1pt} \int \hspace{-1pt}
dx_1dx_2\frac{f_1f_2}{\hat{s}}
\frac{1}{4[(1+\delta )^2-\beta^2]}
\nn\\
&&\hspace{0pt}
\Biggl[
\beta\left[ -1 + (1-21\delta)\beta^2 +\delta (31+13\delta +21\delta^2) \right]
\\
&&\hspace{0pt}
+ 2\left[ 1\hspace{-1pt}-\hspace{-1pt}2\delta(1\hspace{-1pt}-\hspace{-1pt}\delta )\right]
\left[ 3(1\hspace{-1pt}-\hspace{-1pt}\beta^2) \hspace{-1pt}+\hspace{-1pt} 
\delta (2\hspace{-1pt}+\hspace{-1pt}3\delta )\right]
{\rm ln}\,\frac{1\hspace{-1pt}+\hspace{-1pt}\beta\hspace{-1pt} -\hspace{-1pt}\delta}
{1\hspace{-1pt}-\hspace{-1pt}\beta\hspace{-1pt} -\hspace{-1pt}\delta}
\Biggr]
;
\nn
\\
\label{eq:gqsqWinototalcross}
&&\hspace{0pt}
\sigma (gq \to \sq\tilde{W}) = \frac{g_W^2\alpha_s}{48}
\hspace{-1pt} \int \hspace{-1pt}
dx_1dx_2\frac{f_1f_2}{\hat{s}}
\nn\\
&&\hspace{5pt}
\times\frac{1}{2}\left[
\beta -7\delta\beta +4\delta (1-\delta )
{\rm ln}\,\frac{1\hspace{-1pt}+\hspace{-1pt}\beta\hspace{-1pt} -\hspace{-1pt}\delta}
{1\hspace{-1pt}-\hspace{-1pt}\beta\hspace{-1pt} -\hspace{-1pt}\delta}
\right]
,
\ee
where the integrals were performed using the kinematic relations:
\be
\tau_1^{max} &=& \frac{1}{2}(1+\beta - \delta ) \; ;
\\
\tau_1^{min} &=& \frac{1}{2}(1-\beta - \delta ) \; .
\ee

To compare the total cross sections, it is much simpler to consider
the special limit $m = m_W = m_{\tilde{W}}$, i.e., the
case $\delta = 0$.
Then (\ref{eq:gqQWtotalcross}) and (\ref{eq:gqsqWinototalcross})
reduce to:
\be
\label{eq:gqQWtotalcrossdeltazero}
&&\hspace{-15pt}
\sigma (gq \to QW) = 
\nn\\
&&\hspace{-5pt}
\frac{g_W^2\alpha_s}{48}
\hspace{-1pt} \int \hspace{-1pt}
dx_1dx_2\frac{f_1f_2}{\hat{s}}
\frac{1}{4}\left[
-\beta + 6\,{\rm ln}\,\frac{1+\beta}{1-\beta} \right]
\,;
\\
\label{eq:gqsqWinototalcrossdeltazero}
&&\hspace{-15pt}
\sigma (gq \to \sq\tilde{W} ) = 
\nn\\
&&\hspace{-5pt}
\frac{g_W^2\alpha_s}{48}
\hspace{-1pt} \int \hspace{-1pt}
dx_1dx_2\frac{f_1f_2}{\hat{s}}
\frac{1}{2}\,\beta
\;,
\ee  
with $\beta$ given by (\ref{eq:betadef}).

Note that the non-SUSY total cross section is much larger
than the SUSY cross section for partners of the same mass.
The ratio of the cross sections at fixed $\hat{s}$ is
$11/2$ at threshold ($\beta = 0$) and rises monotonically from there,
reaching 9.3 for $\beta = 0.9$ and diverging
logarithmically in the limit $\hat{s}/m^2\to\infty$.



\newpage


\clearpage

\end{document}